\documentclass[9pt]{sigplanconf}

\newif\iftechreport
\techreporttrue

\usepackage{natbib}

\usepackage{graphicx}

\usepackage{paralist}

\usepackage{csquotes}

\usepackage{microtype}

\iftechreport
\usepackage{hyperref}
\else
\usepackage[bookmarksdepth=2]{hyperref}
\fi

\usepackage[all]{hypcap}

\makeatletter
\g@addto@macro{\UrlBreaks}{\UrlOrds}
\makeatother

\usepackage[english]{babel}
\usepackage[l2tabu,orthodox]{nag}
\usepackage{hyperref}
\usepackage{listings}
\usepackage{mathpartir}

\usepackage[cmex10]{amsmath}
\usepackage{array}
\usepackage{mathrsfs}
\usepackage{amssymb}
\usepackage{amsthm}
\usepackage{thmtools}
\usepackage{thm-restate}
\usepackage[only,llbracket,rrbracket,llparenthesis,rrparenthesis,lightning,Lbag,Rbag]{stmaryrd}  
\usepackage{mathtools}
\usepackage{synttree}
\iftechreport
\usepackage{fixltx2e}
\else
\fi
\usepackage{acronym}
\usepackage{pbox}
\usepackage{microtype}
\usepackage{cancel}
\usepackage{mathrsfs}
\usepackage[textsize=small%
]{todonotes}
\usepackage[only,llbracket,rrbracket,llparenthesis,rrparenthesis,lightning,Lbag,Rbag]{stmaryrd}
\usepackage{bbm,dsfont}

\usepackage[normalem]{ulem}

\makeatletter
\newcommand{\@chapapp}{\relax}%
\makeatother
\usepackage[header]{appendix}

\usepackage{xspace}
\usepackage{flushend}
\usepackage{galois}
\usepackage{mdframed}
\usepackage{tikz-cd}
\usepackage{etoolbox}
\usepackage{ellipsis}

\usepackage{balance}

\iftechreport
\usepackage{cleveref}
\else
\usepackage[capitalise,nameinlink]{cleveref}
\fi
\crefname{section}{Sect.}{Sect.}
\Crefname{section}{Section}{Sections}
\crefname{figure}{Fig.}{Fig.}
\Crefname{figure}{Figure}{Figures}

\crefname{appsec}{}{}

\let\oldnewcommand\newcommand
\let\newcommand\newrobustcmd

\newcommand{\myparagraph}[1]{\paragraph{#1.}}

\newcommand{\dt}[1]{\emph{\textbf{#1}}} %

\newcommand{\letexp}{\text{let }}
\newcommand{\inexp}{\text{ in}}
\newcommand{\letin}[1]{\letexp  #1  \inexp }

\newcommand{\justif}[1]{\ensuremath{\Lbag \text{#1} \Rbag}}

\newtheoremstyle{FavStyle}%
  {\topsep}%
  {\topsep}%
  {\rmfamily}%
  {0pt}%
  {\bfseries}%
  {.}%
  { }%
  {\thmname{#1 \thmnumber{#2} #3}}%

\theoremstyle{FavStyle}

\crefname{mylem}{Lemma}{Lemmas} %
\crefname{mytheo}{Theorem}{Theorems}
\crefname{mycor}{Corollary}{Corollaries}
\crefname{appsec}{Appendix}{Appendices}

\newcommand{\skipcom}{\text{{\small\bfseries skip}}\xspace}
\newcommand{\ifcom}{\text{{\small\bfseries if}}\xspace}
\newcommand{\thencom}{\text{{\small\bfseries then}}\xspace}
\newcommand{\elsecom}{\text{{\small\bfseries else}}\xspace}
\newcommand{\whilecom}{\text{{\small\bfseries while}}\xspace}
\newcommand{\docom}{\text{{\small\bfseries do}}\xspace}
\newcommand{\bcmp}{\mathbin{\operatorname{cmp}}}
\newcommand{\tinyifcomoz}{\text{{\bfseries if} b c\textsubscript{1} {\bfseries
      else} c\textsubscript{0}}\xspace} 
\newcommand{\sizeval}{\infty}
\newcommand{\Values}{\ensuremath{\operatorname{\textbf{Val}}}}
\newcommand{\psetValues}{\pset{\Values}}
\newcommand{\psetpsetValues}{\pset{\psetValues}}
\newcommand{\val}{v}
\newcommand{\Val}{V}
\newcommand{\BVal}{\mathbb{\Val}}
\newcommand{\States}{\ensuremath{\operatorname{\textbf{States}}}}
\newcommand{\psetStates}{\pset{\States}}
\newcommand{\psetpsetStates}{\pset{\psetStates}}
\newcommand{\Traces}{\ensuremath{\operatorname{\textbf{Trc}}}}
\newcommand{\InitTraces}{\ensuremath{\operatorname{\textbf{IniTrc}}}}
\newcommand{\pset}[1]{\mathcal{P}(#1)}
\newcommand{\psetTraces}{\pset{\Traces}}
\newcommand{\psetpsetTraces}{\pset{\psetTraces}}
\newcommand{\st}{\sigma}
\newcommand{\stt}{\tau} %
\newcommand{\St}{\Sigma}
\newcommand{\BSt}{\mathbb{S}}
\newcommand{\Statesseq}{\States^\ast}  %
\newcommand{\psetStatesseq}{\pset{\Statesseq}}

\newcommand{\tr}{t}

\newcommand{\Tr}{T}
\newcommand{\BTr}{\mathbb{T}} 
\newcommand{\Trprog}{T_{\prog}}

\newcommand{\cardinal}[1]{\left\vert{#1}\right\vert}
\newcommand{\interv}[2]{\left[ #1, #2 \right]}
\newcommand{\existsu}{\exists!}

\newcommand{\semins}[2]{ \llbracket #1  \rrbracket #2}  %
\newcommand{\semexp}[2]{ \llbracket #1 \rrbracket #2  } %
\newcommand{\seminsc}[2]{\llbrace #1 \rrbrace #2}
\newcommand{\semexpc}[2]{ \llbrace #1 \rrbrace #2}

\newcommand{\guardop}{\operatorname{grd}}
\newcommand{\guardc}[2]{\seminsc{\guardop^{#1}}{#2}}

\newcommand{\seminshc}[2]{\llparenthesis #1 \rrparenthesis #2}
\newcommand{\guardhc}[2]{\seminshc{\guardop^{#1}}{#2}}

\newcommand{\seminshcabs}[2]{\llparenthesis #1 \rrparenthesis^\sharp #2}
\newcommand{\guardhcabs}[2]{\seminshcabs{\guardop^{#1}}{#2}}

\newcommand{\dotpreccurlyeq}{\mathbin{\dot{\preccurlyeq}}}

\newcommand{\concobj}{\boldsymbol{\mathcal{C}}}
\newcommand{\psetconcobj}{\pset{\concobj}}
\newcommand{\psetpsetconcobj}{\pset{\psetconcobj}}
\newcommand{\Obj}{C}
\newcommand{\BObj}{\mathbb{\Obj}}
\newcommand{\absobj}{\boldsymbol{\mathcal{A}}}
\newcommand{\psetabsobj}{\pset{\absobj}}
\newcommand{\aObj}{a}

\newcommand{\hppop}{\operatorname{hpp}}
\newcommand{\alphahpp}{\alpha_{\hppop}}
\newcommand{\gammahpp}{\gamma_{\hppop}}
\newcommand{\eltwop}{\operatorname{elt}}
\newcommand{\alphaelt}{\alpha_{\eltwop}}
\newcommand{\gammaelt}{\gamma_{\eltwop}}
\newcommand{\supremop}{\eltwop}
\newcommand{\alphasup}{\alpha_{\supremop}}
\newcommand{\gammasup}{\gamma_{\supremop}}
\newcommand{\cardvop}{\operatorname{crdval}}
\newcommand{\alphacardv}{\alpha_{\cardvop}}
\newcommand{\gammacardv}{\gamma_{\cardvop}}
\newcommand{\agreeo}{\operatorname{agree}_x} 
\newcommand{\alphaagreeo}{\alpha_{\agreeo}}
\newcommand{\gammaagreeo}{\gamma_{\agreeo}}
\newcommand{\agreev}{\operatorname{agree}}
\newcommand{\alphaagreev}{\alpha_{\agreev}}
\newcommand{\gammaagreev}{\gamma_{\agreev}}

\newcommand{\true}{\operatorname{tt}}
\newcommand{\false}{\operatorname{ff}}
\newcommand{\abstruth}{\{\true,\false\}}
\newcommand{\absbool}{\operatorname{bv}}
\newcommand{\abstruthleq}{\mathbin{\Longleftarrow}}

\newcommand{\prog}{\operatorname{P}}
\newcommand{\varprog}{\operatorname{Var}_{\prog}}
\newcommand{\Ga}{\Gamma}
\newcommand{\mlsLat}{\mathcal{L}}
\newcommand{\mlsl}{l}
\newcommand{\mlslatleq}{\sqsubseteq}
\newcommand{\mlslatjoin}{\sqcup}
\newcommand{\mlslatmeet}{\sqcap}
\newcommand{\dotmlslatleq}{\mathbin{\dot{\mlslatleq}}}
\newcommand{\dotmlslatjoin}{\mathbin{\dot{\mlslatjoin}}}
\newcommand{\mlsbot}{\bot_{\mlsLat}}

\newcommand{\Mod}{\operatorname{Mod}}

\newcommand{\iagreeop}{\models_{\Ga}}
\newcommand{\iagree}[2]{#1 \iagreeop #2}
\newcommand{\alttr}{r} \newcommand{\altTr}{R}
\newcommand{\iagreerl}{\iagree{\altTr}{l}}

\newcommand{\iagreel}[1]{\iagree{#1}{l}}
\newcommand{\semexpi}[2]{ \llbracket #1\rrbracket_{\operatorname{pre}}#2} %
\let\oldshortmid\shortmid

\renewcommand{\shortmid}{\mathord{\oldshortmid}}

\newcommand{\abscardValues}{\interv{0}{\sizeval}}
\newcommand{\abscardval}{n}
\newcommand{\varx}{x}
\DeclareSymbolFont{lasy}{U}{lasy}{m}{n}
\SetSymbolFont{lasy}{bold}{U}{lasy}{b}{n}
\DeclareMathSymbol\leadsto{\mathrel}{lasy}{"3B}

\newcommand{\cardcons}[3]{#1 \leadsto #2 \# #3 }
\newcommand{\cardconslxn}{\cardcons{\mlsl}{\varx}{\abscardval}}
\newcommand{\cardconslx}[1]{\cardcons{\mlsl}{\varx}{#1}}
\newcommand{\cardconsset}{\mathscr{C}}
\newcommand{\cardLat}{\operatorname{Card}}
\newcommand{\cardlatleq}{\mathbin{{\sqsubseteq}^\sharp}}
\newcommand{\cardlatjoin}{\mathbin{{\sqcup}^\sharp}}
\newcommand{\cardlatbigjoin}{\mathop{{\bigsqcup}^\sharp}}
\newcommand{\dotcardlatleq}{\mathbin{\dot{\sqsubseteq}^\sharp}}

\newcommand{\projtomls}[2]{\pi^{#1}{(#2)}}
\newcommand{\projtomlsl}[1]{\projtomls{\mlsl}{#1}}
\newcommand{\cardtrop}{\operatorname{crdtr}}
\newcommand{\alphacardtr}{\alpha_{\cardtrop}}
\newcommand{\gammacardtr}{\gamma_{\cardtrop}}

\newcommand{\obsexpop}{\mathcal{O}}
\newcommand{\obsexpc}[3]{\obsexpop^{#2} \llbrace #1 \rrbrace #3} 
\newcommand{\obsexpcl}[2]{\obsexpop^{\mlsl}\llbrace #1 \rrbrace #2}
\newcommand{\obsexphc}[3]{\obsexpop^{#2} \llparenthesis#1 \rrparenthesis #3} 
\newcommand{\obsexphcl}[2]{\obsexpop^{\mlsl} \llparenthesis #1  \rrparenthesis
  #2}

\newcommand{\absobsexpl}[2]{\obsexpop^{\mlsl}_{C} \llparenthesis #1
  \rrparenthesis^{\sharp} #2}

\newcommand{\cardlatadd}[1]{\mathbin{{\sqcup}^\sharp}_{\operatorname{add}(#1)}}

\newcommand{\dotleq}{\mathbin{\dot{\leq}}}

\newcommand{\depcons}[2]{#1 \leadsto #2}
\newcommand{\depconslx}{\depcons{\mlsl}{\varx}}
\newcommand{\depconsset}{\mathscr{D}}
\newcommand{\depLat}{\operatorname{Dep}}
\newcommand{\deplatleq}{\mathbin{{\sqsubseteq}^\natural}}
\newcommand{\deplatjoin}{\mathbin{{\sqcup}^\natural}}
\newcommand{\deplatbigjoin}{\mathop{{\bigsqcup}^\natural}}
\newcommand{\dotdeplatleq}{\mathbin{\dot{\sqsubseteq}^\natural}}

\renewcommand{\impliedby}{\mathbin{\Longleftarrow}}

\newcommand{\deptrop}{\operatorname{deptr}}
\newcommand{\alphadeptr}{\alpha_{\deptrop}}
\newcommand{\gammadeptr}{\gamma_{\deptrop}}

\newcommand{\seminshcdepabs}[2]{\llparenthesis #1 \rrparenthesis^\natural #2}
\newcommand{\guardhcdepabs}[2]{\seminshcdepabs{\guardop^{#1}}{#2}}

\newcommand{\absobsexpdep}[3]{\obsexpop^{#2}_{D} \llparenthesis#1
  \rrparenthesis^{\natural} #3}  
\newcommand{\absobsexpdepl}[2]{\obsexpop^{\mlsl}_{D} \llparenthesis #1
  \rrparenthesis^{\natural} #2}

\newcommand{\loneopv}{\operatorname{lqone}}
\newcommand{\alphalone}{\alpha_{\loneopv}}
\newcommand{\gammalone}{\gamma_{\loneopv}}
\newcommand{\loneopcc}{\operatorname{lqonecc}}
\newcommand{\alphalonecc}{\alpha_{\loneopcc}}
\newcommand{\gammalonecc}{\gamma_{\loneopcc}}

\newcommand{\intervals}{\operatorname{Int}}
\newcommand{\statesint}{\operatorname{StInt}}
\newcommand{\dotleqintervals}{\mathbin{\dot{\leq}^{\sharp,\intervals}}}
\newcommand{\alphaint}{\alpha^{\intervals}}
\newcommand{\gammaint}{\gamma^{\intervals}}

\newcommand{\redtoint}{\operatorname{toint}}
\newcommand{\redtocard}{\operatorname{tocard}}
\newcommand{\redtodep}{\operatorname{todep}}
\newcommand{\sizeint}{\operatorname{size}}

\newcommand{\alphatohs}{\alpha_{\operatorname{hs}}}
\newcommand{\gammatohs}{\gamma_{\operatorname{hs}}}

\newcommand{\mincap}{\mathcal{ML}}
\newcommand{\mincapl}{\mincap_{\mlsl}}
\newcommand{\kparam}{k}
\newcommand{\secreqop}{\operatorname{SR}}
\newcommand{\secreq}[3]{\secreqop(#1,#2,#3)}
\newcommand{\secreql}[2]{\secreq{\mlsl}{#1}{#2}}
\newcommand{\secreqlkx}{\secreq{\mlsl}{\kparam}{\varx}}

\definecolor{Brown}{cmyk}{0,0.81,1,0.60}
\definecolor{OliveGreen}{cmyk}{0.64,0,0.95,0.40}
\definecolor{CadetBlue}{cmyk}{0.62,0.57,0.23,0}
\definecolor{lightlightgray}{gray}{0.9}

\lstdefinestyle{simple}{
escapeinside={@}{@}, float=htp, frame=Lrb,
language=C,                             %
basicstyle=\ttfamily,                   %
keywordstyle=\bfseries\color{OliveGreen},        %
commentstyle={\color{gray}}, %
numbers=left,                           %
numberstyle=\tiny,                      %
stepnumber=1,                           %
tabsize=2,                              %
captionpos=b,                           %
breakatwhitespace=false,                %
showspaces=false,                       %
showtabs=false,                         %
morekeywords={label,output, stop,then},
literate=%
{<=}{{\tiny$\leq$}}1%
{>=}{{\tiny$\geq$}}1%
}

\newcommand{\hbra}{
  \hbox to \columnwidth{\vrule width0.3mm height 1.8mm depth-0.3mm
    \leaders\hrule height1.8mm depth-1.5mm\hfill
    \vrule width0.3mm height 1.8mm depth-0.3mm}}
\newcommand{\hket}{
  \hbox to \columnwidth{\vrule width0.3mm height1.5mm
    \leaders\hrule height0.3mm\hfill
    \vrule width0.3mm height1.5mm}}

\newcommand{\ratio}{.35}

\newenvironment{tdisplay}[1]{\medskip
    \noindent\textbf{\normalsize #1}\\[-.3ex]
    \hbra\\[-.4ex]
  }{\par\hket
  \medskip}

\RequirePackage{scalerel}
\RequirePackage{accsupp}
\newcommand*{\llbrace}{%
  \BeginAccSupp{method=hex,unicode,ActualText=2983}%
    \textnormal{\usefont{OMS}{lmr}{m}{n}\char102}%
    \mathchoice{\mkern-4.05mu}{\mkern-4.05mu}{\mkern-4.3mu}{\mkern-4.8mu}%
    \textnormal{\usefont{OMS}{lmr}{m}{n}\char106}%
  \EndAccSupp{}%
}
\newcommand*{\rrbrace}{%
  \BeginAccSupp{method=hex,unicode,ActualText=2984}%
    \textnormal{\usefont{OMS}{lmr}{m}{n}\char106}%
    \mathchoice{\mkern-4.05mu}{\mkern-4.05mu}{\mkern-4.3mu}{\mkern-4.8mu}%
    \textnormal{\usefont{OMS}{lmr}{m}{n}\char103}%
  \EndAccSupp{}%
}

\let\newcommand\oldnewcommand

\begin{document}
\iftechreport
\else
\toappear{}
\fi

\setlength{\pdfpageheight}{\paperheight}
\setlength{\pdfpagewidth}{\paperwidth}

\iftechreport
\CopyrightYear{2017} 
\conferenceinfo{POPL '17,}{January 18 - 20, 2017, Paris, France}
\copyrightdata{978-1-4503-4660-3/17/01}
\copyrightdoi{http://dx.doi.org/10.1145/3009837.3009889}
\else
\fi

\iftechreport
\publicationrights{licensed}     %
\else
\fi

\titlebanner{banner above paper title}        %
\preprintfooter{\today} %

\title{Hypercollecting Semantics \\ and its Application
  to Static Analysis of Information Flow}

\authorinfo{Mounir Assaf}{Stevens Institute of
  Technology,\\ Hoboken, US}{first.last@stevens.edu}
\authorinfo{ David A. Naumann}{Stevens Institute of
  Technology,\\ Hoboken, US}{first.last@stevens.edu}
\authorinfo{Julien Signoles}{Software Reliability and Security Lab,\\ CEA LIST,
  Saclay, FR}{first.last@cea.fr}
\authorinfo{{\'E}ric Totel}{CIDRE, CentraleSup{\'e}lec,\\ Rennes,
  FR}{first.last@centralesupelec.fr} 
\authorinfo{Fr{\'e}d{\'e}ric
  Tronel}{CIDRE, CentraleSup{\'e}lec,\\ Rennes, FR}{first.last@centralesupelec.fr}
\hypersetup{
 pdfauthor={Mounir Assaf} {David A. Naumann} {Julien Signoles} {Eric Totel}
 {Frederic Tronel},
 pdftitle={Hypercollecting Semantics and its Application to Static Analysis of
   Information Flow}, 
 pdfkeywords={Information Flow} {Static Analysis} {Hyperproperties} {Security}
 {Abstract Interpretation} {Hypercollecting Semantics} {Calculational derivation}
 {Qualitative Information Flow} {Quantitative Information Flow}, 
}

\maketitle

\begin{abstract}
We show how static analysis for secure information flow can be expressed and
proved correct entirely within the framework of abstract interpretation.
The key idea is to define a Galois connection that directly approximates the
hyperproperty of interest. 
To enable use of such Galois connections, we introduce a fixpoint
characterisation of hypercollecting semantics, i.e. a ``set of sets"
transformer. 
This makes it possible to systematically derive static analyses for
hyperproperties entirely within the calculational framework of abstract
interpretation. 
We evaluate this technique by deriving example static analyses.
For qualitative information flow, we derive a dependence analysis similar to
the logic of Amtoft and Banerjee (SAS'04) and the type system of Hunt and Sands 
(POPL'06). 
For quantitative information flow, we derive a novel cardinality analysis
that bounds the leakage conveyed by a program instead of simply deciding
whether it exists. 
This encompasses problems that are hypersafety but not $k$-safety.
We put the framework to use and introduce variations that achieve precision rivalling the 
most recent and precise static analyses for information flow.
\end{abstract}



\section{Introduction}

Most static analyses tell something about all executions of a program.
This is needed, for example, to validate compiler optimizations. 
Functional correctness is also formulated in terms of 
a predicate on observable behaviours, i.e.\  more or less abstract execution traces:
A program is correct if all its traces satisfy the predicate.  
By contrast with such \emph{trace properties}, extensional definitions of
dependences involve more than one trace. 
To express that the final value of a variable $x$ may
depend only on the initial value of a variable $y$, the requirement---known as
\emph{noninterference} in the security literature \citep{SM03}---is 
that any two traces with the same initial value for $y$ result in the same
final value for $x$.    
Sophisticated information flow policies allow dependences subject to quantitative
bounds---and their formalisations involve more than two traces, sometimes unboundedly
many. 

For secure information flow formulated as decision problems, the theory of
\emph{hyperproperties} classifies the simplest form of noninterference as
\emph{2-safety} and some quantitative flow properties as
\emph{hypersafety properties} \citep{CS10}.   
A number of approaches have been explored for analysis of dependences, including
type systems, program logics, and dependence graphs.
Several works have used abstract interpretation in some way.      
One approach to 2-safety is by forming a \emph{product program} that encodes
execution pairs~\citep{BAR04,TA05,DHS05}, thereby reducing the problem to
ordinary safety which can be checked by abstract interpretation~\citep{KSF13} or
other means.  
Alternatively, a 2-safety property can be checked by dedicated analyses which
may rely in part on ordinary abstract interpretations for trace
properties~\citep{ABB06}.   

The theory of abstract interpretation serves to specify and guide the design of
static analyses. 
It is well known that effective application of the theory requires choosing an
appropriate notion of observable behaviour for the property of interest
\citep{Cou02,Bal.12,Bal.15}. 
Once a notion of ``trace'' is chosen, one has a program semantics and ``all
executions'' can be formalized in terms of \emph{collecting semantics}, which
can be used to define a trace property of interest, and thus to specify an
abstract interpretation \citep{CC77,CC79,Cou99}.  

The foundation of abstract interpretation is quite general, based on Galois
connections between semantic domains on which collecting semantics is defined. 
\citet{CS10} formalize the notion of hyperproperty in a very general
way, as a set of sets of traces.
Remarkably, prior works using abstract interpretation for secure information
flow do not directly address the set-of-sets dimension and instead involve
various ad hoc formulations.
This paper presents a new %
approach of deriving information flow static analyses within the
calculational framework of abstract interpretation.

\textbf{First contribution.} We lift collecting semantics to sets of trace sets,
dubbed \emph{hypercollecting semantics}, in a fixpoint formulation
which is not simply the lifted direct image.
This can be composed with Galois connections that specify hyperproperties beyond
2-safety, without recourse to ad hoc additional notions.
On the basis of this foundational advance, it becomes possible to 
derive static analyses %
entirely within the
calculational framework of abstract interpretation~\citep{CC77,CC79,Cou99}.  

\textbf{Second contribution.} 
We use hypercollecting semantics to derive an analysis for ordinary dependences.
This can be seen as a rational reconstruction of both the type system of \citet{HS06,HS11sd} 
and the logic of \citet{AB04}.
They determine, for each variable $x$, a conservative approximation of the
variables $y$ whose initial values influence the final value of $x$.

\textbf{Third contribution.} 
We derive a novel analysis for quantitative information flow. 
This shows the benefit of taking hyperproperties seriously by means of abstract
interpretation.  
For noninterference, once the variables $y$ on which $x$ depends have fixed
values, there can be only one final value for $x$.  For quantitative information
flow, one is interested in measuring the extent to which other variables
influence $x$: for a given range of variation for the ``high inputs'', what is
the range of variation for the final values of $x$?   
We directly address this question as a hyperproperty:
given a set of traces that agree only on the low inputs, what is the cardinality of the possible final values for $x$? 
Using the hypercollecting semantics, we derive a novel \emph{cardinality
  abstraction}. We show how it can be used for analysis of quantitative
information problems including a bounding problem
which is not $k$-safety for any $k$.

\medskip

The calculational approach disentangles key design decisions and it enabled
us to identify opportunities for improving precision. 
We assess the precision of our analyses and 
provide a formal characterisation of precision for a quantitative 
information flow analysis vis a vis qualitative.
Versions of our analyses 
rival state of the art analyses for qualitative and quantitative information flow.
 
Our technical development uses the simplest programming language and semantic model in which the ideas can be exposed.  
One benefit of working entirely within the framework of abstract interpretation is that a wide range of semantics and analyses are already available for rich programming languages.

\myparagraph{Outline} 
Following the background (\Cref{sec:background}), we
introduce domains and Galois connections for hyperproperties
(\Cref{sec:collecting}) and hypercollecting semantics (\Cref{sec:overview}). 
Hyperproperties for information flow are defined in \Cref{sec:information}.
We use the framework to derive the static analyses in \Cref{sec:dependences} and
\Cref{sec:cardinal}.
\Cref{sec:precision} uses examples to evaluate the precision of the analyses,
and shows how existing analyses can be leveraged to improve precision.
We discuss related work (\Cref{sec:related}) and conclude.
\iftechreport
\emph{Appendices provide detailed proofs for all results, as well as a table of
symbols.}
\else
\emph{An accompanying technical report \citep{Aal16b} contains detailed
  proofs for all results, as well as a table of symbols.} 
\fi

\section{Background: Collecting Semantics, Galois Connections}
\label{sec:background}

The formal development uses deterministic imperative programs over integer variables.
Let $n$ range over literal integers $\mathbb{Z}$, $x$ over variables,
and $\oplus$ (resp.\ $\bcmp$) over some arithmetic (resp.\ comparison) operators.

\medskip
\begin{tabular}{l}
$\textbf{c}  ::=   \skipcom  \mid  x := e
            \mid  c_1; c_2
            \mid  \ifcom~ b~ \thencom~ c_1~ \elsecom~ c_2
            \mid   \whilecom~ b~\docom~ c
           $ \\
$\textbf{e}  ::=    n  \mid  x  \mid  e_1 \oplus e_2
              \mid  b $ \\
$\textbf{b} ::=    e_1 \bcmp e_2  $ 
\end{tabular}
\medskip

Different program analyses may consider different semantic domains as needed to
express a given class of program properties. 
For imperative programs, the usual domains are based on states $\st \in
\States$ that map each variable to a value~\citep{Win93}.
Some program properties require the use of traces that include intermediate states; 
others can use more abstract domains.
For information flow properties involving intermediate outputs,
or restricted to explicit data flow~\citep{SchoepeBPS16}, details about intermediate steps are needed.
By contrast, bounding the range of variables can be expressed in terms of final states.
As another example, consider determining which variables are left unchanged: To
express this, we need both initial and final states.   

\begin{figure}[tp]
\includegraphics{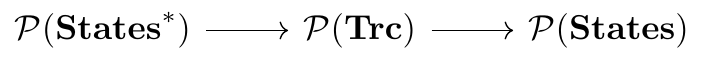}
  \caption{Fragment of the hierarchy of semantic domains
    ($\xrightarrow{\text{abstraction}}$)} 
  \label{fig:prop_hierarchy}
\end{figure}

In this paper we use the succinct term \dt{trace} for elements of 
$\Traces$ defined by $\Traces \triangleq \States \times\States$, interpreting $\tr \in \Traces$ as an
initial and final state.  
In the literature, these are known as \dt{relational traces},
by contrast with \dt{maximal trace} semantics using the set $\States^\ast$ of
finite sequences. %
A uniform framework describes the relationships
and correspondences between these and many other semantic domains
using Galois connections ~\citep{Cou02}. 
Three of these domains are depicted in \Cref{fig:prop_hierarchy}.

Given partially ordered sets $\concobj,\absobj$,
the monotone functions $\alpha\in\concobj\to\absobj$ and $\gamma\in\absobj\to\concobj$
comprise a \dt{Galois connection}, 
a proposition we write 
\( (\concobj,\leq) \galois{\alpha}{\gamma} (\absobj,\sqsubseteq)\),
provided they satisfy
\( \alpha(c)\sqsubseteq a \mbox{ iff }
 c \leq \gamma(a) \mbox{ for all } c\in \concobj, a\in \absobj \).

For example, to specify an analysis that determines which variables are never changed,
let $\absobj$ be sets of variables.  Define 
$\alpha\in\psetTraces\to \pset{\operatorname{Vars}}$  
by 
$\alpha(\Tr) = \{ x \mid \forall (\st,\st')\in\Tr,\: \st(x)=\st'(x) \}$
and $\gamma(X) = \{ (\st,\st') \mid \forall x\in X,\:  \st(x)=\st'(x) \}$.
Then \( (\psetTraces,\subseteq) \galois{\alpha}{\gamma} (\pset{Var},\supseteq)\).

For the hierarchy of usual domains, depicted in \Cref{fig:prop_hierarchy},
the connections are defined by 
an ``element-wise abstraction''. %
Define $\eltwop \in \Statesseq \to \Traces$ by 
$\eltwop(\st_0\st_1\ldots\st_n) \triangleq (\st_0,\st_n)$.
This lifts to an abstraction $\psetStatesseq \to \psetTraces$.

\begin{restatable}[Element-wise abstraction]{mylem}{lemgaloisliftedhyp}
  \label{lem:galoisliftedhyp}
Let $\eltwop \in \concobj \to \absobj$ be a function between sets.
Let $\alphaelt(C) \triangleq \{ \eltwop(c) \mid c \in C \}$ and
  $\gammaelt(A) \triangleq \{c \mid \eltwop(c) \in A\}$. 
Then  $(\psetconcobj,\subseteq)\galois{\alphaelt}{\gammaelt} (\psetabsobj,\subseteq)$. 
\end{restatable} 

The domain $\psetStates$, which suffices to describe the final 
reachable states of a program, is an abstraction of the relational
domain $\psetTraces$, by $\eltwop(\st,\stt) \triangleq \stt$.
In this paper we focus on the domain $\Traces$ because it is the simplest that
can express dependences.  

\myparagraph{Program semantics} %
We define both the denotational semantics 
$\semins{c}{} \in \Traces_{\bot} \to \Traces_{\bot}$ 
of commands and the denotational semantics $\semexp{e}{} \in \Traces \to \Values$ of expressions.
Here $\Values\triangleq \mathbb{Z}$ and $\Traces_{\bot}$ adds bottom element $\bot$ 
using the flat ordering.

\begin{tdisplay}{Standard semantics of commands \hfill 
$\semins{c}{} \in \Traces_{\bot} \to \Traces_{\bot}$} 
\vspace*{-2ex}
  \begin{mathpar}
    \semins{c}{\bot} \triangleq \bot

    \semins{x := e}{(\st,\stt)} \triangleq (\st,\, \stt[ x \mapsto
      \semexp{e}{(\st,\stt)}])

    \semins{c_1; c_2}{\tr} \triangleq \semins{c_2}{} \comp \semins{c_1}{\tr}

    \semins{\skipcom}{\tr} \triangleq \tr

    \semins{\ifcom~ b~ \thencom~ c_1~ \elsecom~ c_2}{\tr} \triangleq
    \begin{cases}
      \semins{c_1}{\tr} & \text{if } \semexp{b}{\tr} = 1 \\
      \semins{c_2}{\tr} &  \text{if } \semexp{b}{\tr} = 0
    \end{cases}

    \semins{\whilecom~ b~\docom~ c}{\tr}
         \triangleq 
         \begin{array}[t]{l}
           (\operatorname{lfp}^{\dotpreccurlyeq}_{(\lambda \tr. \bot)} \mathcal{F})(\tr) 
           \\ 
         \text{where } \mathcal{F}(w)(\tr) \triangleq
            \begin{cases}
              \tr & \text{if } \semexp{b}{\tr} = 0 \\
              w \comp \semins{c}{\tr} & \text{otherwise}
            \end{cases}
         \end{array}
  \end{mathpar}
\end{tdisplay}
Let $\tr$ be a trace $(\st,\stt)$.
The denotation $\semexp{e}{t}$ evaluates $e$ in the ``current state'', $\stt$.
(In \cref{sec:information} we also use $\semexpi{e}{t}$ which evaluates $e$ in the initial state, $\st$.)
The denotation $\semins{c}{t}$ is $(\st,\stt')$ where execution of $c$ in $\stt$ leads to $\stt'$.
The denotation is $\bot$ in case $c$ diverges from $\stt$.
Boolean expressions evaluate to either 0 or 1.
We assume programs do not go wrong. 
We denote by $\dotpreccurlyeq$ the point-wise lifting
to $\Traces_{\bot} \to \Traces_{\bot}$
of the approximation order
$\preccurlyeq$ on $\Traces_{\bot}$.

The terminating computations of $c$ can be written as
its image on the initial traces: 
$\{ \semins{c}{t} \mid t\in\InitTraces \mbox{ and } 
\semins{c}{t} \neq \bot \}$
where
\[ \InitTraces \triangleq \{ (\sigma,\sigma) \mid \sigma\in\States \} \]

To specify properties that hold for all executions we use
\dt{collecting semantics} which lifts the denotational
semantics to arbitrary sets $\Tr \in \psetTraces$ of
traces.
The idea is that $\seminsc{c}{\Tr}$ is the direct image of $\semins{c}{}$ on $\Tr$.
To be precise, in this paper we focus on termination-insensitive properties, and thus 
$\seminsc{c}{\Tr}$ is the set of non-$\bot$ traces $\tr'$ such that 
$\semins{c}{\tr}=\tr'$ for some $\tr\in\Tr$.
Later we also use the collecting semantics of expressions:
$ \semexpc{e}{\Tr} \triangleq \{ \semexp{e}{t} \mid t \in \Tr \} $.

Importantly, the collecting semantics
$\seminsc{c}{} \in \psetTraces \to \psetTraces$
can be defined compositionally using fixpoints~\citep[Sec. 7]{Cou02}.
For conditional guard $b$, write $\guardc{b}{}$ for the filter 
defined by
$\guardc{b}{\Tr} \triangleq \{ \tr \in \Tr \mid \semexp{b}{\tr} = 1 \}$.
\begin{tdisplay}{Collecting semantics \hfill
    $\seminsc{c}{} \in  \psetTraces \to \psetTraces$
  }
\vspace*{-2ex}
  \begin{mathpar}

\seminsc{x := e}{\Tr} 
         \triangleq \{ \semins{x := e}{\tr} \mid \tr \in \Tr \}

      \seminsc{c_1; c_2}{\Tr} \triangleq \seminsc{c_2}{} \comp \seminsc{c_1}{\Tr}

    \seminsc{\skipcom}{\Tr} \triangleq \Tr

     \seminsc{\ifcom~ b~ \thencom~ c_1~ \elsecom~ c_2}{\Tr}  \; \triangleq
          \; \seminsc{c_1}{} \comp  \guardc{b}{\Tr} %
          {} \cup {} \seminsc{c_2}{} \comp  \guardc{\neg b}{\Tr}

     \seminsc{\whilecom~ b~ \docom~ c} \Tr \triangleq
    \guardc{\neg b}{ \left(\operatorname{lfp}_{\Tr}^{\subseteq} \seminsc{\ifcom~
        b~ \thencom~ c~ \elsecom~ \skipcom}{} \right)}

  \end{mathpar}
\vspace*{-3ex}
\end{tdisplay}
The clause for while loops uses the denotation of a constructed conditional
command as a definitional shorthand---its denotation is compositional. 

Given a Galois connection 
$(\psetTraces,\subseteq) \galois{\alpha}{\gamma} (\absobj,\sqsubseteq)$, 
such as the one for unmodified variables,
the desired analysis is specified as $\alpha \comp \seminsc{c} \comp \gamma$. 
Since it is not computable in general, we only require an approximation
$f^\sharp \in \absobj \to \absobj$ that is \dt{sound} in this sense:
\begin{equation}\label{eq:sound}
\alpha \comp \seminsc{c} \comp \gamma \mathbin{\dot{\sqsubseteq}} f^\sharp
\end{equation}
where $\dot{\sqsubseteq}$ denotes the point-wise lifting of the partial order
$\sqsubseteq$.

To explain the significance of this specification, suppose one wishes to
prove program $c$ satisfies a trace property $\Tr\in\pset{\Traces}$, i.e.\ 
to prove that $\seminsc{c}(\InitTraces) \subseteq T$.
Given \cref{eq:sound} it suffices to find an abstract value $a$ that approximates 
$\InitTraces$, i.e.\  $\InitTraces\subseteq \gamma(a)$, and show that 
\begin{equation}\label{eq:soundX}
\gamma(f^\sharp (a)) \subseteq T
\end{equation}
\cref{eq:sound} is equivalent to
$\seminsc{c} \comp \gamma \mathbin{\dot{\subseteq}} \gamma \comp f^\sharp$
by a property of Galois connections.
So \cref{eq:soundX} implies $\seminsc{c}(\gamma(a)) \subseteq T$
which (by monotonicity of $\seminsc{c}$) implies
$\seminsc{c}(\InitTraces) 
\subseteq \seminsc{c}(\gamma(a)) 
\subseteq T$.

The beauty of specification \cref{eq:sound} is that $f^\sharp$ can be obtained
as an abstract interpretation $\seminsc{c}^\sharp$, derived systematically for
all $c$ by calculating from the left side of \cref{eq:sound} as shown by
\citet{Cou99}.

\section{Domains and Galois Connections for Hyperproperties}
\label{sec:collecting}

\begin{figure}[tb]
  \begin{small}
    \includegraphics[width=\columnwidth]{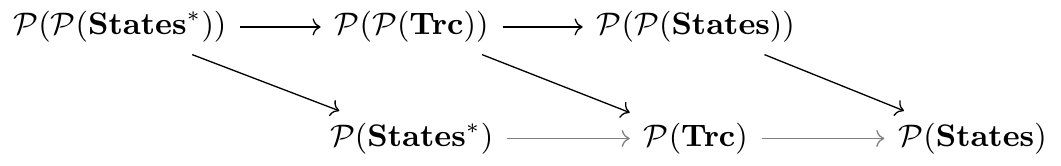}
  \end{small}
  \caption{Extended hierarchy of semantic domains
    ($\xrightarrow{\text{abstraction}}$)} 
  \label{fig:hypprop_hierarchy}
\end{figure}

To express hyperproperties, 
we need Galois connections for domains that involve sets of sets of observable
behaviours. 
This section spells out how such powerset domains form a hierarchy
as illustrated along the top of \Cref{fig:hypprop_hierarchy}.
We describe how dependences and cardinalities for quantitative information flow
can be formulated as Galois connections. 
We spell out a methodology whereby the standard notions and techniques of
abstract interpretation can be applied to specify and  
derive---in the same form as \Cref{eq:sound}---static analyses for hyperproperties.

As a first example, consider the condition: the final value of $x$ depends only
on the initial value of $y$. 
Its expression needs, at least, two traces: If two traces, denoted by
$(\sigma,\sigma')$ and $(\tau,\tau')$, agree on the initial value of $y$ then they
agree on the final value of $x$.  That is,  $\sigma(y)=\tau(y)$ implies
$\sigma'(x)=\tau'(x)$.   
This must hold for any two traces of the program.
This is equivalent to the following: For all sets $T$ of traces, if traces in
$T$ all agree on the initial value of $y$ then they all agree on the final value
of $x$. 
Later we extend this example to an analysis that infers which dependences hold.

Consider the problem of quantifying information flow with min-capacity~\citep{Smi09}.
For a program on two integer variables $h,l$, the problem is to infer how much
information is conveyed via $l$ about $h$:
considering some traces that
agree on the initial value of $l$, how many final values are possible for $l$.
For example, the program  
$l:=(h~\mathord{mod}~ 2) + l$ has two final values for $l$, for each initial $l$,
though there are many possible initial values for $h$.  
This cardinality problem generalizes prior work on quantitative flow analysis, where
typically low inputs are not considered.

Whereas the simple dependence problem can be formulated in terms of 2 traces,
the cardinality problem involves trace sets of unbounded size.   
In the terminology of hyperproperties, it 
is not a $\kparam$-safety hyperproperty for any
$\kparam$~\citep[Sec. 3]{YT11}, although it is hypersafety~\citep{CS10}.
For a fixed $k$, the problem ``variable $l$ has at most $k-1$ final values'' is
$k$-safety, which means it can be formulated in terms of sets with at most $k$
traces.\looseness=-1

It turns out that by using Galois connections on sets of sets, we can 
develop a general theory that encompasses many hyperproperties and which enables
derivation of interesting abstract interpreters.
For our applications, we use relational traces as the notion of observable behavior, 
and thus $\psetpsetTraces$.
The approach works as well for other notions, so there is a hierarchy of domains
as shown at the top of  
\Cref{fig:hypprop_hierarchy}, in parallel with the ordinary hierarchy shown
along the bottom.\looseness=-1

The abstractions of this hierarchy are obtained by lifting 
each abstraction between two standard collecting semantics~\citep{Cou02} to their
hypercollecting versions, by element-wise abstraction (\Cref{lem:galoisliftedhyp}).
For instance, \Cref{lem:galoisliftedhyp} justifies the abstraction between
$\psetpsetTraces$ and $\psetpsetStates$, by lifting the abstraction between
$\psetTraces$ and $\psetStates$~\citep[Sec. 8]{Cou02}. 
Additionally, the diagonal lines in \Cref{fig:hypprop_hierarchy}
represent abstractions between hypercollecting semantics
defined over some form of  observations and the corresponding collecting
semantics defined over the same observations.  

\begin{restatable}[]{mylem}{lemgaloishypprop}
  \label{lem:galoishypprop}
  Let $\concobj$ be a set.
  Define ${\alphahpp(\BObj) \triangleq \cup_{\Obj \in \BObj} \; \Obj}$ and
   $\gammahpp(\Obj) \triangleq \pset{\Obj}$. These form a Galois connection:
\[  (\psetpsetconcobj,\subseteq) \galois{\alphahpp}{\gammahpp}
   (\psetconcobj,\subseteq) \] 
\end{restatable}

It is noted by \citet{CS10} that any trace property can be lifted to a unique
hyperproperty; this lifting is exactly the concretisation $\gammahpp$ of
\Cref{lem:galoishypprop}. 
Although the model of \citet{CS10} is quite general, it does focus on infinite traces.
But hyperproperties can be formulated in terms of other notions of observation,
as illustrated in \Cref{fig:hypprop_hierarchy}.

\myparagraph{Cardinality abstraction}
To lay the groundwork for our quantitative information flow analysis, we
consider abstracting a set of values  by its cardinality.
Cardinality is one ingredient in many quantitative information flow
analyses estimating the  amount of sensitive information a program 
may leak~\cite{Smi09,BKR09,BCP09,KR13,Mal.13,Dal.13}.
The lattice of abstract representations we consider is the set 
\[ \abscardValues
\triangleq \mathbb{N} \cup \{\sizeval\} \] 
where $\sizeval$ denotes an infinite cardinal number. 
We %
use the natural order $\leq$,
and $\max$ as a join. 
Consider the abstraction operator $\cardvop \in \psetValues \to \abscardValues$
computing cardinality and given by $\cardvop(\Val) \triangleq \cardinal{\Val}$.
This operator $\cardvop$ is not \dt{additive}, i.e.\ it does not preserve joins;
e.g.\ $\cardvop(\{1,2\}\cup\{2,3\})\neq\max(\cardvop(\{1,2\}),\cardvop(\{2,3\}))$.   
Thus, there exists no associated concretisation $f$ for which $\cardvop$ is 
the lower adjoint in a Galois connection.
Yet, we can lift the abstraction operator $\cardvop$ to a Galois connection over
$\psetpsetValues$ through what is called a supremus abstraction~\citep[p.52]{Cou02}.
\begin{restatable}[Supremus abstraction]{mylem}{lemgaloissupremus}
  \label{lem:galoissupremus}
  Let $\supremop \in \concobj \to \absobj$ be a function
from a set $\concobj$, with codomain forming a complete lattice
  $(\absobj,\sqsubseteq)$.
  Let $\alphasup(C) \triangleq \sqcup_{c \in C}\supremop(c)$ and 
  $\gammasup(a) \triangleq 
  \{c \in \concobj \mid \supremop(c)\sqsubseteq a\}$.
  Then \[ (\psetconcobj,\subseteq) \galois{\alphasup}{\gammasup}
  (\absobj,\sqsubseteq) \] 
\end{restatable}
For example, define
$\alphacardv(\BVal) \triangleq \max_{\Val \in \BVal}\cardvop(\Val)$
and 
$\gammacardv(\abscardval) \triangleq \{ \Val \mid \cardvop(\Val) \leq n\}$.
Thus we obtain a Galois connection
$(\psetpsetValues,\subseteq)
\galois{\alphacardv}{\gammacardv}(\abscardValues,\leq)$.

As another example let us consider, in simplified form, an ingredient in
dependency or noninterference analysis.   
For program variable $x$, %
$\agreeo \in \psetStates \to \abstruth$ determines whether a set of states
contains only states that all agree on $x$'s value:
\[ \agreeo(\St) \triangleq (\forall \st,\st' \in \St,\: 
\semexp{x}{\st} = \semexp{x}{\st'}) \] 
Function  $\agreeo$ is not additive, so it is not part of a Galois connection
from $\psetStates$ to $\abstruth$.
The same problem arises with agreements on multiple variables, and 
with more concrete domains like the finite maximal trace semantics $\psetStatesseq$.

We lift the operator $\agreeo$ to a Galois connection over $\psetpsetStates$.
A supremus abstraction yields 
\[ \begin{array}{l}
\alphaagreeo(\BSt) \triangleq (\forall \St\in\BSt,\, \agreeo(\St)) \\
\gammaagreeo(\absbool) \triangleq
\{ \St \mid \agreeo(\St) \abstruthleq \absbool \}
\end{array}\]
so that 
$(\psetpsetStates,\subseteq) \galois{\alphaagreeo}{\gammaagreeo} (\abstruth,
\abstruthleq)$. 

These examples are consistent with the many formulations of noninterference
(e.g.~\citep{GM82,VS97,GiacobazziM04,AB04,HS06}) that motivated the  
characterisation of information-flow security requirements as hyperproperties~\citep{CS10}. 
Concretising an abstract value $a$ can be seen as defining the denotation
of a type expression (as in, for instance, \citet[Sec. 3.3.1]{Benton04}
and \citet{HuntS91}),
i.e.\ defining the set of objects that satisfy the description $a$.
Thus, concretising $\true$, when $\true$ is interpreted as 
\enquote{satisfies a property requirement}, naturally yields a set of traces.
Concretising $\true$, where $\true$ is interpreted as \enquote{satisfies
a security requirement}, yields a set of sets of traces.\looseness=-1

Intuitively, the most abstract denotation/concretisation of a property
requirement is defined in terms of a set of traces.
The most abstract concretisation/denotation of a security requirement yields a
set of sets of traces, namely a hyperproperty.
Hints of this intuition appear in the literature~\citep{Mcl94,Vol99,Rus01,ZL97};
e.g.\ security policies \enquote{are predicates on sets of traces 
(i.e. they are higher order)}~\citep[p.2]{Rus01}. 
However, only recently has a comprehensive framework proposed a sharp
characterisation of security policies as hyperproperties~\citep{CS08,CS10}.

\myparagraph{Abstract interpretation of hyperproperties}

The basic methodology for the verification of a hyperproperty
$\operatorname{HP}$, may be described as follows: 
\begin{list}{}{}
\item[\quad Step 1.]  Design approximate representations
  forming a complete lattice $\absobj$, choose a collecting semantics $\concobj$
  among the extended hierarchy (set of sets domains, e.g.\ 
  $\pset{\pset{\Traces}}$), and define $\alpha,\gamma$ for a Galois connection  
$(\concobj,\leq) \galois{\alpha}{\gamma} (\absobj,\sqsubseteq)$.
  
\item[\quad Step 2.] Compute an approximation $\aObj \in \absobj$ of
  the semantics $\Obj \in \concobj$ of the program $\prog$ of interest.

\item[\quad Step 3.] Prove that the inferred approximation 
$\aObj$ %
implies that $\prog$ satisfies
$\operatorname{HP}$.
The concretisation $\gamma(a)$ is a set of trace sets, of which the program's
trace set is a \emph{member}---by contrast to approximations of trace
properties, which infer a single trace set of which the program trace set is a
\emph{subset}. 
Then, it suffices to prove $\gamma(a) \subseteq \operatorname{HP}$.
\end{list}
Step 1 is guided by the need to
have $\gamma(a) \subseteq \operatorname{HP}$, i.e.\ $a$ describes a 
hyperproperty that implies $\operatorname{HP}$. 
The calculational design~\citep{Cou99} of abstract domains greatly
systematises Step 2, by relying on the Galois connection defined in
Step 1. 
Collecting semantics can be adapted to the additional structure of sets, as we
show in \Cref{sec:overview}.

\section{Hypercollecting Semantics}
\label{sec:overview}

In the following, we introduce a hypercollecting semantics defined over sets
$\BTr \in \psetpsetTraces$ of sets of traces.
This is used in subsequent sections to derive static analyses.

Here is Step 2 of the methodology, spelled out in detail.
Given a Galois connection 
$(\psetpsetTraces,\subseteq)
\galois{\alpha}{\gamma}(\absobj,\sqsubseteq^\sharp)$ built by the supremus
abstraction, and an approximation $a$ 
of the initial traces (i.e.\ $\InitTraces$ is in $\gamma(a)$),  find an
approximation $a' \in \absobj$ of the analysed program $c$,
i.e.\ $\seminsc{c}{\InitTraces}$ is in $\gamma(a')$.  
Then  prove that the program satisfies the hyperproperty $\operatorname{HP}$ of 
interest, i.e.\ $\gamma(a') \subseteq  \operatorname{HP}$.
In order to compute $a'$, we define a hypercollecting semantics  
$\seminshc{c}{} \in  \psetpsetTraces \to \psetpsetTraces$.
That will serve to derive---in the manner of \Cref{eq:sound}---a static analysis
that is correct by construction. 

\begin{tdisplay}{Hypercollecting semantics \hfill
    $\seminshc{\,c\,}{} \in  \psetpsetTraces \to \psetpsetTraces$
  }
\vspace*{-2ex}
  \begin{mathpar}

     \seminshc{x := e}{\BTr} \triangleq 
     \left\{
       \seminsc{x := e}{\Tr} \mid \Tr \in \BTr
     \right \}

     \seminshc{c_1; c_2}{\BTr} \triangleq \seminshc{c_2}{} \comp \seminshc{c_1}{\BTr}

     \seminshc{\skipcom}{\BTr} \triangleq \BTr

     \begin{array}{l}
       \seminshc{\ifcom~ b~ \thencom~ c_1~ \elsecom~ c_2}{\BTr}   \triangleq 
       \\
       \hspace*{5em}
       \{ 
       \seminsc{c_1}{} \comp  \guardc{b}{\Tr}  {} \cup {}
       \seminsc{c_2}{} \comp  \guardc{\neg b}{\Tr} 
       \:\mid\:
       \Tr \in \BTr
       \}
     \end{array}

     \seminshc{\whilecom~ b~\docom~ c\,} \BTr \triangleq
     \guardhc{\neg b}{
       \left(\operatorname{lfp}_{\BTr}^{\subseteq}
       \seminshc{\ifcom~ b~ \thencom~ c~ \elsecom~ \skipcom}{} \right)}

     \guardhc{b}{\BTr} \triangleq \{ \guardc{b}{\Tr} \mid \Tr \in \BTr \}
  \end{mathpar}
\vspace*{-3ex}
\end{tdisplay}

Recall from \Cref{sec:background} that standard collecting semantics 
is a fixpoint-based formulation that captures the direct image on sets of the 
underlying program semantics -- this is proved, for example, by~\citet{CP10,AN16}.
The fixpoint formulation at the level of sets-of-sets we use 
is not simply the direct image of the standard collecting semantics.
The direct image of the standard collecting semantics would yield a set of
(inner) fixpoints over sets of traces, whereas an outer fixpoint over 
sets of sets of traces enables straightforward application of the fixpoint
transfer theorem. 

\begin{restatable}[]{mytheo}{theohypercoll}
\label{theo:hypercollsound}
For all $c$ and all $\Tr \in \psetTraces $,
$\, \seminsc{c}{\Tr}$ is in $\seminshc{\,c\,}{\{ \Tr \}}$.
\end{restatable}
For a singleton $\{ \Tr \}$, 
the set $\seminshc{c}{\{ \Tr \}} \in \psetpsetTraces$  is not necessarily a singleton set
containing only the element $\seminsc{c}{\Tr}$. If $c$ is a loop,
$\seminshc{c}{\{ \Tr \}}$ yields a set of sets $\altTr$ of traces, where each
set $\altTr$ of traces contains only traces that exit the loop after less than
$k$ iterations, for $k \in \mathbb{N}$.  
We prove this theorem as corollary of the following:
\[ \forall \BTr \in \psetpsetTraces, \{ \seminsc{c}{\Tr} \mid \Tr
\in \BTr \} \subseteq \seminshc{c}{\BTr} \]
This is proved by structural induction on commands.
For loops, there is a secondary induction on iterations of the
loop body.

In summary, suppose one wishes to 
prove program $c$ satisfies  hyperproperty
$\operatorname{HP} \in\pset{\pset{\Traces}}$, i.e.\  one 
wishes to prove that $\seminsc{c}(\InitTraces) \in \operatorname{HP}$.
Suppose we have an approximation $f^\sharp$ of the hypercollecting semantics, 
similarly to \cref{eq:sound}, i.e.
\begin{equation}\label{eq:soundH}
\alpha \comp \seminshc{c}{} \comp \gamma \mathbin{\dot{\sqsubseteq}^\sharp} 
f^\sharp
\end{equation}
Given \cref{eq:soundH} it suffices to find an abstract value $a$ that approximates 
$\InitTraces$, i.e.\  $\InitTraces\in  \gamma(a)$, and show that:
\begin{equation}\label{eq:soundXH}
\gamma(f^\sharp (a)) \subseteq \operatorname{HP} 
\end{equation}
Why? 
\Cref{eq:soundH} is equivalent to
$\seminshc{c}{} \comp \gamma \mathbin{\dot{\subseteq}} \gamma \comp f^\sharp$
by a property of Galois connections.
So we have
$\seminsc{c}(\InitTraces) 
\in \seminshc{c}{(\gamma(a))} 
\subseteq \gamma(f^\sharp (a))
\subseteq \operatorname{HP} $
using $\InitTraces\in  \gamma(a)$, the Theorem, and \cref{eq:soundXH}.

\section{Information Flow}
\label{sec:information}

This section gives a number of technical definitions which build up to 
the definition of Galois connections with which we specify information flow policies explicitly as hyperproperties.

When a fixed main program is considered, we refer to it as $\prog$
and its variables as $\varprog$. 
Our analyses are parametrised by the program $\prog$ to analyse, and an
initial \dt{typing context} $\Ga \in \varprog \to \mlsLat$ mapping each
variable to a security level $\mlsl \in \mlsLat$ for its initial value.
We assume $(\mlsLat,\sqsubseteq,\sqcup,\sqcap)$ is a finite lattice. %
In the most concrete case, $\mlsLat$ may be defined as the \dt{universal
flow lattice}, i.e.\ the powerset of variables $\pset{\varprog}$, from
which all other information flow types can be inferred through a suitable
abstraction~\citep[Sec. 6.2]{HS06}; the initial typing context is then defined as
$\lambda x. \{ x \}$.

\myparagraph{Initial $\mlsl$-equivalence and variety} 
A key notion in information flow is 
\dt{$\mathbf{\mlsl}$-equivalence}.
Two
states are $\mlsl$-equivalent iff they agree on the values of variables
having security level at most $\mlsl$.
We introduce the same notion over a set of traces, requiring that the
\emph{initial} states are $\mlsl$-equivalent.
Let us first denote by $\semexpi{e} \in \Traces \to \Values$ the evaluation of
expression $e$ in the initial state $\st$ of a trace 
$(\st,\stt) \in \Traces$---unlike $\semexp{e}{} \in \Traces \to \Values$ which evaluates expression $e$ in the final state $\stt$. 
Then, we denote by 
$\iagree{\Tr}{l}$
the judgement that all traces in a set
$\Tr \subseteq \Traces$ are \dt{initially $\mlsl$-equivalent}, i.e. they all
initially agree on the value of variables up to a security level 
$\mlsl \in \mlsLat$. 

For example, in the case that $\mlsLat$ is the universal flow lattice, $\iagree{\Tr}{\{x,y\}}$
means $\forall \tr_1,\tr_2 \in \Tr,
\semexpi{x}{\tr_1} = \semexpi{x}{\tr_2} \wedge 
\semexpi{y}{\tr_1} = \semexpi{y}{\tr_2}$.

\begin{tdisplay}{Initial $\mathbf{\mlsl}$-equivalence  \hfill $\iagree{\Tr}{l}$}
$\iagree{\Tr}{l} \: $ 
iff. 
$\begin{array}[t]{l}
  \forall \tr_1,\tr_2 \in \Tr, \forall x \in \varprog, 
  \\
  \hspace*{3em}
  \Ga(x) \sqsubseteq l \implies 
  \semexpi{x}{\tr_1} = \semexpi{x}{\tr_2}
\end{array}$
\end{tdisplay}

The notion of variety~\citep{Coh77} underlies most definitions of
qualitative and quantitative information flow security.
Information is transmitted from $a$ to $b$ over execution of
program $\prog$ if by \enquote{varying the initial value of $a$ 
(exploring the variety in $a$), the resulting value in $b$ after $\prog$'s
execution will also vary (showing that variety is conveyed to
$b$)}~\citep{Coh77}.
We define the \dt{$\mathbf{\mlsl}$-variety} of expression $e$,
as the set of sets of values $e$ may take, when considering only initially
$l$-equivalent traces.
The variety is defined first as a function
$\obsexpcl{e}{} \in \psetTraces \to \psetpsetValues$
on trace sets, from which we obtain a function
$\obsexphcl{e}{} \in \psetpsetTraces \to \psetpsetValues$, 
on sets of trace sets.
Intuitively, $\mlsl$-variety of expression $e$ is the variety that is conveyed
to $e$ by varying only the input values of variables having a security level
$\mlsl'$ such that $\neg (\mlsl' \sqsubseteq \mlsl)$.

\begin{tdisplay}{$\mathbf{l}$-variety \hfill $\obsexpcl{e}{} \qquad $
    $\obsexphcl{e}{}$} 
\vspace*{-2ex}
\[\begin{array}{l}
  \obsexpcl{e}{} \in \psetTraces \to \psetpsetValues \\[.5ex]
  \obsexpcl{e}{\Tr} \triangleq
     \{ \semexpc{e}{\altTr} \mid \altTr \subseteq \Tr \text{ and } \iagreerl \} \\[1.5ex]
  \obsexphcl{e}{} \in \psetpsetTraces \to \psetpsetValues \\[.5ex]
  \obsexphcl{e}{\BTr} \triangleq \cup_{\Tr \in \BTr} \; \obsexpcl{e}{\Tr}
\end{array}\]
\end{tdisplay}

Each set $\Val \in \obsexpcl{e}{\Tr}$ of values results from
initially $\mlsl$-equivalent traces 
($\iagree{\altTr}{l}$ for $\altTr\subseteq\Tr$).
Thus, expression $e$ does not leak sensitive information to attackers having a
security clearance $\mlsl \in \mlsLat$ if $\obsexpcl{e}{\Tr}$ is a set of
singleton sets.
Indeed, sensitive data for attackers with security clearance
$\mlsl \in \mlsLat$ is all data  having a security level $\mlsl'$ for which
attackers do not have access
(i.e.\ $\neg (\mlsl' \sqsubseteq \mlsl)$~\citep{DD77}). 
Thus, if $\obsexpcl{e}{\Tr}$ is a set of singleton sets, this means that no
matter how sensitive information varies, this variety is not conveyed to
expression $e$. 

Besides a pedagogical purpose, we define $\mlsl$-variety $\obsexpcl{e}{}$
(resp. $\obsexphcl{e}{}$) instead of simply lifting the denotational semantics
$\semexp{e}{}$ of expressions to sets of traces (resp. sets of sets of traces)
since we want to build modular abstractions of traces by relying on underlying
abstractions of values.
Thus, $\mlsl$-variety enables us to pass information about initially
$\mlsl$-equivalent traces to the underlying domain of values by keeping disjoint
values that originate from traces that are not initially $\mlsl$-equivalent.

\myparagraph{Specifying information flow}
We now have the ingredients needed to describe information flow for 
command $c$, with respect to typing context $\Ga \in \varprog \to \mlsLat$.
A quantitative security metric, introduced by~\citet{Smi09,Smi11}, relies
on min-entropy and min-capacity~\citep{Ren61} in order to estimate the leakage of
a program. 
Let us assume a program $\prog$ that is characterized by a set $\Trprog \in
\psetTraces$ of traces, i.e.\ $\Trprog \triangleq \seminsc{\prog}{\InitTraces}$.
For simplicity, assume attackers only observe the value of a single variable
$x \in \varprog$. 
(The generalization to multiple variables is straightforward.)
The leakage of $\prog$, as measured by \dt{min-capacity}, to
attackers  having security clearance $\mlsl \in \mlsLat$ is defined by
\[ \mincapl  \triangleq  \log_2 \comp \alphacardv \comp 
\obsexpcl{x}{\Trprog} \] 
(The definition of $\alphacardv$ follows \cref{lem:galoissupremus}.)
For our purposes, it suffices to know that this quantity aims to measure, in
bits,  the remaining uncertainty about sensitive data for attackers with
security clearance $\mlsl$.
Refer to the original work~\citep{Smi09} for more details.

Leaving aside
the logarithm in the definition of $\mincapl$, a
quantitative security requirement may enforce a limit on the amount of
information leaked to attackers with security clearance $\mlsl \in \mlsLat$, by
requiring that the $\mlsl$-cardinality of variable $x$ is less than or equal to
some non-negative integer $\kparam$.
We denote by $\secreqlkx$ the hyperproperty
that characterises this security
requirement, i.e.\ the set of program denotations satisfying it: 
\[ \secreqlkx \triangleq \{ T \in \psetTraces \mid \alphacardv \comp 
  \obsexpcl{x}{T} \leq \kparam \} \]
Note that $\secreqop$ implicitly depends on
the choice of initial typing $\Ga$, as does $\obsexpcl{x}{T}$.

The termination-insensitive noninterference policy \enquote{the final value of
  $x$ depends only on the initial values of variables labelled at most $\mlsl$} 
corresponds to the hyperproperty $\secreq{\mlsl}{1}{x}$.
Therefore, the program $\prog$ satisfies $\secreq{\mlsl}{1}{x}$ if 
$\alphacardv \comp  \obsexpcl{x}{\Trprog}\leq 1$. 
Let $\BTr = \seminshc{\prog}{\{ \InitTraces \}}$.
Since $\Trprog$ is in $\BTr$ (\cref{theo:hypercollsound}), then $\prog$ satisfies
$\secreq{\mlsl}{1}{x}$ if 
$\alphacardv \comp \obsexphcl{x}{\BTr} \leq 1$, by monotony of $\alphacardv$ and 
by $\obsexpcl{x}{\Trprog} \subseteq \obsexphcl{x}{\BTr}$
from the definition of $\obsexphcl{-}{}$.

\section{Dependences}
\label{sec:dependences}

We rely on abstract interpretation to derive a static analysis similar to
existing ones inferring dependences~\citep{AB04,HS06,ABB06,HS11sd}.

Recall that our analyses are parametrised on a security lattice $\mlsLat$ and program $\prog$.
We denote by $\depconslx$ an atomic dependence constraint, with $\mlsl \in
\mlsLat$ and $\varx \in \varprog$, read as \enquote{agreement up to security
  level $\mlsl$ leads to agreement on $\varx$}. 
It is an atomic pre-post contract 
expressing that the final value of $\varx$ must only depend on initial values having at
most security level $\mlsl$. 
Said otherwise, $\depconslx$ states the
noninterference of variable $\varx$ from data that is sensitive for attackers
with security clearance $\mlsl$, i.e.\  all inputs having security
 level $\mlsl'$ such that $\neg(\mlsl' \sqsubseteq \mlsl)$.

Dependences are similar to information flow types~\citep{HS06} 
and are the %
dual of independences assertions~\citep{AB04}.
Both interpretations are equivalent~\citep[Sec. 5]{HS06}.

\begin{tdisplay}{Lattice of dependence constraints \hfill $\depLat \qquad
    \depconsset \in \depLat$} 
Given a lattice $\mlsLat$ and program $\prog$,
define 
\[ \begin{array}[t]{c}
  \depLat \triangleq \pset{\{ \depconslx \mid \mlsl \in \mlsLat, \varx \in
    \varprog \}} \\[1ex] 
  \depconsset_1 \deplatleq \depconsset_2 \triangleq  \depconsset_1 \supseteq
  \depconsset_2  
\qquad 
  \depconsset_1 \deplatjoin \depconsset_2 \triangleq \depconsset_1 \cap \depconsset_2 
\end{array}\]
\vspace*{-1ex}
\end{tdisplay}

In the rest of this section, $\mlsLat$ and $\prog$ are fixed,
together with a typing context $\Ga \in \varprog \to \mlsLat$.

The semantic characterisation of dependences is tightly linked to variety.
An atomic constraint $\depconslx$ holds if no variety is conveyed to
$\varx$ when the inputs up to security level $\mlsl$ are fixed.
We use this intuition to define the Galois connections linking the 
hypercollecting semantics and the lattice $\depLat$,  %
by instantiating the supremus abstraction in \Cref{lem:galoissupremus}. 

The agreement abstraction approximates a set $\BVal \in \psetpsetValues$ by
determining whether it contains variety.

\begin{tdisplay}{Agreements abstraction  \hfill 
$\agreev \quad \alphaagreev \quad \gammaagreev$} 
\vspace*{-2ex}
\[
\begin{array}{lcl}
\agreev & \in & \psetValues \to \abstruth \\[.5ex]
\agreev(\Val) & \triangleq & (\forall \val_1,\val_2 \in \Val, v_1 = v_2) \\[1ex]

\alphaagreev & \in & \psetpsetValues \to \abstruth \\[.5ex]
\alphaagreev(\BVal) & \triangleq & \wedge_{\Val \in \BVal} \agreev(\Val) \\[1ex]

\gammaagreev & \in & \abstruth \to \psetpsetValues\\[.5ex]
\gammaagreev(\absbool) & \triangleq & \{ \Val \in \psetValues \mid \agreev(V) \abstruthleq \absbool \} 
\end{array}
\] 

\[ (\psetpsetValues,\subseteq)
 \galois{\alphaagreev}{\gammaagreev} (\abstruth,\abstruthleq)
\]
\end{tdisplay}
\noindent Note that $\gammaagreev(\true)$ is
$\{ \Val \in \psetValues \mid \agreev(V)  \}$ and
$\gammaagreev(\false)$ is $\psetValues$.
Also,  $\agreev(\Val)$ iff $\cardinal{\Val} \leq 1$.

The dependence abstraction approximates a set $\BTr \in \psetpsetTraces$ by a
dependence constraint $\depconsset \in \depLat$.
Recall that $\obsexpcl{\varx}{\Tr}$ is the set of final values for variable $\varx$
in traces $\tr \in \Tr$ that agree on inputs of level at most $\mlsl$.
So $\alphaagreev(\obsexpcl{\varx}{\Tr})$ holds just if there is at most one final value.

\begin{tdisplay}{Dependence abstraction \hfill $\deptrop \quad
    \alphadeptr \quad \gammadeptr$}
\vspace*{-2ex}
\[\begin{array}{lcl}
  \deptrop & \in & \psetTraces \to \depLat \\[.3ex]

  \deptrop(T) & \triangleq  & \{ \depconslx \mid \mlsl\in\mlsLat,\,
  \varx\in\varprog,\, \alphaagreev(\obsexpcl{x}{T}) \} \\[1ex] 

  \alphadeptr & \in & \psetpsetTraces \to \depLat \\[.3ex]
  \alphadeptr(\BTr) & \triangleq & \deplatjoin_{\Tr \in \BTr} \deptrop(T) \\[1ex]

  \gammadeptr & \in & \depLat \to \psetpsetTraces \\[.3ex]
  \gammadeptr(\depconsset) & \triangleq &
                \{ \Tr \mid \deptrop(\Tr) \deplatleq \depconsset \} 
\end{array} \]

\[ 
(\psetpsetTraces,\subseteq) 
\galois{\alphadeptr}{\gammadeptr} 
 (\depLat,\deplatleq) 
\]
\vspace*{-2ex}
\end{tdisplay}

Note that $\deptrop(\Tr)$ is the set of dependences 
$\depconslx$ for which $\alphaagreev(\obsexpcl{x}{T})$ holds.
For instance,  the initial typing context $\Ga \in \varprog \to \mlsLat$
determines the initial dependences of a program:
\begin{align*}
	& \alphadeptr(\{\InitTraces\})  \
  \\
  & \quad =   %
  \{ \depconslx \mid \mlsl\in\mlsLat,\, \varx\in\varprog \text{ and }\,
  \alphaagreev(\obsexpcl{x}{\InitTraces}) \} \\
   & \quad = \{ \depconslx \mid \mlsl \in \mlsLat, \varx \in \varprog \text{ and }
  \Ga(x) \mlslatleq \mlsl \}
\end{align*}
We derive an approximation
 $\absobsexpdepl{e}{}$ of  $l$-variety
$\obsexphcl{e}{}$.
This approximation $\absobsexpdepl{e}{} \in \depLat \to \abstruth$, called
$\mlsl$-agreement of expression $e$,  determines whether a set $\depconsset$ of
dependence constraints guarantees that no variety is conveyed to expression $e$
when the inputs up to security level $\mlsl$ are fixed. 
Notice that we use symbol $\natural$ and subscript $D$ here, for
 contrast with similar notation using $\sharp$ and subscript $C$ in later
 sections.

\begin{tdisplay}{$\mlsl$-agreement of expressions \hfill 
$\absobsexpdepl{e}{} \in \depLat \to \abstruth$} 
\vspace{-2ex}
\begin{mathpar}
\absobsexpdepl{n}{\depconsset} \triangleq \true

\absobsexpdepl{x}{\depconsset} \triangleq (\depcons{l}{x} \in \depconsset)

\absobsexpdepl{e_1 \oplus e_2}{\depconsset} \triangleq
\absobsexpdepl{e_1}{\depconsset} \wedge \absobsexpdepl{e_2}{\depconsset}

\absobsexpdepl{e_1 \bcmp e_2}{\depconsset} \triangleq 
\absobsexpdepl{e_1}{\depconsset} \wedge
\absobsexpdepl{e_2}{\depconsset}
\end{mathpar}
\vspace{-2ex}
\end{tdisplay}

Deriving the clauses defining $\absobsexpdepl{-}{}$ amounts to a constructive 
proof of the following.

\begin{restatable}[]{mylem}{lemabsvarietydepsound}
  \label{lem:soundabsvarietydep}
$\absobsexpdepl{e}{}$ is sound:
\[ \forall e, \forall \mlsl , \forall \depconsset , \quad \alphaagreev \comp \obsexphcl{e}{} \comp \gammadeptr (\depconsset)
\abstruthleq \absobsexpdepl{e}{\depconsset} \enspace .  \]
\end{restatable}

\myparagraph{Dependence abstract semantics}
We derive a dependence abstract semantics $\seminshcdepabs{c}{}$
by approximating the hypercollecting semantics $\seminshc{c}{}$.
This abstract semantics $\seminshcdepabs{c}{} \in \depLat \to \depLat$
over-approximates the dependence constraints that hold after execution of a
command $c$, on inputs satisfying initial dependence constraints.

We assume a static analysis approximating the variables
that a command modifies.

\begin{tdisplay}{Modifiable variables \hfill $\Mod \in Com \to \pset{Var}$}
For all $c,x$,
if there exists $\tr,\tr' \in \Traces$ such that
$\semins{c}{\tr} = \tr'$ and $\semexpi{x}{\tr'} \neq \semexp{x}{\tr'}$, 
then $x \in \Mod(c)$.
\end{tdisplay}

The abstract semantics of assignments $x := e$  discards all atomic
constraints related to variable $x$ in the input set $\depconsset$ of
constraints, and adds atomic constraints $\depcons{\mlsl}{x}$  if
$\depconsset$ guarantees $\mlsl$-agreement for expression $e$.
For conditionals, for each security level $\mlsl$, if the input set
$\depconsset$ guarantees $\mlsl$-agreement of the conditional
guard, the abstract semantics computes the join over the dependences of
both conditional branches, after projecting to only those atomic constraints
related to $\mlsl$
(notation $\projtomlsl{-}$).
If $\depconsset$ does not guarantee $\mlsl$-agreement of the
conditional guard, atomic constraints related to both $\mlsl$ and variables  possibly
modified are discarded.
Intuitively, if $\depconsset$ guarantees $\mlsl$-agreement of the conditional
guard, then $l$-agreement over some variable $\varx$ in both branches guarantees
$\mlsl$-agreement over $\varx$ after the conditional command.
Otherwise, the only $\mlsl$-agreements that are guaranteed after the conditional
are those that hold before the conditional for variables that
are not modified.

\begin{tdisplay}{Dependence abstract semantics \hfill 
$\seminshcdepabs{c}{} \in \depLat \to \depLat$}
\vspace{-2ex}
\begin{mathpar}
\seminshcdepabs{\skipcom}{\depconsset} \triangleq \depconsset

\seminshcdepabs{c_1; c_2}{\depconsset} \triangleq 
\seminshcdepabs{c_2}{} \comp \seminshcdepabs{c_1}{\depconsset} 

\begin{array}{l}
\seminshcdepabs{\varx := e}{\depconsset} \triangleq \\
\hspace*{2em}
\{ \depcons{\mlsl}{y} \in \depconsset \mid y \neq \varx
\} \cup
\{ \depcons{\mlsl}{\varx} \mid \mlsl\in\mlsLat,\, \absobsexpdepl{e}{\depconsset} \}
\end{array}

\begin{array}{l}
\seminshcdepabs{\ifcom~ b~ \thencom~ c_1~ \elsecom~ c_2}{\depconsset} \triangleq
\\
\hspace*{3em}
\begin{array}[t]{l}
  \letin{\depconsset_1 = \seminshcdepabs{c_1}{\depconsset}} \\
  \letin{\depconsset_2 = \seminshcdepabs{c_2}{\depconsset}} \\
  \letin{W = \Mod(\ifcom~ b~ \thencom~ c_1~ \elsecom~ c_2)} \\
  \quad
  \bigcup\limits_{\mlsl \in \mlsLat} 
    \begin{cases}
      \projtomlsl{\depconsset_1}
      \deplatjoin   
      \projtomlsl{\depconsset_2}
    & \text{if } \absobsexpdepl{b}{\depconsset}  \\
      \{ \depcons{\mlsl}{x} \in \projtomlsl{\depconsset} \mid x \notin W \} 
    & \text{otherwise}
    \end{cases}
\end{array}
\end{array}

\seminshcdepabs{\whilecom~ b~\docom~ c}{\depconsset} 
\triangleq \operatorname{lfp}^{\cardlatleq}_{\depconsset}
\seminshcdepabs{\ifcom~ b~ \thencom~ c_1~ \elsecom~ c_2}{} 

\projtomlsl{\depconsset} \triangleq 
   \{ \depconslx \in \depconsset \mid \varx \in \varprog \}

\end{mathpar}
\vspace{-4ex}
\end{tdisplay}

\begin{restatable}[]{mytheo}{theoabsdepsem}
\label{thm:soundabsdependences}
The dependence semantics is sound:
\[  \quad \alphadeptr \comp \seminshc{c}{} \comp \gammadeptr
{} \dotdeplatleq {} 
\seminshcdepabs{c}{} \enspace .\]
\end{restatable}

We denote by $\dotdeplatleq$ the point-wise lifting of the partial order
$\deplatleq$. 
We can derive this abstract semantics by directly approximating
the relational hypercollecting semantics $\seminshc{c}{}$ through the 
dependence Galois connection $(\alphadeptr,\gammadeptr)$.
The derivation is by structural induction on commands.
It leverages mathematical properties of Galois connections.
We start with the specification of the best abstract transformer
$\alphadeptr \comp \seminshc{c}{} \comp \gammadeptr \in \depLat \to \depLat$,
and successively approximate it to finally obtain the definition of the
dependence abstract semantics for each form of command.
The derivation is the proof, and the obtained definition of the abstract
semantics is correct by construction.

Let us showcase the simplest derivation for a sequence of commands in order to
illustrate this process:
\begin{align*}
  & \alphadeptr \comp \seminshc{c_1;c_2}{} \comp \gammadeptr \\
  & \quad = \justif{By definition of the hypercollecting semantics} \\
  & \qquad \alphadeptr \comp \seminshc{c_2}{} \comp \seminshc{c_1}{} \comp
  \gammadeptr  \\ 
  & \quad \dotdeplatleq \justif{By $\gammadeptr \comp \alphadeptr$
    is extensive } \\ 
  & \qquad
  \alphadeptr \comp \seminshc{c_2}{} \comp
\gammadeptr \comp \alphadeptr \comp
  \seminshc{c_1}{} \comp
  \gammadeptr  \\
  & \quad \dotdeplatleq \justif{By induction hypothesis
  $\alphadeptr \comp \seminshc{c}{} \comp \gammadeptr  \dotdeplatleq
    \seminshcdepabs{c}{} $} \\
  & \qquad \seminshcdepabs {c_2}{} \comp
  \seminshcdepabs{c_1}{} \\
  & \quad \triangleq \justif{Take this last approximation as the definition.}%
\\
  & \qquad \seminshcdepabs{c_1;c_2}{}
\end{align*}

Alternatively, we can leverage Galois connections to give the analysis as an
approximation of the cardinality analysis.
We work this out by \Cref{lem:alphaagreevdec,lem:alphadeptrdec}, introduced in
\Cref{sec:cardinal}.

\myparagraph{Comparison with previous analyses}
Our dependence analysis is similar to the logic of~\citet{AB04} 
as well as the flow-sensitive type system of~\citet{HS06}.
The relationship between our sets $\depconsset \in \depLat$ of dependence
constraints and the type environments $\Delta \in \varprog \to \mlsLat$ of Hunt and
Sands can be formalised by the abstraction: 
\[ \begin{array}{lcl}
\alphatohs & \in & \depLat \to \varprog \to \mlsLat \\ 
\alphatohs(\depconsset) & \triangleq & 
   \lambda x. \mlslatmeet \{ \mlsl \mid \depconslx \in \depconsset \} \\[1ex]

\gammatohs & \in & (\varprog \to \mlsLat) \to \depLat \\
\gammatohs(\Delta) & \triangleq & \{ \depconslx \mid x \in \varprog,\: \mlsl\in\mlsLat,\: \Delta(x) \sqsubseteq \mlsl \}
\end{array} \]
This is in fact an isomorphism because of the way we interpret dependences.
Indeed, if $\depcons{\mlsl}{\varx}$ holds, then also $\depcons{\mlsl'}{\varx}$
for all $\mlsl' \in \mlsLat$ such that $\mlsl \sqsubseteq \mlsl'$
\iftechreport
(cf. \Cref{cor:gammadepmon} in \Cref{app:precision_proof}).
\else
(cf. report~\citep{Aal16b}).
\fi
This observation 
suggests reformulating the sets 
$\depconsset \in \depLat$ of dependence constraints
to contain only elements with minimal level, but we refrain from doing
so for simplicity of presentation.

Our dependence analysis is at least as precise as 
the type system of  Hunt and Sands.
To state this result, we denote by $\mlsbot$ the bottom element of the lattice
$\mlsLat$.  
We also assume that the modified variables is precise enough to simulate the same
effect as the program counter used in the type system: $\Mod(c)$ is a subset of
the variables that are targets of assignments in $c$.

\begin{restatable}[]{mytheo}{theohsprecision}
\label{thm:theohsprecision}
  For all $c$, $\depconsset_0,\depconsset \in \depLat$, 
  $\Delta_0, \Delta \in \varprog \to \mlsLat$, 
  where $\mlsbot \vdash \Delta_{0} \{ c \} \Delta$, 
  and $\depconsset = \seminshcdepabs{c}{\depconsset_0}$,
  it holds that:  
  \[ 
  \alphatohs(\depconsset_0) \dotmlslatleq \Delta_0
  \implies 
  \alphatohs(\depconsset) \dotmlslatleq \Delta
  \enspace .
  \]
\end{restatable}

\section{Cardinality Abstraction}
\label{sec:cardinal}

Dependence analysis is only concerned with whether variety is conveyed.
We refine this analysis by deriving a cardinality abstraction that
enumerates variety.

We denote by $\cardcons{\mlsl}{\varx}{\abscardval}$ an atomic cardinality
constraint where $\mlsl \in \mlsLat, \varx \in \varprog$ and $ \abscardval \in
\abscardValues$, read as \enquote{agreement up to security level $\mlsl$ leads
  to a variety of at most $\abscardval$ values in variable $\varx$}.

\begin{tdisplay}{Lattice of cardinality constraints \hfill $\cardLat \qquad
    \cardconsset \in \cardLat$}
  For a program $\prog$ and lattice $\mlsLat$, 
  we say
  $\cardconsset$ is a \dt{valid set of constraints} iff
  $\forall \varx \in \varprog, \forall \mlsl \in \mlsLat, \existsu \abscardval
  \in \abscardValues, \cardconslxn \in \cardconsset$.

  Let $\cardLat$ be the set of valid sets of constraints.

  It is a complete lattice:
  \vspace*{0.5ex}
  
 $\begin{array}{l}
\cardconsset_1 \cardlatleq \cardconsset_2   \text{ iff }
   \begin{array}[t]{l}
      \forall \cardconslx{\abscardval_1} \in \cardconsset_1,\ \exists \abscardval_2, \\ 
      \hspace*{3em}
             \cardconslx{\abscardval_2} \in \cardconsset_2 
             \land \abscardval_1 \leq \abscardval_2 
   \end{array}
\\
\cardconsset_1 \cardlatjoin \cardconsset_2   \triangleq
    \begin{array}[t]{l}
      \{ \cardconslx{\max(\abscardval_1,\abscardval_2)} \mid  \\ 
      \hspace*{3em}
           \cardconslx{\abscardval_1} \in \cardconsset_1, \,
            \cardconslx{\abscardval_2} \in \cardconsset_2  \}
    \end{array}
\end{array}$

\vspace*{0.5ex}
\end{tdisplay}

In the rest of this section, $\mlsLat$ and $\prog$ are fixed,
together with a typing context $\Ga \in \varprog \to \mlsLat$.

A valid constraint set is essentially   
a function from $\mlsl$ and $\varx$ to $\abscardval$.
So $\cardlatleq$ is essentially a pointwise order on functions,
and we ensure that $\cardlatleq$ is antisymmetric.

The cardinality abstraction relies on the abstraction
$\alphacardv$, introduced in~\Cref{sec:collecting}, in order to approximate
$\mlsl$-variety of a variable into a cardinality $\abscardval \in
\abscardValues$. 

\pagebreak

\begin{tdisplay}{Cardinality abstraction \hfill $\cardtrop \quad
    \alphacardtr \quad \gammacardtr$}
\vspace*{-2ex}
\[ \begin{array}{lcl}
\cardtrop & \in & \psetTraces \to \cardLat \\[.3ex]
\cardtrop(T) &  \triangleq  & 
     \{ \cardconslx{\abscardval} \mid 
          \begin{array}[t]{l}
          \mlsl \in \mlsLat,\,  \varx \in \varprog, \\
          \abscardval = \alphacardv(\obsexpcl{x}{T}) \: \}
          \end{array}
          \\[1ex]

\alphacardtr & \in & \psetpsetTraces \to \cardLat\\[.3ex]
\alphacardtr(\BTr) & \triangleq & \cardlatjoin_{\Tr \in \BTr} \cardtrop(T) \\[1ex]

\gammacardtr & \in & \cardLat \to \psetpsetTraces \\[.3ex]
\gammacardtr(\cardconsset) & \triangleq &
     \{ \Tr \mid \cardtrop(\Tr) \cardlatleq \cardconsset \} 
\end{array}\]

\[
(\psetpsetTraces,\subseteq) \galois{\alphacardtr}{\gammacardtr} (\cardLat,\cardlatleq)
\]

\end{tdisplay}

The cardinality abstraction enables us to derive an approximation
$\absobsexpl{e}{}$ of $\mlsl$-variety $\obsexphcl{e}{}$.
This approximation $\absobsexpl{e}{} \in \cardLat \to \abscardValues$, called
$\mlsl$-cardinality of expression $e$, enumerates the $\mlsl$-variety conveyed
to expression $e$ assuming a set $\cardconsset \in \cardLat$ of cardinality
constraints holds. 
Note that the infinite cardinal $\sizeval$ is absorbing, i.e.\ 
$\forall n, \sizeval \times n \triangleq \sizeval$.

\begin{tdisplay}{$\mlsl$-cardinality of expressions \hfill 
$\absobsexpl{e}{} \in \cardLat \to \abscardValues$} 
\vspace{-3ex}
\begin{mathpar}
\absobsexpl{n}{\cardconsset} \triangleq 1

\absobsexpl{x}{\cardconsset} \triangleq \abscardval \text{ where }
\cardcons{l}{x}{n} \in \cardconsset

\absobsexpl{e_1 \oplus e_2}{\cardconsset} \triangleq 
 \absobsexpl{e_1}{\cardconsset} \times
  \absobsexpl{e_2}{\cardconsset}

\absobsexpl{e_1 \bcmp e_2}{\cardconsset} \triangleq 
\min\left(2, \absobsexpl{e_1}{\cardconsset} \times
\absobsexpl{e_2}{\cardconsset}\right)
\end{mathpar}
\vspace{-2ex}
\end{tdisplay}

\begin{restatable}[]{mylem}{lemabsvarietysound}
  \label{lem:soundabsvariety}
$\absobsexpl{e}{}$ is sound:
\[ \forall e, \forall \mlsl, \quad \alphacardv \comp \obsexphcl{e}{} \comp \gammacardtr
{} \dotleq {} 
\absobsexpl{e}{} \enspace . \]
\end{restatable}

We now derive a cardinality abstract semantics by approximating
the relational hypercollecting semantics of~\Cref{sec:overview}.
It uses definitions to follow.

\begin{tdisplay}{Cardinality abstract semantics \hfill 
$\seminshcabs{c}{} \in \cardLat \to \cardLat$}
\vspace{-3ex}
\begin{mathpar}
\seminshcabs{\skipcom}{\cardconsset} \triangleq \cardconsset

\seminshcabs{c_1; c_2}{\cardconsset} \triangleq 
\seminshcabs{c_2}{} \comp \seminshcabs{c_1}{\cardconsset} 

\begin{array}{l}
\seminshcabs{x := e}{\cardconsset} 
\triangleq \\
\hspace*{1em}
     \begin{array}[t]{l}
     \{ \cardcons{\mlsl}{y}{\abscardval} \in \cardconsset \mid y \neq x \} \\
     \cup \{ \cardcons{\mlsl}{x}{n} \mid 
              \mlsl \in \mlsLat,\,  \varx \in \varprog,\, 
              n =  \absobsexpl{e}{\cardconsset}  \: \}
   \end{array}
\end{array}

\begin{array}{l}
\seminshcabs{\ifcom~ b~ \thencom~ c_1~ \elsecom~ c_2}{\cardconsset} 
\triangleq
\\
\hspace*{1em}
\begin{array}[t]{l}
\letin{\cardconsset_1 = \seminshcabs{c_1}{\cardconsset}} \\
\letin{\cardconsset_2 =\seminshcabs{c_2}{\cardconsset}} \\
\letin{W = \Mod(\ifcom~ b~ \thencom~ c_1~ \elsecom~ c_2)} \\
\hspace*{1em}
\bigcup\limits_{\mlsl \in \mlsLat}
   \begin{cases}
     \projtomlsl{\cardconsset_1}
     \cardlatjoin   
     \projtomlsl{\cardconsset_2}
   & \text{if } \absobsexpl{b}{\cardconsset} = 1 \\
     \projtomlsl{\cardconsset_1} 
        \cardlatadd{W,\projtomlsl{\cardconsset}}  
     \projtomlsl{\cardconsset_2} 
   & \text{otherwise}
\end{cases}
\end{array}
\end{array} 

\seminshcabs{\whilecom~ b~\docom~ c}{\cardconsset} 
\triangleq \operatorname{lfp}^{\cardlatleq}_{\cardconsset}
             \seminshcabs{\ifcom~ b~ \thencom~ c_1~ \elsecom~ c_2}{}

\end{mathpar}
\vspace{-4ex}
\end{tdisplay}

\[ \begin{array}{l}
\projtomlsl{\cardconsset}  \triangleq 
\{ \cardconslxn \in \cardconsset \mid
                  \varx \in \varprog, \abscardval \in \abscardValues \}
\\
C_1 \cardlatadd{W,C_0} C_2 
\triangleq
  \begin{array}[t]{l} 
    \bigcup_{\varx \in \varprog{\setminus} W } \{  \cardconslxn \in C_0 \} \\
    \cup 
    \bigcup_{\varx \in W}
        \begin{array}[t]{l}
         \{ \cardconslx{(n_1 \mathord{+} n_2)} \mid \\
            \quad \cardconslx{n_j} \in C_j , \: j=1,2\}
        \end{array}
  \end{array}
\end{array}
\]

\smallskip

The abstract semantics of assignments $x := e$ is similar in spirit to the one
for dependences: discard atomic constraints related to $x$, and add new ones by
computing $\mlsl$-cardinality of expression $e$.
The abstract semantics of conditionals is also similar to dependences: if the
conditional guard does not convey $\mlsl$-variety, then all initially
$\mlsl$-equivalent traces follow the same execution path and the join operator
(defined as $\max$ over cardinality) over both conditional branches
over-approximates the $\mlsl$-cardinality after the conditional.
Otherwise, the $\mlsl$-cardinality over both conditional branches have to be
summed---for the variables that may be modified in the conditional
branches---to soundly approximate the $\mlsl$-cardinality after the
conditional. 

\begin{restatable}[]{mytheo}{theoabscardsem}
\label{thm:soundabscardinality}
The cardinality abstract semantics is sound:
\[ \alphacardtr \comp \seminshc{c}{} \comp \gammacardtr
{} \dotcardlatleq {} 
\seminshcabs{c}{} \enspace . \]
\end{restatable}

The lattice $\cardLat$ is complete, although not finite.  
We may define a widening operator
$\nabla \in \cardLat
\times \cardLat \to \cardLat$ to ensure convergence of the analysis
~\citep{CC92}\citep{NNH99}\citep[Sec. 4]{CZ11}. 
\[ \begin{array}{l}

\cardconsset_1 \nabla \cardconsset_2 \triangleq
 \{ \cardconslx{\abscardval} \mid
     \begin{array}[t]{l}
       \cardconslx{\abscardval_1} \in \cardconsset_1, \,
       \cardconslx{\abscardval_2} \in \cardconsset_2, \\
       \abscardval = \abscardval_1 \nabla \abscardval_2 \}
     \end{array}
\\ 

\abscardval_1 \nabla \abscardval_2 \triangleq \ifcom~ 
(\abscardval_2 \leq \abscardval_1)~
\thencom~ \abscardval_1~ \elsecom~ \sizeval
\end{array}
\]

The occurrence of widening depends on the iteration strategy employed by the 
static analyser.
Widening accelerates or forces the convergence of fixpoint
computations.
In the simplest setting, the analyser passes as arguments to the widening
operator the old set $\cardconsset_1$ of cardinality as well as the new set
$\cardconsset_2$ that is computed. 
For each atomic cardinality constraint, the widening operator then compares the
old cardinality $\abscardval_1$ to the new cardinality $\abscardval_2$.
If the cardinality is still strictly increasing
($\abscardval_2 > \abscardval_1$), the widening forces the convergence by
setting it to $\infty$.
If the cardinality is decreasing, the widening operator sets it to the maximum
cardinality $\abscardval_1$ in order to force convergence and ensure the
sequence of computed cardinalities is stationary.

\myparagraph{Min-capacity leakage}
So far, we showed how one can derive static analyses of hyperproperties---the
abstract representations themselves are interpreted as hyperproperties---by
approximating hypercollecting semantics.
Let us now recall the security requirement $\secreqlkx$ introduced in
\Cref{sec:overview} in order to illustrate how these analyses may prove that a
program satisfies a hyperproperty, i.e.\ Step 3 of the methodology
in \Cref{sec:collecting} (see also~\Cref{eq:soundXH}).

Consider a program $\prog$ characterised by a set $\Trprog \in \psetTraces$ of traces,
i.e.\  $\Trprog$ is $\seminsc{\prog}{\InitTraces}$.
How do we prove that $\prog$ satisfies the hyperproperty $\secreqlkx$?
We can use the cardinality analysis to prove that variable $x$ has a
$\mlsl$-cardinality that is at most $\kparam$.
Indeed, if $\cardconsset$ approximates $\Trprog$
(i.e.\  $\alphacardtr(\{ \Trprog \}) \cardlatleq \cardconsset$) then 
\( \alphacardv \comp  \obsexpcl{x}{\Trprog} \leq \absobsexpl{x}{\cardconsset} \).
Thus, if the inferred $\mlsl$-cardinality of $\cardconsset$ is 
at most $k$ 
then program $\prog$ is guaranteed to satisfy
the hyperproperty $\secreqlkx$.
We have $\{ \Trprog \} \subseteq \gammacardtr (\cardconsset)$ 
since $\cardconsset$ approximates $\Trprog$ 
(i.e.\  $\alphacardtr(\{ \Trprog \}) \cardlatleq \cardconsset$).
And we have $\gammacardtr (\cardconsset) \subseteq  \secreqlkx$
by assumption $\absobsexpl{x}{\cardconsset} \leq k$.
Hence $\Trprog \in  \secreqlkx$.

The hyperproperty $\secreqlkx$ is a \dt{$\mathbf{(\kparam+1)}$-safety
  hyperproperty}~\citep{CS10}, i.e.\ it requires exhibiting at most $\kparam+1$ traces
in order to prove that a program does not satisfy $\secreqlkx$. 
For example, termination-insensitive noninterference for security level $\mlsl$,
which corresponds to the hyperproperty $\secreql{1}{x}$, is 2-safety.
A $k$-safety hyperproperty of a program can be reduced to a safety property of a
$k$-fold product program~\citep{BAR04,TA05,DHS05,CS10}.

Various quantitative information flow properties are not $k$-safety. 
For example, the bounding problem that the cardinality analysis targets, namely
min-capacity leakage
is not a $\kparam$-safety hyperproperty for any
$\kparam$~\citep[Sec. 3]{YT11}. %
Instead, this bounding problem is hypersafety~\citep{CS10}.

\myparagraph{Cardinalities vs.\ dependences}
Just as quantitative security metrics are the natural generalisations of
qualitative metrics such as noninterference, the cardinality abstraction is a
natural generalisation of dependence analysis.  
Instead of deciding if variety is conveyed, the cardinality analysis
enumerates this variety.  
In other words, dependences are abstractions of cardinalities. 
We can factor the Galois connections, e.g.\   %
$(\alphaagreev,\gammaagreev)$ is $(\alphalone \comp \alphacardv,\gammacardv
\comp \gammalone)$ for suitable $(\alphalone,\gammalone)$.

\begin{restatable}[]{mylem}{lemalphaagreevdec}
\label{lem:alphaagreevdec}
 $(\alphaagreev,\gammaagreev)$ is the composition 
 of two  Galois connections
$(\alphacardv,\gammacardv)$ and
$(\alphalone,\gammalone)$ :
\[ 
(\psetpsetValues,\subseteq) \galois{\alphacardv}{\gammacardv} 
(\abscardValues, \leq)
 \galois{\alphalone}{\gammalone} (\abstruth, \abstruthleq) \]

with:
\[
\begin{array}{l}
\alphalone(\abscardval) \triangleq 
\begin{cases}
\true & \text{if } \abscardval \leq 1 \\
\false & \text{otherwise.}
\end{cases}, \text{ and } 
\\
  \gammalone(\absbool) \triangleq 
  \begin{cases}
    1 & \text{if } \absbool = \true \\
    \sizeval & \text{otherwise.}
  \end{cases} 
\end{array} \]
\end{restatable}

\begin{restatable}[]{mylem}{lemalphadeptrdec}
\label{lem:alphadeptrdec}
$(\alphadeptr,\gammadeptr)$ is the composition of two
Galois connections
$(\alphacardtr,\gammacardtr)$ and
$(\alphalonecc,\gammalonecc)$ :
\[
(\psetpsetTraces,\subseteq) \galois{\alphacardtr}{\gammacardtr}
(\cardLat,\cardlatleq)
 \galois{\alphalonecc}{\gammalonecc}
(\depLat,\deplatleq)
\]

with:
\[
\begin{array}{l}
\alphalonecc(\cardconsset) \triangleq \{ \depconslx \mid \cardconslxn \in 
 \cardconsset \text{ and }  \alphalone(n)  \}
\\
\gammalonecc(\depconsset)
\triangleq 
\bigcup\limits_{\mlsl \in \mlsLat, \varx \in \varprog}
\{ \cardconslxn \mid 
 n =
\gammalone( \depconslx \in \depconsset) \}
\end{array}
\]
\end{restatable}

We use \Cref{lem:alphaagreevdec,lem:alphadeptrdec} to abstract further the
cardinality abstract semantics and derive the correct by construction
dependence analysis of \Cref{sec:dependences}. 
This derivation, which can be found in
\iftechreport
\Cref{ap:dep_reloaded},
\else
\citet{Aal16b},
\fi
proves
\Cref{lem:soundabsvarietydep} and 
\Cref{thm:soundabsdependences} stated earlier.

As a corollary and by \Cref{thm:theohsprecision}, this also proves the precision
of the cardinality analysis relative to Amtoft and Banerjee's logic~\citep{AB04}
as well as Hunt and Sands' type system~\citep{HS06,HS11sd}. 
\begin{restatable}[No leakage for well-typed programs]{mycor}{corprecision}
  \label{cor:precision}
 For all $c$, 
$\cardconsset_0, \cardconsset \in \cardLat$, 
  $\Delta_0, \Delta \in \varprog \to \mlsLat$, 
  where $\mlsbot \vdash \Delta_{0} \{ c \} \Delta$, 
  and 
  $\cardconsset = \seminshcabs{c}{\cardconsset_0}$,
  it holds that:  %
\begin{align*} 
& \alphatohs \comp \alphalonecc (\cardconsset_0)  
\dotmlslatleq \Delta_0
 \implies \\
& \qquad \left( 
\forall \varx \in \varprog, \mlsl \in \mlsLat, 
\quad \Delta(x) \mlslatleq \mlsl 
\implies
\absobsexpl{x}{} \leq 1
\right)
\end{align*}
\end{restatable}

The cardinality analysis determines that there is no leakage for programs that are
``well-typed'' by the flow-sensitive type system of Hunt and Sands.
By ``well-typed'', we mean that the final typing environment that is computed by
the type system allows attackers with security clearance $\mlsl \in \mlsLat$ to
observe a variable $\varx \in \varprog$.  

To the best of our knowledge, the cardinality abstraction is the first
approximation-based analysis for quantitative information flow that 
provides a formal precision guarantee wrt.\ traditional analyses for
qualitative information flow.
This advantage makes the cardinality analysis appealing even when interested
in proving a qualitative security policy such as non-interference, since the
cardinality abstraction provides quantitative information that may assist
in making better informed decisions if declassification is necessary.
Nonetheless, we need further experimentation to compare to other quantitative
analyses ---see \Cref{sec:related}.

\section{Towards More Precision}
\label{sec:precision}

This section introduces examples to evaluate the precision of the analyses, and
shows how existing analyses can be leveraged to improve precision. 
For simplicity, we consider a two point lattice $\{L,H\}$  and an initial typing
context  where variables $y_i$ are the only low
variables ($\Ga(y_i) = L$).
As is usual, low may flow to high ($L \mlslatleq H $).

Consider the following program.
{\begin{lstlisting}[style=simple,float=false,label={lst:example1},caption={Leaking
    1 bit of {\ttfamily secret}}]
if (@y\textsubscript{1}@ >= secret) then
   x := @\text{y\textsubscript{2}}@
else
   x := @\text{y\textsubscript{3}}@
\end{lstlisting}}

The cardinality abstraction determines that $x$ has
at most 2 values after the execution of the program in \Cref{lst:example1}, for
initially $L$-equivalent traces.
For fixed low inputs,  $x$ has one value in the then branch and one value
in the else branch, and these cardinalities get summed after the conditional since the
conditional guard may evaluate to 2~different values.
Thus, the cardinality abstraction proves that this example program satisfies the
hyperproperty  $\secreq{L}{2}{x}$.

\myparagraph{Stronger trace properties} Another way
of proving a hyperproperty is by proving a stronger trace property. 
If a program is proven to satisfy a trace property
$\Tr \in \psetTraces$, then proving that $\Tr$ is stronger than 
hyperproperty $H \in \psetpsetTraces$---in the sense that $\gammahpp(\Tr) \subseteq H$---guarantees
the program satisfies the hyperproperty~$H$.
For instance, by proving for some program that an output variable $x$ ranges over
an interval of integer values whose size is $\kparam$, we can prove that program
satisfies $\secreq{L}{k}{x}$. 

However, approximating a hyperproperty by a trace property may be too coarse
for some programs, as we can illustrate with an interval analysis~\citep{CC77}
on the example program in \Cref{lst:example1}.
Such an interval analysis loses too much precision in the initial state of this
program, since it maps all low input variables $y_1$, $y_2$ and $y_3$ to
$[-\infty,+\infty]$. 
After the conditional, it determines that $x$ belongs to the interval
$[-\infty,+\infty]$, which is a coarse over-approximation.
Also, a polyhedron~\citep{CH78} does not capture the disjunction that is needed
for this example program ($x = y_2$ or $x = y_3$).
Both abstract domains and many more existing ones are not suitable for
the task of inferring cardinalities or dependences because they are
convex.   
Using them as a basis to extract counting information delivers an
over-approximation of the leakage, but a coarse one, especially in the
presence of low inputs.

A disjunction of two polyhedra ---through powerset domains, disjunctive postconditions, or
partitioning~\citep{Bou92}--- is as precise as the cardinality analysis for this example.
However, disjunctions are not tractable in general.
As soon as one fixes a maximum number of disjunctive
elements (as in the quantitative information flow analysis  of
\citet{Mal.11,Mal.13}) or defines a widening operator to guarantee  
convergence, one loses the relative precision wrt.\ classical dependence
analyses~\citep{AB04,HS06} that the cardinality analysis guarantees
(Cf.\ \Cref{cor:precision}).
Future work will investigate relying on cardinality analysis as a strategy
guiding trace partitioning~\citep{RM07}.
Combining our analyses with existing domains will also deliver better precision.

Consider the following program.
{\begin{lstlisting}[style=simple,float=false,label={lst:example2},caption={Leaking
    x}]
if (@y\textsubscript{1}@ >= secret) then x := @\text{y\textsubscript{2}}@ else x := @\text{y\textsubscript{3}}@;
o := x * @y\textsubscript{4}@ 
\end{lstlisting}}
The cardinal abstraction determines that variable $o$ leaks the two possible
values of $x$: for fixed low inputs, x has two possible values whereas $y_4$ has
one possible value.
Relational abstract domains such as polyhedra~\citep{CH78} or
octogons~\citep{Min06} do not support non-linear expressions, and therefore are
unable to compute a precise bound of the leakage for variable~$o$.
Consider an analysis with a disjunction $\{ x = y_2 \vee x=y_3 \}$
of polyhedra and linearisation over intervals~\citep{Min06b}.
Linearisation of expressions $y_2 * y_4$ and $y_3 * y_4$  will compute
the following constraints for variable $o$:
$\{(o = y_2*[-\infty,+\infty]) \vee (o = y_3*[-\infty,+\infty])\}$ if
linearisation happens for the right side of expressions, or
constraint $\{(o = [-\infty,+\infty]*y_4)  \vee (o=[-\infty,+\infty]*y_4)\}$ if
linearisation happens for the left side expressions.
Two more combinations of constraints are possible, but none will deduce that variable $o$
has at most 2 values, because the underlying domain of intervals lacks the
required precision.
Linearisation over both intervals and cardinalities delivers better precision.

\myparagraph{Scaling to richer languages}
We can rely on existing abstract domains to support richer language constructs,
e.g. pointers and aliasing.  
Consider the following variation of \Cref{lst:example1}.

\begin{minipage}{0.9\columnwidth}
{\begin{lstlisting}[style=simple,float=false,numbers=none,label={lst:example3},caption={Leaking
    1 bit of {\ttfamily secret}}]
if (@y\textsubscript{1}@ >= secret) then
   p := &@\text{y\textsubscript{2}}@
else
   p := &@\text{y\textsubscript{3}}@
o := *p
\end{lstlisting}}
\end{minipage}

The cardinality abstraction determines that initially $L$-equivalent memories
lead to a variety of at most 2 in the pointer $p$ after the conditional, whereas
both $y_2$ and $y_3$ have a variety of 1.
Assuming an aliasing analysis determines that $p$ may point to $y_2$ or $y_3$,
the cardinality analysis determines that variable $o$ has a variety of at most
2, for initially $L$-equivalent memories.

\myparagraph{Improving precision}
To improve precision of the cardinality abstraction, we can augment it with
existing abstract domains. %
One shortcoming of the cardinality analysis is the fact
that it is not relational.
Assuming attackers with security clearance $L$ observe both variables $x$ and
$o$ after execution of the program in \Cref{lst:example2}, the cardinality abstraction
leads us to compute a leakage of two bits: four different possible values, instead of
only 2 possible values for initially L-equivalent memories.
Relying on a relational domain with linearisation~\citep{Min06b} over
cardinalities captures the required constraints 
$\{\cardcons{L}{x}{2},\cardcons{L}{o}{1*x} \}$ to compute a leakage of only one
bit; these constraints are to be interpreted as ``initially $L$-equivalent
memories result in $o$ being equal to one fixed integer times $x$, and $x$
having at most 2 values''.

We leave these extensions of cardinality analysis ---and its abstraction
as dependence analysis--- for future work. 
In the following, we focus on one particular improvement to both previous analyses
in order to gain more precision.
We uncovered this case while deriving the analyses, by relying on the
\dt{calculational framework of abstract interpretation}.
Indeed, notice that the following holds:
\begin{align*}
  \alphacardv \comp \obsexphcl{\varx_1}{} \comp \guardhc{\varx_1 == \varx_2}{}
  \comp
  \gammacardtr(\cardconsset)
  \leq \absobsexpl{\varx_2}{\cardconsset} \\
   \alphacardv \comp \obsexphcl{\varx_2}{} \comp \guardhc{\varx_1 == \varx_2}{}
   \comp
  \gammacardtr(\cardconsset)
  \leq \absobsexpl{\varx_1}{\cardconsset} \\
\end{align*}

Therefore, we can deduce that:
\begin{align*}
& \alphacardtr \comp \guardhc{\varx_1 == \varx_2}{} \comp
\gammacardtr(\cardconsset)
\\ 
& \quad 
\cardlatleq
\{ \cardconslxn \in \cardconsset \mid \varx \neq \varx_1,\varx \neq \varx_2 \}
\\
& \qquad
{} \cup {}
\{ 
\cardcons{\mlsl}{\varx_1}{\min(\abscardval_1,\abscardval_2)}, 
\cardcons{\mlsl}{\varx_2}{\min(\abscardval_1,\abscardval_2)} \mid 
\\
& \qquad \qquad
\cardcons{\mlsl}{\varx_1}{\abscardval_1} \in \cardconsset,
\cardcons{\mlsl}{\varx_2}{\abscardval_2} \in \cardconsset
\} 
\\
& \quad \triangleq 
\guardhcabs{x_1 == x_2}{\cardconsset}
\end{align*}
For other comparison operators, we use as before
$\guardhcabs{b}{\cardconsset} \triangleq \cardconsset$.

We can now also improve the dependence abstraction:
\begin{align*} %
& \alphalonecc \comp \guardhcabs{x_1 == x_2}{} \comp
\gammalonecc(\depconsset)
\\ 
& \quad 
\deplatleq
\alphalonecc\left(
\{ \cardconslxn \in \gammalonecc(\depconsset) \mid \varx \neq \varx_1,\varx \neq
\varx_2 \}\right)
\\
& \qquad
{} \cup {}
\alphalonecc(
\{ 
\cardcons{\mlsl}{\varx_1}{\min(\abscardval_1,\abscardval_2)}, 
\cardcons{\mlsl}{\varx_2}{\min(\abscardval_1,\abscardval_2)} \mid 
\\
& \qquad \qquad
\cardcons{\mlsl}{\varx_1}{\abscardval_1} \in \gammalonecc(\depconsset),
\cardcons{\mlsl}{\varx_2}{\abscardval_2} \in \gammalonecc(\depconsset)
\}  )
\\
&
\quad \deplatleq
\{ \depconslx \in \depconsset \mid \varx \neq \varx_1, \varx \neq \varx_2 \} 
\\
&
\qquad
{} \cup {}
\{ \depcons{\mlsl}{\varx_1},\depcons{\mlsl}{\varx_2} \mid 
\depcons{\mlsl}{\varx_1} \in \depconsset \text{ or }
\depcons{\mlsl}{\varx_2} \in \depconsset \} \\
& \quad \triangleq 
\guardhcdepabs{x_1 == x_2}{\depconsset}
\end{align*}
For other comparison operators, we also use
$\guardhcdepabs{b}{\depconsset} \triangleq \depconsset$.

With these new definitions, we can update the abstract semantics of conditionals
and loops, for both dependences and cardinalities,
to leverage the transfer functions $\guardhcdepabs{-}{}$ and
$\guardhcabs{-}{}$.  

\begin{tdisplay}{Improved dependences abstract semantics \hfill 
$\seminshcdepabs{c}{} \in \depLat \to \depLat$}
\vspace{-2ex}
\begin{mathpar}
\begin{array}{l}
\seminshcdepabs{\ifcom~ b~ \thencom~ c_1~ \elsecom~ c_2}{\depconsset} \triangleq
\\
\hspace*{3em}
\begin{array}[t]{l}
  \letin{\depconsset_1 = \guardhcdepabs{b}{} \comp \seminshcdepabs{c_1}{\depconsset}} \\
  \letin{\depconsset_2 = \guardhcdepabs{\neg b}{} \comp\seminshcdepabs{c_2}{\depconsset}} \\
  \letin{W = \Mod(\ifcom~ b~ \thencom~ c_1~ \elsecom~ c_2)} \\
  \quad
  \bigcup\limits_{\mlsl \in \mlsLat} 
    \begin{cases}
      \projtomlsl{\depconsset_1}
      \deplatjoin   
      \projtomlsl{\depconsset_2}
    & \text{if } \absobsexpdepl{b}{\depconsset}  \\
      \{ \depcons{\mlsl}{x} \in \projtomlsl{\depconsset} \mid x \notin W \} 
    & \text{otherwise}
    \end{cases}
\end{array}
\end{array}

\seminshcdepabs{\whilecom~ b~\docom~ c}{\depconsset} 
\triangleq \guardhcdepabs{\neg b}{} \comp \operatorname{lfp}^{\cardlatleq}_{\depconsset}
\seminshcdepabs{\ifcom~ b~ \thencom~ c_1~ \elsecom~ c_2}{}

\end{mathpar}
\vspace{-2ex}
\end{tdisplay}

\begin{tdisplay}{Improved cardinality abs. semantics 
\hfill 
$\seminshcabs{c}{} \in \cardLat \to \cardLat$}
\vspace{-3ex}
\begin{mathpar}
\begin{array}{l}
\seminshcabs{\ifcom~ b~ \thencom~ c_1~ \elsecom~ c_2}{\cardconsset} 
\triangleq
\\
\hspace*{1em}
\begin{array}[t]{l}
\letin{\cardconsset_1 = \guardhcabs{b}{} \comp \seminshcabs{c_1}{\cardconsset}} \\
\letin{\cardconsset_2 = \guardhcabs{\neg b}{} \comp \seminshcabs{c_2}{\cardconsset}} \\
\letin{W = \Mod(\ifcom~ b~ \thencom~ c_1~ \elsecom~ c_2)} \\
\hspace*{1em}
\bigcup\limits_{\mlsl \in \mlsLat}
   \begin{cases}
     \projtomlsl{\cardconsset_1}
     \cardlatjoin   
     \projtomlsl{\cardconsset_2}
   & \text{if } \absobsexpl{b}{\cardconsset} = 1 \\
     \projtomlsl{\cardconsset_1} 
        \cardlatadd{W,\projtomlsl{\cardconsset}}  
     \projtomlsl{\cardconsset_2} 
   & \text{otherwise}
\end{cases}
\end{array}
\end{array} 

\seminshcabs{\whilecom~ b~\docom~ c}{\cardconsset} 
\triangleq \guardhcdepabs{\neg b}{} \comp \operatorname{lfp}^{\cardlatleq}_{\cardconsset}
             \seminshcabs{\ifcom~ b~ \thencom~ c_1~ \elsecom~ c_2}{}

\end{mathpar}
\vspace{-4ex}
\end{tdisplay}

To illustrate the benefits of this improvement, consider the following example.
{\begin{lstlisting}[style=simple,float=false,label={lst:example4},caption={Improved
    precision}]
while (secret != @y\textsubscript{3}@) do {
  x := x+1;
  secret := secret - 1;
}
o := secret;
\end{lstlisting}}

The cardinality analysis determines that initially $L$-equivalent
memories result in  $x$ having an infinity of values: the $L$-cardinality of $x$ grows
until it is widened to $\infty$.
In contrast, cardinalities also determine that variables $o$ and $secret$
have only 1 value, assuming $L$-equivalent memories. 
This is because of the reduction that concerns variable $secret$ after the while
loop,
specifically 
$\guardhcdepabs{secret==y_3}{}$.
Similarly, the improved dependence analysis also determines that both variables
$secret$ and $o$ are low.
These are sound precision gains for termination-insensitive noninterference;
\citet{Aal.08} discusses the guarantees provided by this security requirement.

Remarkably, this has been overlooked by many previous analyses. 
In fact, this simple improvement makes our dependence analysis strictly more
precise than \citet{AB04}'s and \citet{HS06,HS11sd}'s analyses and incomparable
to the more recent dependence analysis of \citet{MKS15}.
\begin{minipage}{0.9\columnwidth}
\myparagraph{Combination with intervals}
Consider now the following example inspired from \citet{MKS15}.
{\begin{lstlisting}[style=simple,float=false,label={lst:example5},caption={Example
      program from \citet{MKS15}}]
if (secret == 0) then {
   x := 0;
   y := y + 1;
}
else {
   x := 0;
}
\end{lstlisting}}
\end{minipage}

The analysis of \citet{MKS15} determines that $x$ is low, whereas
the cardinality abstraction determines that $L$-equivalent memories result in
at most 2 values for variable $x$, because it does not track the actual values of
variables.
We can combine cardinality with an interval analysis to
be more precise in such cases, through a reduced
product~\citep{CC79,Gran92,CCP13}.

Assume a set $\statesint$ of interval environments provided with the usual
partial order that we denote by $\dotleqintervals$. 
Assume also a Galois connection $(\alphaint,\gammaint)$ enabling the derivation of an interval
analysis as an approximation of a standard collecting semantics
defined over $\psetTraces$. 
We can lift this Galois connection to $\psetpsetTraces$ to obtain a Galois
connection by compositing with $(\alphahpp,\gammahpp)$, to obtain
$(\alpha',\gamma') \triangleq (\alphaint \comp \alphahpp,\gammaint \comp
\gammahpp)$ with:
\[
(\psetpsetTraces,\subseteq)
\galois{\alphahpp}{\gammahpp}
(\psetTraces,\subseteq) \galois{\alphaint}{\gammaint}
(\statesint,\dotleqintervals)
\]

A Granger's reduced product~\citet{Gran92} for the cardinality abstraction and
an interval analysis may be defined as a pair of functions 
$\redtoint \in \cardLat \times \statesint \to \statesint$ and
$\redtocard \in \cardLat \times \statesint \to \cardLat$ verifying the following
conditions:
\begin{enumerate}
\item soundness:
  
\qquad$\begin{array}[t]{ll}
  \gamma'(\redtoint(\cardconsset,\imath)) \cap \gammacardtr(\cardconsset) &
  = \; \gamma'(\imath) \cap  \gammacardtr(\cardconsset) \\[0.5ex]
    \gamma'(\imath) \cap \gammacardtr(\redtocard(\cardconsset,\imath)) &
  = \; \gamma'(\imath) \cap \gammacardtr(\cardconsset)
\end{array}$
  
\item reduction:
  
 \qquad$\begin{array}[t]{ll}
      \redtoint(\cardconsset,\imath) & \dotleqintervals \; \imath \\
    \redtocard(\cardconsset,\imath) & \cardlatleq \; \cardconsset
  \end{array}$
\end{enumerate}

Let us denote by $\sizeint$ the function that returns the size of an interval. 
One such Granger's reduced product can be defined as:
\[\begin{array}{lcl}
  \redtocard & \in &\cardLat \times \statesint \to \cardLat \\[.5ex]
 \redtocard(\cardconsset,\imath) & \triangleq &
 \{ \cardconslx{\abscardval'} \mid 
      \begin{array}[t]{l}
        \cardconslx{\abscardval} \in \cardconsset \: \text{and} \\
        \abscardval' =  \min\left(\abscardval, \sizeint~ \imath(\varx) \right)
        \} 
      \end{array}
\\[.5ex]
 \redtoint & \in& \cardLat \times \statesint \to \cardLat \\[.5ex]
 \redtoint(\cardconsset,\imath) & \triangleq &\imath
\end{array}\]

Once enhanced with this reduced product, the cardinality analysis determines
for the program in \Cref{lst:example5}, that $L$-equivalent memories result in
at most one possible value for variable $x$.

The dependence analysis can be improved similarly, with a reduction function
defined as follows:
\begin{align*}
  \redtodep & \in \depLat \times \statesint \to \depLat \\
 \redtodep(\depconsset,\imath) & \triangleq 
 \depconsset \cup \{ \depconslx \mid \mlsl \in \mlsLat \text{ and }
 \sizeint~ \imath(\varx) = 1 \}
\end{align*}
Once extended with a reduced product with intervals, the dependence analysis is
also able to determine that variable $x$ is low for the program in \Cref{lst:example5}.

\myparagraph{More reduced products}
As a final example, let us consider \Cref{lst:example6}, inspired
by~\citet[program 7]{BBJ16}, that we annotate with the result of the improved cardinality
abstraction.
To the best of our knowledge, no existing automated static analysis determines
that variable $o$ is low at the end of this program.
Also, no prior monitor but the one recently presented by~\citet{BBJ16} accepts
all executions of this program, assuming attackers with clearance $L$ can
observe variable $o$. 
{\begin{lstlisting}[style=simple,float=t,label={lst:example6},caption={No
      leakage for variable $o$},firstnumber=0]
//@$\color{gray}\small \cardcons{L}{h}{\infty},\cardcons{L}{y_1}{1},\cardcons{L}{y_2}{1},\cardcons{L}{y_3}{1}$@
@y\textsubscript{1}@ := 1;//@$\color{gray}\small\cardcons{L}{y_1}{1}$@
if (h == @y\textsubscript{1}@) then {
   skip; //@$\color{gray}\small \cardcons{L}{h}{1},\cardcons{L}{y_1}{1},\cardcons{L}{y_2}{1}$@
} 
else {
   @y\textsubscript{2}@ := 5;  //@$\color{gray}\small \cardcons{L}{y_1}{1},\cardcons{L}{y_2}{1}$@
   while (@y\textsubscript{2}@ != 1) do {
      @y\textsubscript{2}@ := @y\textsubscript{2}@-1;//@$\color{gray}\small \cardcons{L}{y_2}{1}$@
      @y\textsubscript{1}@ := @y\textsubscript{2}@;//@$\color{gray}\small \cardcons{L}{y_1}{1}$@
   }//@$\color{gray}\small \cardcons{L}{y_1}{1},\cardcons{L}{y_2}{1}$@
}
//@$\color{gray}\small  \cardcons{L}{h}{\infty},\cardcons{L}{y_1}{2},\cardcons{L}{y_2}{2},\cardcons{L}{y_3}{1}$@
o := @y\textsubscript{1} * y\textsubscript{3}@;//@$\color{gray}\small  \cardcons{L}{o}{2}$@
\end{lstlisting}}

For initially $L$-equivalent memories, the cardinality abstraction determines
that variables $y_1$, $y_2$ and $o$ have at most two values.
This result is precise for $y_2$, but not precise for $y_1$ and $o$.
As a challenge, let us see what is required to gain more precision to
determine that both variables $y_1$ and $o$ have at most 1 possible value
-- they are low.

To tackle this challenge, we need to consider cardinality combined with an
interval analysis and a simple relational domain tracking equalities.
With the equality $y_1 = y_2$ at the exit of the loop, both $y_1$ and $y_2$ will be
reduced to the singleton interval $[1,1]$.
After the conditional, we still deduce that $y_2$ has at most 2 different values
thanks to the cardinality abstraction.
Using intervals, we deduce that variable $y_1$ has only one value (singleton
interval $[1,1]$).
And finally, at the last assignment the cardinalities abstraction determines
that variable $o$ has only one possible value.
Similarly, this same combination of analyses can be put to use to let the
dependence analysis reach the desired precision.

\section{Related Work}
\label{sec:related}

Although noninterference has important applications, for many security
requirements it is too strong. 
That is one motivation for research in quantitative information flow analysis.
In addition, a number of works investigate weakenings
of noninterference and downgrading policies that are conditioned on events or
data values~\citep{AskarovS07,BNR08,SS09,MB11}.
\citet[Chapter~4]{Ass15} proposes to take the guarantees provided by
termination-insensitive noninterference~\citep{Aal.08} as an explicit definition
for security; this \emph{Relative Secrecy} requirement is inspired
by~\citet{VS00} who propose a type-system preventing batch-job programs from
leaking secrets in polynomial time.
\citet{GiacobazziM04} introduce \emph{abstract noninterference}, which
generalizes noninterference by means of abstract interpretations that specify,
for example, limits on the attacker's power and the extent of partial releases (declassification). 
The survey by \citet{Mastroeni13} further generalizes the notion and highlights,
among other things, its applicability to a range of underlying semantics. 
The Galois connections in this work are at the level of trace sets, not sets of sets. 
Abstract noninterference retains the explicit 2-run formulation~\citep{VIS96,SM03}:
from two related initial states, two executions lead to related final states.
The relations are defined in terms of abstract interpretations of the individual
states/executions.
\citet{MB11} show how to infer indistinguishability relations---modelling
attackers' observations---to find the best abstract noninterference policy that
holds. 
The inference algorithm iteratively refines the relation
by using counter-examples and abstract domain completion~\citep{CC79}.

Set-of-sets structures occur in work on abstraction for nondeterministic programs, 
but in those works one level of sets are powerdomains for nondeterminacy; the
properties considered are trace properties~\citep{Schmidt09,Schmidt12}. 
\citet{HuntS91} develop a binding time analysis and a strictness analysis
\citep{Hunt90} based on partial equivalence relations: 
Their concretisations are sets of equivalence classes.
\citet{CC94} point out that this analysis could be achieved by a 
collecting semantics over sets-of-sets, defined simply as a direct image.
To the best of our knowledge this has not been explored further in the
literature, except in unpublished work on which this paper
builds~\citep{Ass15,Aal16a}.

\citet{ClarksonFKMRS14,FinkbeinerRS15} extend temporal logic with means to quantify over multiple traces in order to express hyperproperties, 
and provide model checking algorithms for finite space systems.
\citet{AgrawalB16} introduce a technique for runtime verification of $k$-safety properties.

The dependences analysis we derive is similar to the information flow logic of
\citet{AB04} and the equivalent  
flow-sensitive type system of \citet{HS06}. 
Amtoft and Banerjee use the domain $\pset{\Traces}$  
and on the basis of a relational logic they validate a forward analysis.
In effect their interpretation of ``independences'' is a Galois connection with sets of sets,
but the analysis is not formulated or proved correct as an abstract interpretation.
To deal with dynamically allocated state, \citet{ABB06}
augment the relational assertions of information flow logic with
region assertions,  which can be computed by abstract interpretation.
This is used both to express agreement relations between the two executions 
and to approximate modifiable locations.
This approach is generalized in~\citet{BanerjeeNN16} to a relational Hoare logic for object-based programs
that encompasses information flow properties with conditional downgrading~\citep{BNR08}. 

\citet{MKS15} give a %
backwards analysis that infers dependencies and is proved strictly more precise
than~\citep{HS06,AB04}.  This is achieved by product construction that
facilitates inferring relations between variables in executions that follow
different control paths.   
Correctness of the analysis is proved by way of a relational Hoare logic.  
The variations of our proposed analyses, in \Cref{sec:precision}, rivals theirs
in terms of precision---they are incomparable. 

Our dependence analysis relies on  an approximation of the
modifiable variables, to soundly track implicit flows due to
control flow, instead of labelling a program counter variable $pc$ to account
for implicit flows~\citep{SM03}.  
\citet{ZC11} also derive a similar analysis through a
syntactic Galois connection---a syntactic assignment  $z := x * y$ is
abstracted into a propositional formula  $x \to z \wedge y \to z$ denoting an
information flow from variables $x$ and $y$ to variable $z$.
The soundness of this analysis wrt.\ a semantic property such as
noninterference requires more justification, though it is remarkable that the
concretisation of propositional formula yields, roughly speaking, a set of
program texts. 
\citet{Zan02} also provides an abstract interpretation account of
a flow-insensitive type system~\citep{VIS96} enforcing noninterference by
guaranteeing a stronger safety property, namely that sensitive locations should
not influence public locations~\citep{Bou08}.\looseness=-1

\citet{KSF13}
explicitly formulate termination-insensitive noninterference as an abstract interpretation,
namely the ``merge over all twin computations'' that makes explicit both the
2-safety aspect and the need for an analysis to relate some aligned intermediate
states. 
Their analysis, like many others, is based on reducing the problem to a safety
property of product programs. 
\citet{SousaD16} implement an algorithm that automates reasoning in a Hoare logic 
for $k$-safety, implicitly constructing product programs;
the performance compares favorably with explicit construction of product programs.
Program dependency graphs are another approach to dependency,
shown to be correct for noninterference 
by~\citet{Wasserrab09} using slicing and a simulation argument.\looseness=-1

\citet[Chap. 5]{Den82} proposes the first quantitative measure of a
program's leakage in terms of Shannon entropy~\citep{Sha48}.
Other quantitative metrics emerge in the
literature~\citep{BCP09,CMS09,Smi09,Dwo11,Smi11,CCS12}.
These quantitative security metrics model different scenarios suitable
for different policies.
Most existing static analyses for quantitative information flow leverage
existing model checking tools and abstract domains for safety;
they prove that a program satisfies a quantitative security requirement by
proving a stronger safety property.
In contrast, the cardinal abstraction proves a hyperproperty by inferring a 
stronger hyperproperty satisfied by the analysed program.
This is key to target quantitative information flow in mutlilevel security
lattices, beyond the 2-point lattice $\{L,H\}$.

\citet{BKR09} synthesize equivalence classes induced by outputs over
low equivalent memories by relying on software model checkers, in order to bound
various quantitative metrics.
\citet{HM09} also rely on a similar technique to quantify
information flow for database queries.
\citet{KR10} note that the exact computation of
information-theoretic characteristics is prohibitively hard, and propose
to rely on approximation-based analyses, among which are randomisation
techniques and abstract interpretation ones. 
They also propose to rely on a self-composed product program to model a scenario where
attackers may refine their knowledge by influencing the low inputs.
\citet{Kle14} relies on similar techniques to handle programs with
 low inputs, and uses polyhedra to synthesize linear constraints~\citep{CH78}
 over  variables.
 \citet{Mal.13} decide whether answering a query on sensitive
 data augments attackers'  knowledge  beyond a  certain threshold, by using
 probabilistic polyhedra.

\section{Conclusion}

Galois connection-based semantic characterisations of program analyses
provide new perspectives and insights that lead to improved techniques.
We have extended the framework to fully encompass hyperproperties,
through a remarkable form of hypercollecting semantics that 
enables calculational derivation of analyses.
This new foundation raises questions too numerous to list here.

One promising direction is to 
combine dependence and cardinality analysis with existing abstract domains, 
e.g.\ through advanced symbolic methods~\citep{Min06b}, and
partitioning~\citep{HT98,RM07}.

Static analysis of secure information flow has yet to catch up with recent
advances in dynamic information flow
monitoring~\citep{BBJ13,BelloHS15,HedinBS15,AN16,BBJ16}. 
We discussed, in \Cref{sec:precision}, how existing static analyses may
be of use to statically secure information flow. 
It seems likely that hypercollecting semantics will also be of use for dynamic analyses.

\acks
Thanks to Anindya Banerjee and the anonymous reviewers for thoughtful comments and
helpful feedback.   
This work was partially supported by NSF awards CNS-1228930 and CCF-1649884,
ANR project AnaStaSec ANR-14-CE28-0014 and a CFR CEA Phd Fellowship.

\bibliographystyle{abbrvnat}

\balance
\bibliography{biblio}

\begin{thebibliography}{90}
\providecommand{\natexlab}[1]{#1}
\providecommand{\url}[1]{\texttt{#1}}
\expandafter\ifx\csname urlstyle\endcsname\relax
  \providecommand{\doi}[1]{doi: #1}\else
  \providecommand{\doi}{doi: \begingroup \urlstyle{rm}\Url}\fi

\bibitem[Agrawal and Bonakdarpour(2016)]{AgrawalB16}
S.~Agrawal and B.~Bonakdarpour.
\newblock Runtime verification of k-safety hyperproperties in {HyperLTL}.
\newblock In \emph{IEEE Computer Security Foundations Symposium}, pages
  239--252, 2016.

\bibitem[Alvim et~al.(2012)Alvim, Chatzikokolakis, Palamidessi, and
  Smith]{CCS12}
M.~S. Alvim, K.~Chatzikokolakis, C.~Palamidessi, and G.~Smith.
\newblock Measuring information leakage using generalized gain functions.
\newblock In \emph{IEEE Computer Security Foundations Symposium}, pages
  265--279, 2012.

\bibitem[Amtoft and Banerjee(2004)]{AB04}
T.~Amtoft and A.~Banerjee.
\newblock Information flow analysis in logical form.
\newblock In \emph{Static Analysis Symposium}, pages 100--115, 2004.

\bibitem[Amtoft et~al.(2006)Amtoft, Bandhakavi, and Banerjee]{ABB06}
T.~Amtoft, S.~Bandhakavi, and A.~Banerjee.
\newblock A logic for information flow in object-oriented programs.
\newblock In \emph{ACM Symposium on Principles of Programming Languages}, pages
  91--102, 2006.

\bibitem[Askarov and Sabelfeld(2007)]{AskarovS07}
A.~Askarov and A.~Sabelfeld.
\newblock Gradual release: Unifying declassification, encryption and key
  release policies.
\newblock In \emph{IEEE Symposium on Security and Privacy}, 2007.

\bibitem[Askarov et~al.(2008)Askarov, Hunt, Sabelfeld, and Sands]{Aal.08}
A.~Askarov, S.~Hunt, A.~Sabelfeld, and D.~Sands.
\newblock Termination-insensitive noninterference leaks more than just a bit.
\newblock In \emph{European Symposium on Research in Computer Security}, volume
  5283 of \emph{LNCS}, 2008.

\bibitem[Assaf(2015)]{Ass15}
M.~Assaf.
\newblock \emph{From Qualitative to Quantitative Program Analysis : Permissive
  Enforcement of Secure Information Flow}.
\newblock PhD thesis, Universit{\'e} de Rennes 1, May 2015.
\newblock \url{https://hal.inria.fr/tel-01184857}.

\bibitem[Assaf and Naumann(2016)]{AN16}
M.~Assaf and D.~Naumann.
\newblock Calculational design of information flow monitors.
\newblock In \emph{IEEE Computer Security Foundations Symposium}, pages
  210--224, 2016.

\bibitem[Assaf et~al.(2016{\natexlab{a}})Assaf, Naumann, Signoles, Totel, and
  Tronel]{Aal16b}
M.~Assaf, D.~Naumann, J.~Signoles, {\'E}.~Totel, and F.~Tronel.
\newblock Hypercollecting semantics and its application to static analysis of
  information flow.
\newblock Technical report, Apr. 2016{\natexlab{a}}.
\newblock URL \url{https://arxiv.org/abs/1608.01654}.

\bibitem[Assaf et~al.(2016{\natexlab{b}})Assaf, Signoles, Totel, and
  Tronel]{Aal16a}
M.~Assaf, J.~Signoles, {\'E}.~Totel, and F.~Tronel.
\newblock The cardinal abstraction for quantitative information flow.
\newblock In \emph{Workshop on Foundations of Computer Security (FCS)}, June
  2016{\natexlab{b}}.
\newblock \url{https://hal.inria.fr/hal-01334604}.

\bibitem[Backes et~al.(2009)Backes, K{\"o}pf, and Rybalchenko]{BKR09}
M.~Backes, B.~K{\"o}pf, and A.~Rybalchenko.
\newblock Automatic discovery and quantification of information leaks.
\newblock In \emph{IEEE Symposium on Security and Privacy}, pages 141--153.
  IEEE, 2009.

\bibitem[Banerjee et~al.(2008)Banerjee, Naumann, and Rosenberg]{BNR08}
A.~Banerjee, D.~A. Naumann, and S.~Rosenberg.
\newblock Expressive declassification policies and modular static enforcement.
\newblock In \emph{IEEE Symposium on Security and Privacy}, pages 339--353,
  2008.

\bibitem[Banerjee et~al.(2016)Banerjee, Naumann, and Nikouei]{BanerjeeNN16}
A.~Banerjee, D.~A. Naumann, and M.~Nikouei.
\newblock Relational logic with framing and hypotheses.
\newblock In \emph{36th {IARCS} Annual Conference on Foundations of Software
  Technology and Theoretical Computer Science}, 2016.
\newblock To appear.

\bibitem[Barthe et~al.(2004)Barthe, D'Argenio, and Rezk]{BAR04}
G.~Barthe, P.~R. D'Argenio, and T.~Rezk.
\newblock Secure information flow by self-composition.
\newblock In \emph{IEEE Computer Security Foundations Workshop}, pages
  100--114, 2004.

\bibitem[Bello et~al.(2015)Bello, Hedin, and Sabelfeld]{BelloHS15}
L.~Bello, D.~Hedin, and A.~Sabelfeld.
\newblock Value sensitivity and observable abstract values for information flow
  control.
\newblock In \emph{Logic for Programming, Artificial Intelligence, and
  Reasoning {(LPAR)}}, pages 63--78, 2015.

\bibitem[Benton(2004)]{Benton04}
N.~Benton.
\newblock Simple relational correctness proofs for static analyses and program
  transformations.
\newblock In \emph{ACM Symposium on Principles of Programming Languages}, pages
  14--25, 2004.

\bibitem[Bertrane et~al.(2012)Bertrane, Cousot, Cousot, Feret, Mauborgne,
  Min{\'e}, and Rival]{Bal.12}
J.~Bertrane, P.~Cousot, R.~Cousot, J.~Feret, L.~Mauborgne, A.~Min{\'e}, and
  X.~Rival.
\newblock Static analysis and verification of aerospace software by abstract
  interpretation.
\newblock In \emph{AIAA Infotech@Aerospace 2010}, 2012.

\bibitem[Bertrane et~al.(2015)Bertrane, Cousot, Cousot, Feret, Mauborgne,
  Min{\'e}, and Rival]{Bal.15}
J.~Bertrane, P.~Cousot, R.~Cousot, J.~Feret, L.~Mauborgne, A.~Min{\'e}, and
  X.~Rival.
\newblock Static analysis and verification of aerospace software by abstract
  interpretation.
\newblock \emph{Foundations and Trends in Programming Languages}, 2\penalty0
  (2-3):\penalty0 71--190, 2015.

\bibitem[Besson et~al.(2013)Besson, Bielova, and Jensen]{BBJ13}
F.~Besson, N.~Bielova, and T.~Jensen.
\newblock Hybrid information flow monitoring against web tracking.
\newblock In \emph{IEEE Computer Security Foundations Symposium}, pages
  240--254. IEEE, 2013.

\bibitem[Besson et~al.(2016)Besson, Bielova, and Jensen]{BBJ16}
F.~Besson, N.~Bielova, and T.~Jensen.
\newblock Hybrid monitoring of attacker knowledge.
\newblock In \emph{IEEE Computer Security Foundations Symposium}, pages
  225--238, 2016.

\bibitem[Boudol(2008)]{Bou08}
G.~Boudol.
\newblock Secure information flow as a safety property.
\newblock In \emph{Formal Aspects in Security and Trust}, pages 20--34, 2008.

\bibitem[Bourdoncle(1992)]{Bou92}
F.~Bourdoncle.
\newblock Abstract interpretation by dynamic partitioning.
\newblock \emph{Journal of Functional Programming}, 2\penalty0 (04):\penalty0
  407--435, 1992.

\bibitem[Braun et~al.(2009)Braun, Chatzikokolakis, and Palamidessi]{BCP09}
C.~Braun, K.~Chatzikokolakis, and C.~Palamidessi.
\newblock Quantitative notions of leakage for one-try attacks.
\newblock In \emph{Mathematical Foundations of Programming Semantics ({MFPS})},
  volume 249, pages 75--91, 2009.

\bibitem[Cachera and Pichardie(2010)]{CP10}
D.~Cachera and D.~Pichardie.
\newblock A certified denotational abstract interpreter.
\newblock In \emph{Interactive Theorem Proving {(ITP)}}, pages 9--24, 2010.

\bibitem[Clarkson and Schneider(2008)]{CS08}
M.~R. Clarkson and F.~B. Schneider.
\newblock Hyperproperties.
\newblock In \emph{IEEE Computer Security Foundations Symposium}, pages 51--65,
  2008.

\bibitem[Clarkson and Schneider(2010)]{CS10}
M.~R. Clarkson and F.~B. Schneider.
\newblock Hyperproperties.
\newblock \emph{Journal of Computer Security}, 18\penalty0 (6):\penalty0
  1157--1210, 2010.

\bibitem[Clarkson et~al.(2009)Clarkson, Myers, and Schneider]{CMS09}
M.~R. Clarkson, A.~C. Myers, and F.~B. Schneider.
\newblock Quantifying information flow with beliefs.
\newblock \emph{Journal of Computer Security}, 17:\penalty0 655--701, 2009.

\bibitem[Clarkson et~al.(2014)Clarkson, Finkbeiner, Koleini, Micinski, Rabe,
  and S{\'{a}}nchez]{ClarksonFKMRS14}
M.~R. Clarkson, B.~Finkbeiner, M.~Koleini, K.~K. Micinski, M.~N. Rabe, and
  C.~S{\'{a}}nchez.
\newblock Temporal logics for hyperproperties.
\newblock In \emph{Principles of Security and Trust}, volume 8414 of
  \emph{LNCS}, pages 265--284, 2014.

\bibitem[Cohen(1977)]{Coh77}
E.~Cohen.
\newblock Information transmission in computational systems.
\newblock In \emph{Proceedings of the sixth ACM Symposium on Operating Systems
  Principles}, pages 133--139, 1977.

\bibitem[Cortesi and Zanioli(2011)]{CZ11}
A.~Cortesi and M.~Zanioli.
\newblock Widening and narrowing operators for abstract interpretation.
\newblock \emph{Computer Languages, Systems {\&} Structures}, pages 24--42,
  2011.

\bibitem[Cortesi et~al.(2013)Cortesi, Costantini, and Ferrara]{CCP13}
A.~Cortesi, G.~Costantini, and P.~Ferrara.
\newblock A survey on product operators in abstract interpretation.
\newblock In \emph{Semantics, Abstract Interpretation, and Reasoning about
  Programs: Essays Dedicated to David A. Schmidt on the Occasion of his
  Sixtieth Birthday}, volume 129 of \emph{{EPTCS}}, pages 325--336, 2013.

\bibitem[Cousot(1999)]{Cou99}
P.~Cousot.
\newblock The calculational design of a generic abstract interpreter.
\newblock In M.~Broy and R.~Steinbr{\"u}ggen, editors, \emph{Calculational
  System Design}, volume 173, pages 421--506. NATO ASI Series F. IOS Press,
  Amsterdam, 1999.

\bibitem[Cousot(2002)]{Cou02}
P.~Cousot.
\newblock Constructive design of a hierarchy of semantics of a transition
  system by abstract interpretation.
\newblock \emph{Theoretical Computer Science}, 277\penalty0 (1-2):\penalty0
  47--103, 2002.

\bibitem[Cousot and Cousot(1977)]{CC77}
P.~Cousot and R.~Cousot.
\newblock Abstract interpretation: a unified lattice model for static analysis
  of programs by construction or approximation of fixpoints.
\newblock In \emph{ACM Symposium on Principles of Programming Languages}, pages
  238--252, 1977.

\bibitem[Cousot and Cousot(1979)]{CC79}
P.~Cousot and R.~Cousot.
\newblock Systematic design of program analysis frameworks.
\newblock In \emph{ACM Symposium on Principles of Programming Languages}, pages
  269--282, 1979.

\bibitem[Cousot and Cousot(1992)]{CC92}
P.~Cousot and R.~Cousot.
\newblock Comparing the galois connection and widening/narrowing approaches to
  abstract interpretation.
\newblock In \emph{Programming Language Implementation and Logic Programming
  {(PLILP)}}, pages 269--295, 1992.

\bibitem[Cousot and Cousot(1994)]{CC94}
P.~Cousot and R.~Cousot.
\newblock Higher-order abstract interpretation (and application to comportment
  analysis generalizing strictness, termination, projection and per analysis of
  functional languages).
\newblock In \emph{International Conference on Computer Languages {(ICCL)}},
  pages 95--112, 1994.

\bibitem[Cousot and Halbwachs(1978)]{CH78}
P.~Cousot and N.~Halbwachs.
\newblock Automatic discovery of linear restraints among variables of a
  program.
\newblock In \emph{ACM Symposium on Principles of Programming Languages}, pages
  84--96, 1978.

\bibitem[Darvas et~al.(2005)Darvas, H{\"a}hnle, and Sands]{DHS05}
{\'A}.~Darvas, R.~H{\"a}hnle, and D.~Sands.
\newblock A theorem proving approach to analysis of secure information flow.
\newblock In \emph{Security in Pervasive Computing}, pages 193--209, 2005.

\bibitem[Denning(1982)]{Den82}
D.~E.~R. Denning.
\newblock \emph{Cryptography and Data Security}.
\newblock Addison-Wesley Longman Publishing Co., Inc., 1982.

\bibitem[Denning and Denning(1977)]{DD77}
D.~E.~R. Denning and P.~J. Denning.
\newblock Certification of programs for secure information flow.
\newblock \emph{Communications of ACM}, 20\penalty0 (7):\penalty0 504--513,
  1977.

\bibitem[Doychev et~al.(2013)Doychev, Feld, K{\"o}pf, Mauborgne, and
  Reineke]{Dal.13}
G.~Doychev, D.~Feld, B.~K{\"o}pf, L.~Mauborgne, and J.~Reineke.
\newblock Cacheaudit: A tool for the static analysis of cache side channels.
\newblock In \emph{USENIX Security Symposium}, pages 431--446, 2013.

\bibitem[Dwork(2011)]{Dwo11}
C.~Dwork.
\newblock A firm foundation for private data analysis.
\newblock \emph{Communications of ACM}, pages 86--95, 2011.

\bibitem[Finkbeiner et~al.(2015)Finkbeiner, Rabe, and
  S{\'{a}}nchez]{FinkbeinerRS15}
B.~Finkbeiner, M.~N. Rabe, and C.~S{\'{a}}nchez.
\newblock Algorithms for model checking {HyperLTL} and {HyperCTL {\^{}}*}.
\newblock In \emph{Computer Aided Verification}, volume 9206 of \emph{LNCS},
  pages 30--48, 2015.

\bibitem[Giacobazzi and Mastroeni(2004)]{GiacobazziM04}
R.~Giacobazzi and I.~Mastroeni.
\newblock Abstract non-interference: parameterizing non-interference by
  abstract interpretation.
\newblock In \emph{ACM Symposium on Principles of Programming Languages}, pages
  186--197, 2004.

\bibitem[Goguen and Meseguer(1982)]{GM82}
J.~A. Goguen and J.~Meseguer.
\newblock Security policies and security models.
\newblock In \emph{IEEE Symposium on Security and Privacy}, pages 11--20, 1982.

\bibitem[Granger(1992)]{Gran92}
P.~Granger.
\newblock Improving the results of static analyses programs by local decreasing
  iteration.
\newblock In \emph{Foundations of Software Technology and Theoretical Computer
  Science}, volume 652, pages 68--79, 1992.

\bibitem[Handjieva and Tzolovski(1998)]{HT98}
M.~Handjieva and S.~Tzolovski.
\newblock Refining dtatic analyses by trace-based partitioning using control
  flow.
\newblock In \emph{International Static Analysis Symposium}, 1998.

\bibitem[Hedin et~al.(2015)Hedin, Bello, and Sabelfeld]{HedinBS15}
D.~Hedin, L.~Bello, and A.~Sabelfeld.
\newblock Value-sensitive hybrid information flow control for a
  {JavaScript-Like} language.
\newblock In \emph{IEEE Computer Security Foundations Symposium}, pages
  351--365, 2015.

\bibitem[Heusser and Malacaria(2009)]{HM09}
J.~Heusser and P.~Malacaria.
\newblock Applied quantitative information flow and statistical databases.
\newblock In \emph{Formal Aspects in Security and Trust}, pages 96--110, 2009.

\bibitem[Hunt(1990)]{Hunt90}
S.~Hunt.
\newblock {PER}s generalize projections for strictness analysis (extended
  abstract).
\newblock In \emph{Proceedings of the Third Annual Glasgow Workshop on
  Functional Programming}, 1990.

\bibitem[Hunt and Sands(1991)]{HuntS91}
S.~Hunt and D.~Sands.
\newblock Binding time analysis: {A} new {PERspective}.
\newblock In \emph{Proceedings of the Symposium on Partial Evaluation and
  Semantics-Based Program Manipulation, PEPM'91, Yale University, New Haven,
  Connecticut, USA, June 17-19, 1991}, pages 154--165, 1991.

\bibitem[Hunt and Sands(2006)]{HS06}
S.~Hunt and D.~Sands.
\newblock On flow-sensitive security types.
\newblock In \emph{ACM Symposium on Principles of Programming Languages}, pages
  79--90, 2006.

\bibitem[Hunt and Sands(2011)]{HS11sd}
S.~Hunt and D.~Sands.
\newblock From exponential to polynomial-time security typing via principal
  types.
\newblock In \emph{ACM Workshop on Programming Languages and Analysis for
  Security}, pages 297--316, 2011.

\bibitem[Klebanov(2014)]{Kle14}
V.~Klebanov.
\newblock Precise quantitative information flow analysis - a symbolic approach.
\newblock \emph{Theoretical Computer Science}, 538:\penalty0 124--139, 2014.

\bibitem[K{\"o}pf and Rybalchenko(2010)]{KR10}
B.~K{\"o}pf and A.~Rybalchenko.
\newblock Approximation and randomization for quantitative information-flow
  analysis.
\newblock In \emph{IEEE Computer Security Foundations Symposium}, pages 3--14,
  2010.

\bibitem[K{\"o}pf and Rybalchenko(2013)]{KR13}
B.~K{\"o}pf and A.~Rybalchenko.
\newblock Automation of quantitative information-flow analysis.
\newblock In \emph{Formal Methods for Dynamical Systems - 13th International
  School on Formal Methods for the Design of Computer, Communication, and
  Software Systems}, volume 7938 of \emph{LNCS}, pages 1--28, 2013.

\bibitem[Kov{\'a}cs et~al.(2013)Kov{\'a}cs, Seidl, and Finkbeiner]{KSF13}
M.~Kov{\'a}cs, H.~Seidl, and B.~Finkbeiner.
\newblock Relational abstract interpretation for the verification of
  2-hypersafety properties.
\newblock In \emph{ACM SIGSAC conference on Computer and Communications
  Security}, pages 211--222, 2013.

\bibitem[Mardziel et~al.(2011)Mardziel, Magill, Hicks, and Srivatsa]{Mal.11}
P.~Mardziel, S.~Magill, M.~Hicks, and M.~Srivatsa.
\newblock Dynamic enforcement of knowledge-based security policies.
\newblock In \emph{IEEE Computer Security Foundations Symposium}, pages
  114--128. IEEE, 2011.

\bibitem[Mardziel et~al.(2013)Mardziel, Magill, Hicks, and Srivatsa]{Mal.13}
P.~Mardziel, S.~Magill, M.~Hicks, and M.~Srivatsa.
\newblock Dynamic enforcement of knowledge-based security policies using
  probabilistic abstract interpretation.
\newblock \emph{Journal of Computer Security}, 21\penalty0 (4):\penalty0
  463--532, 2013.

\bibitem[Mastroeni(2013)]{Mastroeni13}
I.~Mastroeni.
\newblock Abstract interpretation-based approaches to security - {A} survey on
  abstract non-interference and its challenging applications.
\newblock In \emph{Semantics, Abstract Interpretation, and Reasoning about
  Programs: Essays Dedicated to David A. Schmidt on the Occasion of his
  Sixtieth Birthday}, volume 129 of \emph{{EPTCS}}, pages 41--65, 2013.

\bibitem[Mastroeni and Banerjee(2011)]{MB11}
I.~Mastroeni and A.~Banerjee.
\newblock Modelling declassification policies using abstract domain
  completeness.
\newblock \emph{Mathematical Structures in Computer Science}, 21\penalty0
  (06):\penalty0 1253--1299, 2011.

\bibitem[McLean(1994)]{Mcl94}
J.~McLean.
\newblock A general theory of composition for trace sets closed under selective
  interleaving functions.
\newblock In \emph{IEEE Symposium on Security and Privacy}, pages 79--93, 1994.

\bibitem[Min{\'e}(2006{\natexlab{a}})]{Min06}
A.~Min{\'e}.
\newblock The octagon abstract domain.
\newblock \emph{Higher-order and symbolic computation}, 19\penalty0
  (1):\penalty0 31--100, 2006{\natexlab{a}}.

\bibitem[Min{\'e}(2006{\natexlab{b}})]{Min06b}
A.~Min{\'e}.
\newblock Symbolic methods to enhance the precision of numerical abstract
  domains.
\newblock In \emph{Verification, Model Checking, and Abstract Interpretation},
  pages 348--363, 2006{\natexlab{b}}.

\bibitem[M{\"u}ller et~al.(2015)M{\"u}ller, Kov{\'a}cs, and Seidl]{MKS15}
C.~M{\"u}ller, M.~Kov{\'a}cs, and H.~Seidl.
\newblock An analysis of universal information flow based on self-composition.
\newblock In \emph{IEEE Computer Security Foundations Symposium}, pages
  380--393, 2015.

\bibitem[Nielson et~al.(1999)Nielson, Nielson, and Hankin]{NNH99}
F.~Nielson, H.~R. Nielson, and C.~Hankin.
\newblock \emph{Principles of Program Analysis}.
\newblock Springer, 1999.

\bibitem[R{\'e}nyi(1961)]{Ren61}
A.~R{\'e}nyi.
\newblock On measures of entropy and information.
\newblock In \emph{the Fourth Berkeley Symposium on Mathematical Statistics and
  Probability}, 1961.

\bibitem[Rival and Mauborgne(2007)]{RM07}
X.~Rival and L.~Mauborgne.
\newblock The trace partitioning abstract domain.
\newblock \emph{ACM Transactions on Programming Languages and Systems},
  29\penalty0 (5):\penalty0 26, 2007.

\bibitem[Rushby(2001)]{Rus01}
J.~Rushby.
\newblock Security requirements specifications: How and what.
\newblock In \emph{Symposium on Requirements Engineering for Information
  Security (SREIS)}, 2001.

\bibitem[Sabelfeld and Myers(2003)]{SM03}
A.~Sabelfeld and A.~C. Myers.
\newblock Language-based information-flow security.
\newblock \emph{IEEE Journal on Selected Areas in Communications}, 21\penalty0
  (1):\penalty0 5--19, 2003.

\bibitem[Sabelfeld and Sands(2009)]{SS09}
A.~Sabelfeld and D.~Sands.
\newblock Declassification: Dimensions and principles.
\newblock \emph{Journal of Computer Security}, 17\penalty0 (5), 2009.

\bibitem[Schmidt(2009)]{Schmidt09}
D.~A. Schmidt.
\newblock Abstract interpretation from a topological perspective.
\newblock In \emph{Static Analysis, 16th International Symposium}, volume 5673
  of \emph{LNCS}, pages 293--308, 2009.

\bibitem[Schmidt(2012)]{Schmidt12}
D.~A. Schmidt.
\newblock Inverse-limit and topological aspects of abstract interpretation.
\newblock \emph{Theoretical Computer Science}, 430:\penalty0 23--42, 2012.

\bibitem[Schoepe et~al.(2016)Schoepe, Balliu, Pierce, and
  Sabelfeld]{SchoepeBPS16}
D.~Schoepe, M.~Balliu, B.~C. Pierce, and A.~Sabelfeld.
\newblock Explicit secrecy: {A} policy for taint tracking.
\newblock In \emph{{IEEE} European Symposium on Security and Privacy}, pages
  15--30, 2016.

\bibitem[Shannon(1948)]{Sha48}
C.~E. Shannon.
\newblock A mathematical theory of communication.
\newblock \emph{The Bell System Technical Journal}, 27:\penalty0 379--423,
  1948.

\bibitem[Smith(2009)]{Smi09}
G.~Smith.
\newblock On the foundations of quantitative information flow.
\newblock In \emph{International Conference on Foundations of Software Science
  and Computational Structures}, pages 288--302, 2009.

\bibitem[Smith(2011)]{Smi11}
G.~Smith.
\newblock Quantifying information flow using min-entropy.
\newblock In \emph{Quantitative Evaluation of Systems (QEST), 2011 Eighth
  International Conference on}, pages 159--167. IEEE, 2011.

\bibitem[Sousa and Dillig(2016)]{SousaD16}
M.~Sousa and I.~Dillig.
\newblock Cartesian {Hoare} logic for verifying k-safety properties.
\newblock In \emph{ACM Conference on Programming Language Design and
  Implementation}, pages 57--69, 2016.

\bibitem[Terauchi and Aiken(2005)]{TA05}
T.~Terauchi and A.~Aiken.
\newblock Secure information flow as a safety problem.
\newblock In \emph{Static Analysis Symposium}, pages 352--367, 2005.

\bibitem[Volpano and Smith(1997)]{VS97}
D.~Volpano and G.~Smith.
\newblock Eliminating covert flows with minimum typings.
\newblock In \emph{IEEE Computer Security Foundations Workshop}, pages
  156--168, 1997.

\bibitem[Volpano and Smith(2000)]{VS00}
D.~Volpano and G.~Smith.
\newblock Verifying secrets and relative secrecy.
\newblock In \emph{ACM Symposium on Principles of Programming Languages}, pages
  268--276, 2000.

\bibitem[Volpano et~al.(1996)Volpano, Irvine, and Smith]{VIS96}
D.~Volpano, C.~Irvine, and G.~Smith.
\newblock A sound type system for secure flow analysis.
\newblock \emph{Journal of Computer Security}, 4\penalty0 (2-3):\penalty0
  167--187, 1996.

\bibitem[Volpano(1999)]{Vol99}
D.~M. Volpano.
\newblock Safety versus secrecy.
\newblock In \emph{Static Analysis Symposium}, pages 303--311, 1999.

\bibitem[Wasserrab et~al.(2009)Wasserrab, Lohner, and Snelting]{Wasserrab09}
D.~Wasserrab, D.~Lohner, and G.~Snelting.
\newblock On {PDG}-based noninterference and its modular proof.
\newblock In \emph{ACM Workshop on Programming Languages and Analysis for
  Security}, pages 31--44, 2009.

\bibitem[Winskel(1993)]{Win93}
G.~Winskel.
\newblock \emph{The Formal Semantics of Programming Languages: an
  Introduction}.
\newblock Cambridge, 1993.

\bibitem[Yasuoka and Terauchi(2011)]{YT11}
H.~Yasuoka and T.~Terauchi.
\newblock On bounding problems of quantitative information flow.
\newblock \emph{Journal of Computer Security}, 19\penalty0 (6):\penalty0
  1029--1082, 2011.

\bibitem[Zakinthinos and Lerner(1997)]{ZL97}
A.~Zakinthinos and S.~Lerner.
\newblock A general theory of security properties.
\newblock In \emph{IEEE Symposium on Security and Privacy}, pages 94--102,
  1997.

\bibitem[Zanioli and Cortesi(2011)]{ZC11}
M.~Zanioli and A.~Cortesi.
\newblock Information leakage analysis by abstract interpretation.
\newblock In \emph{SOFSEM 2011: Theory and Practice of Computer Science}, pages
  545--557, 2011.

\bibitem[Zanotti(2002)]{Zan02}
M.~Zanotti.
\newblock Security typings by abstract interpretation.
\newblock In \emph{Static Analysis Symposium}, volume 2477, pages 360--375,
  2002.

\end{thebibliography}

\iftechreport
\onecolumn
\newpage

\onecolumn
\begin{appendices}
\renewcommand{\thesection}{\appendixname~\Alph{section}}

\crefalias{section}{appsec}
\crefalias{subsection}{appsec}
\crefname{appsec}{Appendix}{Appendices}

\setcounter{tocdepth}{1}

\bigskip
\section{Symbols}

\begin{tabular}{ll}
$\Values$ & a set of integers \\
$\sizeval$ & an infinite cardinal number \\
$\val \in \Values$ & an integer \\
$\Val \in \psetValues$ & a set of  values \\
$\BVal \in \psetpsetValues$ & a set of sets of values \\ 
$\Traces$ & the set of (relational) traces \\
$t  \in \Traces$ & a trace \\
$T  \in \psetTraces$ & a set of traces \\
$\BTr \in \psetpsetTraces$ & a set of sets of traces \\
$\States$ & the set of states  \\
$\st \in \States$ & a state \\
$\St \in \psetStates$ & a set of states \\
$\BSt \in \psetpsetStates$ & a set of sets of states \\
$\Statesseq \triangleq \cup_{n \in \mathbb{N}} \States^n$ & the set of finite
  sequence of states \\
\\
$\varprog$ & the set of variables of a program $\prog$\\
$\mlsLat$ & a multilevel security lattice \\
$\mlsl \in \mlsLat$ & a security level \\
$\Ga \in \varprog \to \mlsLat$ & an initial typing context \\
\\
$\concobj \galois{\alpha}{\gamma} \absobj$ & a Galois connection
  \\
\\
$\semins{c}{} \in \Traces \to \Traces$ & denotational semantics of commands \\
$\semexp{e}{} \in \Traces \to \Values$ & value of $e$ in final state \\ 
$\semexpi{e}{} \in  \Traces \to \Values$ & value of $e$ in initial state \\

$\seminsc{c}{} \in \psetTraces\to\psetTraces$ & collecting semantics\\
$\seminshc{c}{} \in \psetpsetTraces\to\psetpsetTraces$ & hypercollecting
semantics \\
\\
$\depconslx$ & atomic dependence: ``agreement up to security \\
& \quad level $\mlsl$ leads to agreement on $\varx$"\\ 
$\depconsset \in \depLat$ & A set of atomic dependency constraints \\
$\cardconslxn$ & atomic cardinality: ``agreement up to security \\
& \quad level $\mlsl$ leads to an $\mlsl$-cardinality of $n$ values for $\varx$'' \\
$\cardconsset \in \cardLat$ & A valid set of atomic cardinality constraints
\end{tabular}

\clearpage

\section{Background: Collecting Semantics, Galois Connections}

\lemgaloisliftedhyp*

\textit{Proof.}

Let $C \in \psetconcobj$ and $A \in \psetabsobj$.
\begin{align*}
  \alphaelt(C) \subseteq A &
  \iff \{ \eltwop(c) \mid c \in C \} \subseteq A \\
  & \iff \forall c \in C, \; \eltwop(c) \in A \\
  & \iff C \subseteq \{ c \mid \eltwop(c) \in A \} \\
  & \iff C \subseteq \gammaelt(A)
\end{align*}

\hfill $\qed$

\section{Domains and Galois Connections for Hyperproperties}

\lemgaloishypprop*

\textit{Proof.}

This is a special case of the supremus abstraction~\cite[p.52]{Cou02},  that is defined in 
\Cref{lem:galoissupremus}. 
Indeed, we can instantiate a supremus abstraction by taking
$\hppop \triangleq \operatorname{id} \;(\in \psetconcobj \to \psetconcobj)$.
We thus obtain a Galois connection
$\psetpsetconcobj\galois{\alphahpp}{\gammahpp}\psetconcobj$, with  
$\alphahpp(\BObj) = \cup_{\Obj \in \BObj} \;\Obj$
and 
$\gammahpp(\Obj) = \{ \Obj' \in \psetconcobj \mid \Obj' \subseteq \Obj \} \; 
( = \pset{\Obj})$.
Notice here that the powerset of a set $\concobj$, provided with set inclusion
as a partial order, is a complete lattice as required by the supremus
abstraction.

\hfill $\qed$

\lemgaloissupremus*

\textit{Proof.}

Notice that the assumption that the lattice $(\absobj,\sqsubseteq,\sqcup)$ is complete
guarantees that $\alphasup(C)$ is well-defined: the set
$\{ \supremop(c) \mid c \in C \}$ does have a supremum.

Let $C \in \psetconcobj$ and $a \in \absobj$.  The proof goes by definitions.

\begin{align*}
  \alphasup(C) \sqsubseteq a &  \iff   \sqcup_{c \in C} \supremop(c) \sqsubseteq a \\
  & \iff \forall c \in C, \; \supremop(c) \sqsubseteq a \\
  & \iff C \subseteq  \{c \in \concobj \mid \supremop(c)\sqsubseteq a\} \\
  & \iff C \subseteq \gammasup(a) 
\end{align*}

\hfill $\qed$

\clearpage

\section{Hypercollecting Semantics}

Before proving the main result of this section in \Cref{theo:hypercollsound},
we will first prove \Cref{lem:collectingdirect}.

Both proofs of \Cref{lem:collectingdirect} and  \Cref{theo:hypercollsound} are
by structural induction.
Most cases follow from definitions.
The important cases are for while loops and the proof technique is a classical
one when using a denotational semantics.
E.g., in order to prove equality of  two denotations characterised as a fixpoint,
it suffices to introduce two sequences that converge towards the fixpoint
characterisations and  prove equality of these sequences.
This ensures that their limits -- the denotations characterised as a fixpoint --
are equal.

Let us now prove \Cref{lem:collectingdirect} -- this lemma is used later in the
proof case of while loops for  \Cref{theo:hypercollsound}.

\begin{restatable}[]{mylem}{lemcollectingsemantics}
  \label{lem:collectingdirect}
  For all commands $c$, for all sets of traces $\Tr \in \psetTraces$,
  the standard collecting semantics (\Cref{sec:background}) can be expressed as
  the direct image of the 
  denotational semantics :
  \[ \seminsc{c}{\Tr} = \{ \semins{c}{\tr} \in \Traces \mid \tr \in \Tr \}  \]
\end{restatable}

\textit{Proof.}

The proof proceeds by structural induction on commands. The most important case
is the case of while loops.

1 -- Case $\skipcom$:

\[ \seminsc{\skipcom}{\Tr} = \Tr = \{ \semins{\skipcom}{\tr} \mid \tr \in \Tr \} \]

2 -- Case $x:= e$:

\[ \seminsc{x := e}{\Tr} = \{ \semins{x := e}{\tr} \mid \tr \in \Tr \} \]

3 -- Case $c_1;~c_2$:

\begin{align*}
  \seminsc{c_1;~c_2}{\Tr} & = \seminsc{c_2}{} \comp \seminsc{c_1}{\Tr} \\
  & = \justif{By induction on $c_1$ } \\
  & \qquad
  \seminsc{c_2}{( \{ \semins{c_1}{\tr} \in \Traces \mid \tr \in \Tr \}) }  \\
  & = \justif{By induction on $c_2$} \\
  & \qquad
  \{\semins{c_2}{} \comp  \semins{c_1}{\tr} \in \Traces \mid \tr \in \Tr \} \\
  & = \{ \semins{c_1;~c_2}{\tr} \in\Traces \mid \tr \in \Tr \}
\end{align*}

4 -- Case $\ifcom~ (b)~\thencom~c_1~\elsecom~c_2$:

\begin{align*}
  \seminsc{\ifcom~ (b)~ \thencom~ c_1~ \elsecom~ c_2}{\Tr} & =
  \seminsc{c_1}{} \comp  \guardc{b}{\Tr} \; \cup \; \seminsc{c_2}{} \comp
  \guardc{\neg b}{\Tr}  \\
  & = \justif{By induction hypothesis on both $c_1$ and $c_2$} \\
  & \qquad \{ \semins{c_1}{\tr} \in \Traces \mid \tr \in  \guardc{b}{\Tr} \}
  \; \cup \;
  \{ \semins{c_2}{\tr} \in \Traces \mid \tr \in  \guardc{\neg b}{\Tr} \} \\
  & = \{ \semins{\ifcom~ (b)~ \thencom~ c_1~ \elsecom~ c_2}{\tr} \in \Traces
  \mid  \tr \in \Tr  \}
\end{align*}

5 -- Case $\whilecom~(b)~\docom~ c$:

 5.1 -- Let us first prove the following intermediate result:
  \[
   \forall \Tr \in \psetTraces, \quad
   \{ \semins{\whilecom~(b)~\docom~c}{\tr} \in \Traces \mid \tr \in \Tr \} = 
  \guardc{\neg b}{}
  \left( \operatorname{lfp_{\emptyset}^{\subseteq}
    \lambda X. \Tr \cup \seminsc{c}{}\comp \guardc{b}{X} }\right).
  \]

  Indeed, let the sequence $(x_n^{\Tr})_{n \geq 0}$ be defined as:
  \[
  x_{n}^{\Tr}  \triangleq \{ \mathcal{F}^{(n)}(\bot)(\tr) \in \Traces \mid
  \tr \in \Tr \}
  \]

    with $\mathcal{F}$ defined as
\[
\mathcal{F}(w)(\tr) \triangleq
   \begin{cases}
        \tr & \text{if } \semexp{b}{\tr} = 0 \\
        w \comp \semins{c}{\tr} & \text{otherwise}
 \end{cases}
\]

  Notice that for all $\tr \in \Tr$, the sequence
  $(\mathcal{F}^{(n)}(\bot)(\tr))_{n \geq 0}$ converges and is equal to the
  evaluation of
  the while loop in the state $\tr$
  (i.e.\  $\semins{\whilecom~ b~\docom~ c}{\tr} = \mathcal{F}^{(\infty)}(\bot)(\tr)$), by
  definition of the denotational semantics of loops; thus, the sequence
  $x_n^\Tr$ converges to
  $\{ \semins{\whilecom~ b~\docom~ c}{\tr} \in \Traces \mid \tr \in \Tr\}$.

  Let also the sequences $(y_n^\Tr)_{n \geq 0}$ and $(g_n^\Tr)_{n \geq 0}$ be
  defined as:
  \begin{align*}
    y_{n}^{\Tr} & \triangleq \guardc{\neg b}{g_n^{\Tr}} \\
          g_{n+1}^{\Tr} & \triangleq \Tr \cup \seminsc{c}{} \comp \guardc{b}{g_n^{\Tr}} \\
    g_0^{\Tr} & \triangleq \emptyset
  \end{align*}
  Notice that for all $\Tr \in \psetTraces$, the sequence $g_n^T$ converges to
  $\operatorname{lfp_{\emptyset}^{\subseteq}
    \lambda X. \Tr \cup  \seminsc{c}{}\comp \guardc{b}{X} }$ (or written
  otherwise: $\operatorname{lfp_{\Tr}^{\subseteq}  \lambda X. \seminsc{c}{}\comp
    \guardc{b}{X} }$).
  This also means that the sequence $y_n^\Tr$ converges to
  $\guardc{\neg b}{}( \;
  \operatorname{lfp_{\emptyset}^{\subseteq}  \lambda X. \Tr \; \cup \; \seminsc{c}{} \comp
    \guardc{b}{X} } \; )$.
  
 Thus, it  suffices to prove that:
  \[ \forall \Tr \in \psetTraces, \forall n \in \mathbb{N},
  x_n^\Tr = y_n^\Tr. \]
  The proof proceeds by induction on $n$.

  - $ x_0^\Tr = \emptyset = y_0^\Tr$

  - Let $n \in \mathbb{N}$ such that:
  $\forall \Tr \in \psetTraces, x_n^\Tr = y_n^\Tr$.
  Then:
  \begin{align*}
    x_{n+1}^\Tr & = \{ \mathcal{F}^{(n+1)}(\bot)(\tr) \in \Traces \mid \tr \in \Tr \} \\
    & = \guardc{\neg b}{\Tr} \; \cup \;
    \{ \mathcal{F}^{(n)}(\bot)(\semins{c}{\tr}) \in \Traces \mid
    \st \in \guardc{b}{\Tr}   \}  \\
    & = \guardc{\neg b}{\Tr} \; \cup \;  \{ \mathcal{F}^{(n)}(\bot)(\tr) \in \Traces \mid
    \tr \in \seminsc{c}{} \comp \guardc{b}{\Tr}   \}  \\
    & =  \guardc{\neg b}{\Tr} \; \cup \; x_n^{ \seminsc{c}{} \comp \guardc{b}{\Tr}} \\
    & = \justif{By induction hypothesis} \\
    & \quad \quad \guardc{\neg b}{\Tr} \; \cup \;
    y_n^{\seminsc{c}{} \comp \guardc{b}{\Tr}} \\
    & = \justif{By definition of $y_n^{\seminsc{c}{} \comp \guardc{b}{\Tr}}$ } \\
    & \quad \quad \guardc{\neg b}{\Tr} \; \cup \;
    \guardc{\neg b}{g_n^{\seminsc{c}{} \comp \guardc{b}{\Tr}}} \\
    & = \guardc{\neg b}{}  \left(\Tr \cup g_n^{\seminsc{c}{} \comp
      \guardc{b}{\Tr}} \right) \\
    & = 
    \justif{Because for all $\Tr$, $ g_{n}^{\Tr} =
            \bigcup\limits_{0 \leq k \leq n-1} (\seminsc{c}{} \comp \guardc{b}{})^{(k)} (\Tr)$ } \\
    & \quad \quad \guardc{\neg b}{g_{n+1}^{\Tr}} \\
    & = y_{n+1}^{\Tr} 
  \end{align*}

   5.2 -- Let us now prove that :
  \[ \operatorname{lfp_{\emptyset}^{\subseteq} \lambda X. \Tr \cup
    \seminsc{c}{}\comp \guardc{b}{X} } = 
  \operatorname{lfp_{\emptyset}^{\subseteq}} \lambda X. \Tr \cup
  \seminsc{\ifcom~ b~ \thencom~ c~ \elsecom~ \skipcom}{X}
  \]

  Indeed, let the sequence $(f_n^\Tr)_{n \geq 0}$ be defined as:
  \begin{align*}
    f_0^\Tr & \triangleq \emptyset \\
    f_{n+1}^\Tr & \triangleq \Tr \cup
    \seminsc{\ifcom~ b~ \thencom~ c~ \elsecom~ \skipcom}{f_n^\Tr}
  \end{align*}
  Therefore, by induction on $n \in \mathbb{N}$, it holds that $f_n = g_n$:

  - $f_0^\Tr = g_0^\Tr = \emptyset$.

  - let $n \in \mathbb{N}$, such that $f_n^\Tr = g_n^\Tr$. Then:
  \begin{align*}
    g_{n+1}^\Tr & = \St \cup \seminsc{c}{} \comp \guardc{b}{g_n^\Tr} \\
    &  = \justif{Since $\guardc{\neg b}{g_n^\Tr} \subseteq g_n^\Tr \subseteq g_{n+1}^\Tr$} \\
    & \quad \quad
    \Tr \cup \seminsc{c}{} \comp \guardc{b}{g_n^\Tr} \cup \guardc{\neg b}{g_n^\Tr} \\
    & = \Tr \cup \seminsc{\ifcom~ b~ \thencom~ c~ \elsecom~ \skipcom}{g_n^\Tr} \\
    & = \justif{By induction hypothesis} \\
    & \quad \quad
    \Tr \cup  \seminsc{\ifcom~ b~ \thencom~ c~ \elsecom~  \skipcom}{f_n^\Tr} \\
    & = f_{n+1}^\Tr 
  \end{align*}
This concludes our induction on $n$.
  
Thus, by passing to the limit of both sequences, we obtain the desired result.
  
  5.3 -- Finally, we can conclude:
  \begin{align*}
    \seminsc{\whilecom~ b~\docom~ c}{\Tr} = &
    \guardc{\neg b}{ \left(\operatorname{lfp}_{\Tr}^{\subseteq}
      \seminsc{\ifcom~    b~ \thencom~ c~ \elsecom~ \skipcom}{} \right)} \\
    & = \justif{ by intermediate result 5.2 } \\
    & \qquad
    \guardc{\neg b}{ \left(\operatorname{lfp}_{\Tr}^{\subseteq}
      \seminsc{c}{} \guardc{b}{} \right)} \\
    & = \justif{ by intermediate result 5.1} \\
    & \qquad
    \{ \seminsc{\whilecom~(b)~\docom~c}{\tr} \in \Traces \mid \tr \in \Tr \}
  \end{align*}

  We conclude this proof by structural induction, and Cases 1 to 5.
    \hfill $\qed$

 \clearpage
    
\theohypercoll*

\textit{Proof.}

We prove the theorem as a corollary of this more general result:
\[ \forall \BTr \in \psetpsetTraces,
\{ \seminsc{c}{\Tr} \mid \Tr \in \BTr \}
\subseteq
\seminshc{c}{\BTr}
\]

This proof proceeds by structural induction on commands. The most important case
is the one for while loops; the other ones follow from definition.
\newcommand{\seqx}[2]{\mathbb{X}_{#2}^{#1}}
\newcommand{\seqxt}[1]{\seqx{\BTr}{#1}}
\newcommand{\seqxtn}{\seqx{\BTr}{n}}
\newcommand{\seqy}[2]{\mathbb{Y}_{#2}^{#1}}
\newcommand{\seqyt}[1]{\seqy{\BTr}{#1}}
\newcommand{\seqytn}{\seqy{\BTr}{n}}
\newcommand{\seqg}[2]{\mathbb{G}_{#2}^{#1}}
\newcommand{\seqgt}[1]{\seqg{\BTr}{#1}}
\newcommand{\seqgtn}{\seqg{\BTr}{n}}

1 -- Case $\skipcom$ :

\[ \seminshc{\skipcom}{\BTr}
= \{ \seminsc{\skipcom}{\Tr} \mid \Tr \in \BTr \} 
\supseteq \{ \seminsc{\skipcom}{\Tr} \mid \Tr \in \BTr \}  \]

2 -- Case $x := e$:

\[ \seminshc{x:=e}{\BTr} = \{ \seminsc{x:=e}{\Tr} \mid \Tr \in \BTr \}
\supseteq \{ \seminsc{x:=e}{\Tr} \mid \Tr \in \BTr \} \]

3 -- Case $c_1;~c_2$:
\begin{align*}
  \seminshc{c_1;~c_2}{\BTr} & = \seminshc{c_2}{} \comp \seminshc{c_1}{\BTr} \\
  & \supseteq \justif{by structural induction on $c_1$, and monotonicity of the
    hypercollecting semantics $\seminshc{c}{}$} \\
  & \qquad \seminshc{c_2}{ (\{ \seminsc{c_1}{\Tr} \mid \Tr \in \BTr \} ) } \\
  & \supseteq \justif{by structural induction on $c_2$} \\
  & \qquad \{ \seminsc{c_2}{\Tr'} \mid \Tr' \in \{ \seminsc{c_1}{\Tr} \mid \Tr
  \in \BTr \} \} \\
  & = \{ \seminsc{c_2}{} \comp \seminsc{c_1}{\Tr} \mid \Tr \in \BTr \} \\
  & = \{ \seminsc{c_1;~c_2}{\Tr} \mid \Tr \in \BTr \}
  \end{align*}

4 -- Case $\ifcom~(b)~\thencom~c_1~\elsecom~c_2$:
\begin{align*}
  \seminshc{\ifcom~(b)~\thencom~c_1~\elsecom~c_2}{\BTr} & =
  \{ \seminsc{\ifcom~(b)~\thencom~c_1~\elsecom~c_2}{\Tr} \mid \Tr \in \BTr \} \\
  & \supseteq \{ \seminsc{\ifcom~(b)~\thencom~c_1~\elsecom~c_2}{\Tr} \mid \Tr \in \BTr \}
\end{align*}

5 -- Case $\whilecom~(b)~\docom~c$:

Let $(\seqxtn)_{n \in \mathbb{N}}$ be the sequence defined as
\[ \seqxtn \triangleq \left\{ \{ \mathcal{F}^{(n)}(\bot)(\tr) \in \Traces \mid \tr \in \Tr \}  \mid
\Tr \in \BTr \right\} \text{for } n \geq 1, \quad \seqxt{0} = \emptyset \]
where:
\[  \mathcal{F}(w)(\tr) \triangleq
   \begin{cases}
     \tr & \text{if } \semexp{b}{\tr} = 0 \\
     w \comp \semins{c}{\tr} & \text{otherwise}
   \end{cases}
\]
Notice that the limit of the sequence
$x_n^{\Tr} \triangleq
\{ \mathcal{F}^{(n)}(\bot)(\tr) \in \Traces \mid \tr \in \Tr \} $  is the
ordinary collecting semantics $\seminsc{\whilecom~(b)~\docom~c}{\Tr}$ of the
while loop, as proved in \Cref{lem:collectingdirect}.
Thus, the sequence $\seqxtn$ converges to
$\{ \seminsc{\whilecom~(b)~\docom~c}{\Tr} \mid \Tr \in \BTr \}$.

Let also $(\seqytn)_{n \in \mathbb{N}}$ and   $(\seqgtn)_{n \in \mathbb{N}}$ be
the sequences defined as
 \[ \seqgt{n+1} \triangleq \BTr \cup
 \seminshc{\ifcom~(b)~\thencom~c~\elsecom~\skipcom}{\seqgtn}
 \text{ for } n \geq 0,
 \qquad
\seqgt{0} \triangleq \emptyset
\]
\[ \seqytn \triangleq \guardhc{\neg b}{\seqgtn} \]
Notice that the limit of $\seqytn$ is the hypercollecting semantics of the while
loop ($\seminshc{\whilecom~(b)~\docom~c}{\BTr}$).

Thus, it suffices to prove that the sequences $\seqxt{n}$ and $\seqyt{n}$ verify the
following result
$\forall \BTr \in \psetpsetTraces, \forall n \in \mathbb{N} ,  \seqxt{n+1}
\subseteq  \seqyt{n+1}$;
passing to the limit in this inequality leads to the required result
$\forall \BTr \in \psetpsetTraces,  \{ \seminsc{\whilecom~(b)~\docom~c}{\Tr}
\mid \Tr \in \BTr \}
\subseteq
\seminshc{\whilecom~(b)~\docom~c}{\BTr}$.

We prove the following more precise characterisation of the sequences $\seqxt{n}$ and
$\seqyt{n}$ (this implies $\seqxt{n+1} \subseteq  \seqyt{n+1}$):
\[
\forall n \in \mathbb{N} ,
\forall \BTr \in \psetpsetTraces, \seqyt{n+1} =
\seqytn \cup \seqxt{n+1}
\]

The remaining of this proof proceeds by induction on $n \in \mathbb{N}$.

- case $n=0$:

\begin{align*}
  \seqyt{1} & =  \guardhc{\neg b}{\seqgt{1}} \\
  & = \guardhc{\neg b}{\BTr} \\
  & = \{ \guardc{\neg b}{\Tr} \mid \Tr \in \BTr \} \\
  & = \{  \{ \mathcal{F}^{(1)}(\bot)(\tr) \in \Traces \mid \tr \in \Tr \}
  \mid \Tr \in \BTr \} \\
  & = \justif{since $\seqyt{0} = \emptyset$ and by definition of $\seqxt{1}$} \\
  & \qquad \seqyt{0} \cup \seqxt{1} 
\end{align*}

- Let $n \in \mathbb{N}$ such that $\seqyt{n+1} = \seqyt{n} \cup \seqxt{n+1}$.
Then:

\begin{align*}
  \seqyt{n+2} & = \guardhc{\neg b}{} \seqgt{n+2} \\
  & = \guardhc{\neg b}{}
  \left(
  \BTr \cup \seminshc{\ifcom~(b)~\thencom~c_1~\elsecom~\skipcom}{\seqgt{n+1}}
  \right) \\
  & = \guardhc{\neg b}{\BTr}
  \; \cup \; \guardhc{\neg b}{} \comp
  \seminshc{\ifcom~(b)~\thencom~c_1~\elsecom~\skipcom}{\seqgt{n+1}} \\
  & = \justif{ $\forall n \in \mathbb{N}, \seqgt{n+1} = \cup_{ 0 \leq k \leq n}
    \seminshc{\ifcom~(b)~\thencom~c_1~\elsecom~c_2}{}^{(k)} \BTr $}  \\
  & \qquad
  \guardhc{\neg b}{\BTr}
  \; \cup \; \guardhc{\neg b}{} \left(
  \cup_{ 1 \leq k \leq n+1}
    \seminshc{\ifcom~(b)~\thencom~c~\elsecom~\skipcom}{}^{(k)} \BTr 
    \right) \\
   & =  \cup_{ 0 \leq k \leq n+1}  \guardhc{\neg b}{} \comp
    \seminshc{\ifcom~(b)~\thencom~c~\elsecom~\skipcom}{}^{(k)} \BTr \\
    & = \left( \cup_{ 0 \leq k \leq n}  \guardhc{\neg b}{} \comp
    \seminshc{\ifcom~(b)~\thencom~c~\elsecom~\skipcom}{}^{(k)} \BTr \right)
    \; \cup \;
     \guardhc{\neg b}{} \comp
     \seminshc{\ifcom~(b)~\thencom~c~\elsecom~\skipcom}{}^{(n+1)} \BTr \\
     & = \guardhc{\neg b}{} \comp \left( \cup_{ 0 \leq k \leq n}  
    \seminshc{\ifcom~(b)~\thencom~c~\elsecom~\skipcom}{}^{(k)} \BTr \right)
    \; \cup \;
     \guardhc{\neg b}{} \comp
     \seminshc{\ifcom~(b)~\thencom~c~\elsecom~\skipcom}{}^{(n+1)} \BTr \\
     & = \guardhc{\neg b}{} \seqgt{n+1}
     \; \cup \;
     \guardhc{\neg b}{} \comp
     \seminshc{\ifcom~(b)~\thencom~c~\elsecom~\skipcom}{}^{(n+1)} \BTr \\
     & = \seqyt{n+1}
     \; \cup \;
     \guardhc{\neg b}{} \comp
     \seminshc{\ifcom~(b)~\thencom~c~\elsecom~\skipcom}{}^{(n+1)} \BTr \\
     & = \seqyt{n+1}
     \; \cup \;
     \{
      \guardc{\neg b}{} \comp
      \seminsc{\ifcom~(b)~\thencom~c_1~\elsecom~\skipcom}{}^{(n+1)} \Tr \mid \Tr \in \BTr
     \} \\
     & = \justif{the set $\guardc{\neg b}{} \comp
       \seminsc{\ifcom~(b)~\thencom~c_1~\elsecom~c_2}{}^{(n+1)} \Tr$
       is the set of traces exiting the loop body after n+1 or less iterations:}
       \\
       & \qquad \justif{ it is equal to
         $ \{ \mathcal{F}^{(n+2)}(\bot)(\tr) \mid \tr \in \Tr \}$
         by definition of $\mathcal{F}$}
       \\
     & \qquad
      \seqyt{n+1}
     \; \cup \;
     \left\{
     \{ \mathcal{F}^{(n+2)}(\bot)(\tr) \mid \tr \in \Tr \}
     \mid \Tr \in \BTr \right\} \\
     & = \seqyt{n+1} \; \cup \; \seqxt{n+2}
\end{align*}
This concludes our induction on $n$.

We conclude this proof by structural induction, and Cases 1 to 5.
 \hfill $\qed$

\clearpage

\section{Dependences}

\begin{restatable}[]{mylem}{lemdependencesgalois}
$(\alphadeptr,\gammadeptr)$ yields a Galois connection:  
$\psetpsetTraces \galois{\alphadeptr}{\gammadeptr} \depLat$.
\end{restatable}

\textit{Proof.}

The lattice $\depLat$ is finite, therefore complete. Thus, this is a Galois
connection since it is an instance of the supremus abstraction presented in
\Cref{lem:galoissupremus}.
\hfill $\qed$

The same reasoning applies for $\psetpsetValues
\galois{\alphaagreev}{\gammaagreev} \{\true,\false\}$.

\bigskip

The proofs of of both \Cref{lem:soundabsvarietydep} and \Cref{thm:soundabsdependences} are
deferred to \Cref{ap:dep_reloaded}: as we explain after
\Cref{lem:alphaagreevdec,lem:alphadeptrdec}, we derive the dependence abstract
semantics as an approximation of the cardinality semantics.

\section{Cardinality Abstraction}
\label{ap:cardinality}

\begin{restatable}[]{mylem}{lemcardinalitygalois}
$(\alphacardtr,\gammacardtr)$ yields a Galois connection:  
$\psetpsetTraces \galois{\alphacardtr}{\gammacardtr} \cardLat$.
\end{restatable}

\textit{Proof.}
The lattice $\cardLat$ is complete, since all subsets of $\cardLat$ have an
infimum and a supremum wrt.\ partial order $\cardlatleq$, notably because the closed interval
$\interv{0}{\infty}$ is complete wrt.\ partial order $\leq$.
Thus, this is an instance of  the supremus abstraction~\Cref{lem:galoissupremus}.  %

\lemabsvarietysound*

\text{Proof.}

The derivation proof is by structural induction on expressions.
In each case we start from the left side and derive the definition on the right side.
The interesting case is for binary arithmetic operations.

1 -- Case : integer literal $n$

Let $\mlsl \in \mlsLat$, and
$\cardconsset \in \cardLat$.
\begin{align*}
  &  \alphacardv \comp \obsexphcl{n}{} \comp \gammacardtr(\cardconsset)
  \\
  &  \quad    = 
  \alphacardv \comp \obsexphcl{n}{}
  \left( \{ \Tr \mid \cardtrop(\Tr) \cardlatleq
  \cardconsset \} \right) \\
  & \quad  =  \alphacardv \left(
  \cup_{\Tr \in \gammacardtr(\cardconsset)} \;  \{ \semexpc{n}{\altTr} \mid
  \altTr \subseteq \Tr \text{ and } \iagreerl \} \right) \\
  & \quad = \alphacardv \left(
   \{ \semexpc{n}{\altTr} \mid
  \altTr \subseteq \Tr, \iagreerl \text{ and } \Tr \in \gammacardtr(\cardconsset) \} 
  \right) \\
 & \quad  \leq \justif{NB: precision loss for simplicity of presentation, when $\cardconsset$
    is bottom} \\
  & \qquad  \alphacardv \left( \{ \{ n \} \} \right) \\
  & \quad = \max_{\Val \in {} \{ \{ n \} \} } \cardvop(\Val) \\
  & \quad = 1 \\
  & \quad \triangleq \absobsexpl{n}{\cardconsset} 
\end{align*}
Here we use $\triangleq$ to indicate that $\absobsexpl{n}{\cardconsset}$ is being defined.  

2 -- Case : variable $id$

Let $\mlsl \in \mlsLat$, and
$\cardconsset \in \cardLat$.

\begin{align*}
  &  \alphacardv \comp \obsexphcl{id}{} \comp \gammacardtr(\cardconsset)
  \\
  & \quad = \alphacardv \left(
  \cup_{\Tr \in  \gammacardtr(\cardconsset)} \;
   \obsexpcl{id}{\Tr}
   \right) \\
   & \quad = \justif{$\alphacardv$ preserves joins} \\
   & \qquad 
   \max_{\Tr \in  \gammacardtr(\cardconsset)} \alphacardv \left(
   \obsexpcl{id}{\Tr}
   \right) \\
   & \quad  = n \quad \justif{ where $\cardcons{id}{l}{n} \in
     \alphacardtr \comp \gammacardtr(\cardconsset)$} \\
   & \quad \leq
   \justif{$ \alphacardtr \comp \gammacardtr$ is reductive : $\alphacardtr
     \comp \gammacardtr(\cardconsset) \cardlatleq \cardconsset$} \\
   &  \qquad n \quad \justif{ where $\cardcons{id}{l}{n} \in \cardconsset$}  \\
   & \quad \triangleq  \absobsexpl{id}{\cardconsset}
\end{align*}

3 -- Case : $e_1 \oplus e_2$

Let $\mlsl \in \mlsLat$, and $\cardconsset \in \cardLat$.

\begin{align*}
  &  \alphacardv \comp \obsexphcl{e_1 \oplus e_2}{} \comp \gammacardtr(\cardconsset)
  \\
  & \quad = \alphacardv \left(
   \{ \semexpc{e_1 \oplus e_2}{\altTr} \mid
  \altTr \subseteq \Tr, \iagreerl \text{ and } \Tr \in \gammacardtr(\cardconsset) \} 
  \right) \\
  & \quad \leq %
  \alphacardv \left(
   \{ \semexpc{e_1 }{\altTr} \mid
  \altTr \subseteq \Tr, \iagreerl \text{ and } \Tr \in \gammacardtr(\cardconsset) \} 
  \right) \times \\
   & \quad \qquad \alphacardv \left(
   \{ \semexpc{ e_2}{\altTr} \mid
  \altTr \subseteq \Tr, \iagreerl \text{ and } \Tr \in \gammacardtr(\cardconsset) \} 
  \right) %
  \\
  & \quad = %
  \alphacardv \comp \obsexphcl{e_1}{}
  \comp \gammacardtr(\cardconsset) \times
  \alphacardv \comp \obsexphcl{e_2}{}
  \comp \gammacardtr(\cardconsset) %
  \\
  & \quad \leq \justif{By induction hypothesis} \\
  & \quad \qquad
 \absobsexpl{e_1}{\cardconsset}  \times \absobsexpl{e_2}{\cardconsset}
  \\
  & \quad \triangleq  \absobsexpl{e_1 \oplus e_2}{\cardconsset}
\end{align*}

4 -- Case : $e_1 \bcmp e_2$

This derivation is similar to case $e_1 \oplus e_2$, with the difference that
booleans evaluate to at most $2$ different values, 1 or 0.

\begin{align*}
  &  \alphacardv \comp \obsexphcl{e_1 \bcmp e_2}{} \comp
  \gammacardtr(\cardconsset) \\
  & \quad \leq  \min\left(2,   \absobsexpl{e_1}{\cardconsset}  \times
  \absobsexpl{e_2}{\cardconsset} \right) \\
  & \quad
  \triangleq  \absobsexpl{e_1 \bcmp e_2}{\cardconsset}
\end{align*}

5 -- Case : conclusion

We conclude by structural induction on expressions, and cases 1 to 4. \hfill $\qed$

\theoabscardsem*

\textit{Proof.}

The derivation proof is by structural induction on commands.
The interesting case is for conditionals.

1 -- Case : $\skipcom$

Let $\cardconsset \in \cardLat$.

\begin{align*}
  & \alphacardtr \comp \seminshc{\skipcom}{} \comp \gammacardtr(\cardconsset)\\
  & \quad = \alphacardtr \comp \gammacardtr(\cardconsset) \\
  & \quad \cardlatleq \justif{$\alphacardtr \comp \gammacardtr$ is reductive:
    $\alphacardtr \comp \gammacardtr(\cardconsset) \cardlatleq \cardconsset$} \\
  &  \qquad \cardconsset \\
  & \quad \triangleq  \seminshcabs{\skipcom}{\cardconsset} 
\end{align*}

2 -- Case : $c_1; c_2$

\begin{align*}
  & \alphacardtr \comp \seminshc{c_1; c_2}{} \comp \gammacardtr(\cardconsset) \\
  & \quad = \alphacardtr \comp \seminshc{c_2}{} \comp \seminshc{c_1}{}
  \comp \gammacardtr(\cardconsset) \\
  & \quad \cardlatleq \justif{$\gammacardtr  \comp  \alphacardtr$ is extensive,
    $\seminshc{c_2}{}$ and $\alphacardtr $ are monotone} \\
  & \qquad
  \alphacardtr \comp \seminshc{c_2}{} \comp
   \gammacardtr  \comp  \alphacardtr \comp
   \seminshc{c_1}{} \comp \gammacardtr(\cardconsset) \\
   & \quad \cardlatleq \justif{By induction hypothesis} \\
   & \qquad
   \seminshcabs{c_2}{} \comp  \seminshcabs{c_1}{\cardconsset} \\
   & \quad \triangleq \seminshcabs{c_1;c_2}{\cardconsset}
\end{align*}

3 -- Case : $id := e$

3.1 -- We first proceed towards an intermediate derivation:
\begin{align*}
  & \alphacardtr \comp \seminshc{id := e}{} \comp \gammacardtr(\cardconsset) \\
  & \quad
  =  \cardlatbigjoin_{\Tr \in  \seminshc{id := e}{} \comp
    \gammacardtr(\cardconsset)} \; \cardtrop(T) \\
  & \quad =
  \cardlatbigjoin_{\Tr \in  \seminshc{id := e}{} \comp
    \gammacardtr(\cardconsset)} \;
  \left(
   \bigcup_{\mlsl \in \mlsLat,\varx \in \varprog}
   \{ \cardconslx{\abscardval} \mid \abscardval  = 
   \alphacardv(\obsexpcl{x}{T}) \; \} \right) \\
   & \quad =
   \bigcup_{\mlsl \in \mlsLat,\varx \in \varprog} \;
  \left(
   \cardlatbigjoin_{\Tr \in\seminshc{id := e}{}\comp\gammacardtr(\cardconsset)} \;
   \{ \cardconslx{\abscardval} \mid \abscardval  = 
   \alphacardv(\obsexpcl{x}{T}) \; \} \right) \\
   & \quad =
   \bigcup_{\mlsl \in \mlsLat,\varx \in \varprog} \;
   \left(
   \left\{  \cardconslx{\abscardval} \mid n
    = 
   \max_{\Tr \in\seminshc{id := e}{}\comp\gammacardtr(\cardconsset)}
   \alphacardv(\obsexpcl{x}{T}) 
  \right\}
   \right)
\end{align*}

We now consider two cases: variables that are not modified by the assignment,
and variable that are.

3.2 -- Case $x \neq id$:

Notice that $\forall \mlsl \in \mlsLat, \forall \varx \in \varprog$, such that
$ \varx \neq id$,
$\forall \Tr \in \gammacardtr(\cardconsset)$:
\[ \obsexpcl{x}{T} =  \obsexpcl{x}{\big(\seminsc{id := e}{T}\big)} \]

Thus:
\begin{align*}
  \max_{\Tr \in\seminshc{id := e}{}\comp\gammacardtr(\cardconsset)} 
  & \alphacardv(\obsexpcl{\varx}{T})  \\
   &\quad =  \max_{\Tr \in \gammacardtr(\cardconsset)}  \alphacardv(\obsexpcl{\varx}{T}) \\
 & \quad  = \justif{$\alphacardv$ preserves joins} \\
  & \qquad  \alphacardv\left(
  \bigcup_{\Tr \in\gammacardtr(\cardconsset)} \;
  \obsexpcl{\varx}{T}\right) \\ 
  & \quad = \justif{By definition of $\obsexphcl{\varx}{}$} \\
  & \qquad \alphacardv\left(
  \obsexphcl{\varx}{\left(\gammacardtr(\cardconsset)\right)}\right) \\
  & \quad =
  \alphacardv \comp  \obsexphcl{\varx}{} \comp \gammacardtr(\cardconsset) \\
  & \quad \leq  \justif{By soundness of $ \absobsexpl{\varx}{\cardconsset}$,
    \Cref{lem:soundabsvariety}} \\
  &  \qquad  \absobsexpl{\varx}{\cardconsset} \\
  & = \abscardval \text{ where } \cardconslxn \in \cardconsset
  \end{align*}

3.3 -- Case $x$ is $id$ :

$ \forall \mlsl \in \mlsLat$, we have :
\begin{align*}
  \max_{\Tr \in\seminshc{id := e}{}\comp\gammacardtr(\cardconsset)} 
  & \alphacardv(\obsexpcl{id}{T})  \\
  & \quad  =  \max_{\Tr \in \gammacardtr(\cardconsset)}  \alphacardv(\obsexpcl{e}{T})
  \\
  & \quad  = \alphacardv \comp \obsexphcl{e}{} \comp   \gammacardtr(\cardconsset) \\
  & \quad \leq \justif{By soundness of $ \absobsexpl{\varx}{\cardconsset}$,
    \Cref{lem:soundabsvariety}} \\
  & \qquad   \absobsexpl{e}{\cardconsset} 
\end{align*}

3.4 -- Final derivation:

\begin{align*}
  & \alphacardtr \comp \seminshc{id := e}{} \comp \gammacardtr(\cardconsset) \\
  & \quad = \justif{Recall the intermediate derivation in Case 3.1} \\
  & \qquad 
   \bigcup_{\mlsl \in \mlsLat,\varx \in \varprog} \;
   \left(
   \left\{  \cardconslx{\abscardval} \mid n
    = 
   \max_{\Tr \in\seminshc{id := e}{}\comp\gammacardtr(\cardconsset)}
   \alphacardv(\obsexpcl{x}{T}) 
   \right \} \right) \\
   &  \quad \cardlatleq\justif{By  Cases 3.2 and 3.3} \\
   & \qquad
   \bigcup_{\mlsl \in \mlsLat} \;
   \left( \big( \bigcup_{\varx \in \varprog\setminus \{id\}} \left\{ \cardconslxn  \in
   \cardconsset  \} \big) 
   \cup \{ \cardcons{\mlsl}{id}{\absobsexpl{e}{\cardconsset} }  \right\} \right)\\
   & \qquad \justif{NB: this set of constraints remains valid, owing to exclusion of $id$ on the left} \\
   & \quad =  \left\{ \cardconslxn  \in \cardconsset \mid \varx \neq id \}
   \cup \{ \cardcons{\mlsl}{id}{\absobsexpl{e}{\cardconsset} }
   \mid \mlsl \in \mlsLat  \right\} \\
   & \quad \triangleq \seminshcabs{id := e}{\cardconsset}
\end{align*}

4 -- Case $\ifcom~ b~\thencom~c_1~\elsecom~c_0$:

4.1 -- Intermediate derivation:

\begin{align*}
  & \alphacardtr \comp \seminshc{\ifcom~b~\thencom~c_1~\elsecom~c_0}{} \comp
  \gammacardtr(\cardconsset) \\ 
  &  \quad =
  \cardlatbigjoin_{\Tr \in  \seminshc{\tinyifcomoz}{} \comp  \gammacardtr(\cardconsset)}
  \; \cardtrop(T) \\ 
  & \quad =
   \cardlatbigjoin\limits_{\Tr \in  \seminshc{\tinyifcomoz}{} \comp
    \gammacardtr(\cardconsset)} \;
  \left(
   \bigcup_{\mlsl \in \mlsLat,\varx \in \varprog}
   \left\{ \cardconslx{ \alphacardv(\obsexpcl{x}{T})}  \; \right\} \right) \\
   & \quad =
   \bigcup_{\mlsl \in \mlsLat,\varx \in \varprog}
   \left(
    \cardlatbigjoin\limits_{\Tr \in  \seminshc{\tinyifcomoz}{} \comp
      \gammacardtr(\cardconsset)} \;
    \left\{ \cardconslx{ \alphacardv(\obsexpcl{x}{T})}  \; \right\} 
    \right) \\
    & \quad =
      \bigcup_{\mlsl \in \mlsLat,\varx \in \varprog}
    \left\{ \cardconslx{
      \max_{\Tr \in
        \seminshc{\tinyifcomoz}{}
        \comp\gammacardtr(\cardconsset)}
   \alphacardv(\obsexpcl{x}{T})  }
    \; \right\} 
\end{align*}

4.2 -- Case $\absobsexpl{b}{\cardconsset} = 1$ :

Let $\mlsl \in \mlsLat$, and assume $\absobsexpl{b}{\cardconsset} = 1$.
Let $\varx \in \varprog$.

$\forall \Tr' \in \seminshc{\tinyifcomoz}{} \comp \gammacardtr(\cardconsset)$,
exists $\Tr \in \gammacardtr(\cardconsset)$
\text{ such that } 
$\Tr' = \seminsc{\tinyifcomoz}{\Tr}$.

(Since $\seminshc{c}{}$ is not just the lifting
of $\seminsc{c}{}$ to a set of sets (semantics of loops is not),
in general if $\Tr' \in \seminshc{c}{\BTr}$, we only have  the existence of
$\Tr \in \BTr$ such that $\Tr' \subseteq \seminsc{c}{\Tr}$.
Here, we also rely on the fact that $\gammacardtr(\cardconsset)$ is a
subset-closed.
This is merely a convenient shortcut to avoid lengthy details; it should be
possible to use only the fact that $\Tr' \subseteq \seminsc{c}{\Tr}$ to perform
the same derivation. )

Let $\Tr'  \in \seminshc{\tinyifcomoz}{} \comp \gammacardtr(\cardconsset),$ and
$\Tr \in \gammacardtr(\cardconsset)$ \text{ such that } 
$\Tr' = \seminsc{\tinyifcomoz}{\Tr}$.

Since
$\alphacardv \comp \obsexphcl{b} \comp \gammacardtr(\cardconsset)
\dotleq \absobsexpl{b}{\cardconsset} (= 1) $,
$\forall \altTr \subseteq \Tr$ such that $ \iagreerl$, the traces
$\alttr \in \altTr$ all evaluate $b$ to 1 or (exclusively) to 0; i.e. the sets
$\altTr \subseteq \Tr$ such that $ \iagreerl$ are partitioned into the sets
evaluating $b$ to 1, and those evaluating $b$ to 0.

Therefore, $\forall \altTr' \subseteq \Tr'$ such that $ \iagreel{\altTr'}$,
exists $\altTr \in \Tr$ and $j\in \{ 0, 1\}$ such that
$\semexpc{b}{\altTr} = \{ j \}$ and  $\seminsc{c_j}{\altTr} = \altTr'$. 

Thus,

\begin{align*}
  & \alphacardv(\obsexpcl{x}{\Tr'})  \\
  & \quad =
  \alphacardv(  \{ \semexpc{x}{\altTr'} \mid \altTr' \subseteq \Tr' \text{ and }
  \iagreel{\altTr'} \}  ) \\
  & \quad = \alphacardv \left(
  \bigcup_{ \altTr \subseteq \Tr  \text{ and }  \iagreel{\altTr} }
  \{ \semexpc{x}{ (\seminsc{\tinyifcomoz}{\altTr}) } \}  \right) \\
  & \quad = \alphacardv \left(
  \bigcup_{j \in \{ 0,1\}}
  \bigcup_{\altTr \subseteq \Tr  \text{ and }  \iagreel{\altTr}  \text{ and }
    \semexpc{b}{\altTr} = \{ j \}
}
  \left\{ \semexpc{x}{( \seminsc{c_j}{\altTr})}   \right\}  \right) \\
  & \quad = \max_{j \in \{0,1 \}}
  \left( \alphacardv \left(
  \bigcup_{ \altTr \subseteq \Tr  \text{ and }  \iagreel{\altTr}  \text{ and }
    \semexpc{b}{\altTr} = \{ j \}}
  \left\{  \semexpc{x}{( \seminsc{c_j}{\altTr})}   \right\} 
  \right) \right) \\
  & \quad \leq \justif{$\alphacardv$ is monotone} \\
  & \qquad
  \max_{j \in \{0,1 \}}
  \left( \alphacardv \comp
  \obsexphcl{x}{} \comp \seminshc{c_j}{} \comp \gammacardtr( \cardconsset )
  \right) \\
  & \quad \leq \justif{$\alphacardv\comp\obsexphcl{x}{}$ is monotone,
    $\gammacardtr \comp \alphacardtr$ extensive } \\
  & \qquad
   \max_{j \in \{0,1 \}}
  \left( \alphacardv \comp
  \obsexphcl{x}{} \comp \gammacardtr \comp \alphacardtr \seminshc{c_j}{} \comp
  \gammacardtr( \cardconsset ) 
  \right) \\
  & \quad \leq \justif{By induction hypothesis} \\
  & \qquad
  \max_{j \in \{0,1 \}}
   \left( \alphacardv \comp
  \obsexphcl{x}{} \comp \gammacardtr \comp \seminshcabs{c_j}{ \cardconsset }
  \right) \\
  & \quad \leq \justif{By soundness of abstract variety, \Cref{lem:soundabsvariety}} \\
  & \qquad
    \max_{j \in \{0,1 \}}
    \left(  \absobsexpl{x}{} \comp \seminshcabs{c_j}{ \cardconsset }
    \right) \\
    & \quad  = \max_{j \in \{0,1 \}}
    \left( n_j \text { where } \cardconslx{n_j} \in
    \seminshcabs{c_j}{ \cardconsset } \right) \\
\end{align*}

4.3 -- Case $\absobsexpl{b}{\cardconsset} > 1$, $\varx \notin \Mod(\tinyifcomoz)$ :

Let $\mlsl \in \mlsLat$, and assume $\absobsexpl{b}{\cardconsset} > 1$.
Let $\varx \in \varprog$.

Let $\Tr'  \in \seminshc{\tinyifcomoz}{} \comp \gammacardtr(\cardconsset),$ and
$\Tr \in \gammacardtr(\cardconsset)$ \text{ such that } 
$\Tr' = \seminsc{\tinyifcomoz}{\Tr}$.

Notice first that if $\varx \notin \Mod(\tinyifcomoz)$, then:
\begin{align*}
  \alphacardv(\obsexpcl{\varx}{\Tr'}) & =  \alphacardv(\obsexpcl{\varx}{\Tr}) \\
  & \leq \justif{$\alphacardv$ is monotone} \\
  & \quad
  \alphacardv \comp \obsexphcl{\varx} \comp \gammacardtr(\cardconsset) \\
  & \leq \absobsexpl{\varx}{\cardconsset} \\
  & = \abscardval \text{ s.t }\cardconslxn \in \cardconsset
\end{align*}

4.4 -- Case $\absobsexpl{b}{\cardconsset} > 1$, $\varx \in \Mod(\tinyifcomoz)$ :

Let $\mlsl \in \mlsLat$, and assume $\absobsexpl{b}{\cardconsset} > 1$.
Let $\varx \in \varprog$.

Let $\Tr'  \in \seminshc{\tinyifcomoz}{} \comp \gammacardtr(\cardconsset),$ and
$\Tr \in \gammacardtr(\cardconsset)$ \text{ such that } 
$\Tr' = \seminsc{\tinyifcomoz}{\Tr}$.

\begin{align*}
  & \alphacardv(\obsexpcl{\varx}{\Tr'}) \\
  & \quad =
  \alphacardv(\obsexpcl{\varx}{} \comp \seminsc{\tinyifcomoz}{\Tr}) \\
  & \quad \leq 
  \alphacardv(\obsexpcl{\varx}{} \left(
  \seminsc{c_1}{}\comp \guardc{b}{\Tr}
  \cup
    \seminsc{c_0}{}\comp \guardc{\neg b}{\Tr}
  \right ) \\
  & \quad \leq
   \alphacardv\left(\obsexpcl{\varx}{} \comp 
   \seminsc{c_1}{}\comp \guardc{b}{\Tr}
   \right)
   +
   \alphacardv\left(\obsexpcl{\varx}{} \comp
   \seminsc{c_0}{}\comp \guardc{\neg b}{\Tr}
   \right ) \\
   & \quad \leq \justif{By monotonicity, $T \in \gammacardtr(\cardconsset)$ and
     \Cref{theo:hypercollsound}} \\ 
   & \qquad
    \alphacardv \comp \obsexphcl{\varx}{} \comp 
   \seminshc{c_1}{} \comp \guardhc{\neg b}{} \comp \gammacardtr(\cardconsset)
   +
   \alphacardv \comp \obsexphcl{\varx}{} \comp
   \seminshc{c_0}{}\comp \guardhc{\neg b}{} \comp \gammacardtr(\cardconsset)
    \\
    & \quad \leq
   \absobsexpl{\varx}{} \comp 
   \seminshcabs{c_1}{} \comp \guardhcabs{\neg b}{\cardconsset}
   +
   \absobsexpl{\varx}{} \comp
   \seminshcabs{c_0}{} \comp \guardhcabs{\neg b}{\cardconsset}
    \\
   & \quad \leq \justif{As a first approximation, we simply use
     $\guardhcabs{b}{\cardconsset} \cardlatleq \cardconsset$. We refine this
      in \Cref{sec:precision}} \\
 & \qquad 
   \absobsexpl{\varx}{} \comp \seminshcabs{c_1}{\cardconsset}
   +
   \absobsexpl{\varx}{} \comp \seminshcabs{c_2}{\cardconsset} \\
   & \quad =
   \abscardval_1 + \abscardval_2 \text{ s.t }
   \cardconslx{\abscardval_1} \in
   \seminshcabs{c_1}{\cardconsset} \text{ and }
    \cardconslx{\abscardval_2} \in
   \seminshcabs{c_2}{\cardconsset} 
\end{align*}

4.4 -- Final derivation:

\begin{align*}
 & \alphacardtr \comp \seminshc{\ifcom~b~\thencom~c_1~\elsecom~c_0}{} \comp
  \gammacardtr(\cardconsset) \\
  & \quad = \justif{By the intermediate derivation in case 4.1} \\
  & \qquad 
      \bigcup_{\mlsl \in \mlsLat,\varx \in \varprog}
    \left\{ \cardconslx{
      \max_{\Tr \in
        \seminshc{\tinyifcomoz}{}
        \comp\gammacardtr(\cardconsset)}
   \alphacardv(\obsexpcl{x}{T})  }
    \; \right\}
    \\
    & \quad \leq
    \bigcup_{\mlsl \in \mlsLat}
    \begin{cases}
      \projtomlsl{\seminshcabs{c_1}{\cardconsset}} \cardlatjoin
      \projtomlsl{\seminshcabs{c_2}{\cardconsset} }
      & \text{ if } \absobsexpl{b} = 1 \\
      \projtomlsl{\seminshcabs{c_1}{\cardconsset}}
      \cardlatadd{\tinyifcomoz,\projtomlsl{\cardconsset}} 
      \projtomlsl{\seminshcabs{c_1}{\cardconsset}}
      & \text{ otherwise }
    \end{cases}
\end{align*}

with
\[ \projtomlsl{\cardconsset} \triangleq \{ \cardconslxn \in \cardconsset \mid
\varx \in \varprog, \abscardval \in \abscardValues \} \]

and

\begin{multline*}
C_1 {}  \cardlatadd{com,C_0} {} C_2
\triangleq \\
\bigcup_{\varx \in \Mod(com)}
\{ \cardconslx{n} \mid n \triangleq n_1 + n_2 \text{ s.t } \cardconslx{n_j} \in
C_j , j=1,2\} \\
\bigcup_{\varx \in \varprog \setminus \Mod(com) } \{  \cardconslxn \in C_0 \} 
\end{multline*}

5 -- Case $\whilecom~ b~\docom~c$:

\begin{align*}
& \alphacardtr \comp \seminshcabs{\whilecom~b~\docom~c}{} \comp 
\gammacardtr(\cardconsset) \\
& \quad = 
\alphacardtr \comp 
\guardhc{\neg b}{}
\left(\operatorname{lfp}_{\gammacardtr(\cardconsset)}^{\subseteq}
  \seminshc{\ifcom~ b~ \thencom~ c~ \elsecom~ \skipcom}{} \right) \\
& \quad \cardlatleq \justif{$\alphacardtr$,$\guardhc{\neg b}{}$ are
  monotone, $\gammacardtr \comp \alphacardtr$ is extensive} \\
& \qquad 
\alphacardtr \comp 
\guardhc{\neg b}{} \comp \gammacardtr \comp \alphacardtr 
\left(\operatorname{lfp}_{\gammacardtr(\cardconsset)}^{\subseteq}
  \seminshc{\ifcom~ b~ \thencom~ c~ \elsecom~ \skipcom}{} \right) \\
& \quad \cardlatleq \justif{By assuming $\guardhcabs{b}{}$ is sound} \\
& \qquad \guardhcabs{\neg b}{} \comp \alphacardtr
\left(\operatorname{lfp}_{\gammacardtr(\cardconsset)}^{\subseteq}
  \seminshc{\ifcom~ b~ \thencom~ c~ \elsecom~ \skipcom}{} \right) \\
& \quad \cardlatleq \justif{By the fixpoint transfer theorem} \\
& \qquad  
\guardhcabs{\neg b}{} \comp 
\operatorname{lfp}^{\cardlatleq}_{\cardconsset}
\seminshcabs{\ifcom~ b~ \thencom~ c_1~ \elsecom~ c_2}{} \\
& \quad \cardlatleq 
\justif{precision loss for simplicity as a first approximation, $\guardhcabs{\neg b}{}
  \dotcardlatleq \operatorname{id}$} \\
& \qquad \operatorname{lfp}^{\cardlatleq}_{\cardconsset}
\seminshcabs{\ifcom~ b~ \thencom~ c_1~ \elsecom~ c_2}{} \\
& \quad \triangleq 
\seminshcabs{\whilecom~b~\docom~c}{\cardconsset}
\end{align*}

6 -- Case : conclusion

We conclude by structural induction on commands and cases 1 to 5. \hfill $\qed$

\clearpage

\section{Dependencies reloaded}
\label{ap:dep_reloaded}

\subsection{Soundness proof for dependences semantics}
\label{ap:dep_reloaded:soundsess}

As noted in the text, we can derive the dependency analysis by calculation from 
its specification.
The derivation looks similar to the one
in \Cref{ap:cardinality} for the cardinality abstraction.
So here we choose a different way of proving soundness for dependency analysis.
We formulate it as an abstraction of the cardinality abstraction.
This is another illustration of the benefit gained from working with 
hyperproperties entirely within the framework of abstract interpretation.

This proof of soundness also implies that the cardinality abstraction is at least
as precise as the type system of Hunt and Sands~\cite{HS06} and the logic of
Amtoft and Banerjee~\cite{AB04}, as a corollary of  
\Cref{thm:theohsprecision}.

\lemalphaagreevdec*

\textit{Proof.}

Notice that:
\[ 
\agreev(\Val)  \triangleq (\forall \val_1,\val_2 \in \Val, v_1 = v_2)  = (\cardvop(\Val) \leq 1)
\]

Also,
\begin{align*}
\alphaagreev(\BVal) & \triangleq \wedge_{\Val \in \BVal} \agreev(\Val) \\
& = \wedge_{\Val \in \BVal} (\cardvop(\Val) \leq 1)\\
& = \left( \max_{\Val \in \BVal} \cardvop(\Val) \right) \leq 1  \\
& =  \alphacardv(\BVal) \leq 1 \\
& = \alphalone \comp \alphacardv(\BVal) \qquad  \text{where } 
\alphalone(\abscardval) \triangleq 
\begin{cases}
\true & \text{if } \abscardval \leq 1 \\
\false & \text{otherwise}
\end{cases}
\end{align*}
And,
\begin{align*}
  \gammaagreev(\absbool) & \triangleq \{ \Val \in \psetValues \mid \agreev(\Val)
  \impliedby \absbool  \} \\
  & = \{ \Val \in \psetValues \mid \
  \cardvop(\Val) \leq  \gammalone (\absbool) \} \quad 
  \text{where } 
  \\
  & \qquad \qquad
  \gammalone(\absbool) \triangleq 
  \begin{cases}
    1 & \text{if } \absbool = \true \\
    \sizeval & \text{otherwise}
  \end{cases} \\
  & = \gammacardv \comp \gammalone (\absbool)
\end{align*}

Notice that $\abscardValues \galois{\alphalone}{\gammalone} \abstruth$:
\[ 
\forall \abscardval \in \abscardValues, \forall \absbool \in \abstruth,
\qquad
\alphalone(\abscardval) \impliedby \absbool 
\text{ iff. } 
\abscardval \leq \gammalone(\absbool) 
\]

Thus, we obtain $\alphaagreev = \alphalone \comp \alphacardv$, as well as
$\gammaagreev = \gammacardv \comp \gammalone$:
\[ 
(\psetpsetValues,\subseteq) \galois{\alphacardv}{\gammacardv} 
(\abscardValues, \leq)
 \galois{\alphalone}{\gammalone} (\abstruth, \impliedby) \]

\hfill $\qed$

\lemalphadeptrdec*

\textit{Proof.}

First,
\begin{align*}
  \alphadeptr(\BTr) & =  \deplatbigjoin_{\Tr \in \BTr} \deptrop(T) \\
  & =
  \deplatbigjoin_{\Tr \in \BTr} \;
  \bigcup_{\mlsl \in \mlsLat,\varx \in \varprog}
  \{ \depconslx \mid  \alphaagreev(\obsexpcl{x}{T})
  \} \\
  & = \justif{By the decomposition in Case 1} \\
  & \qquad  \deplatbigjoin_{\Tr \in \BTr} \;
  \bigcup_{\mlsl \in \mlsLat,\varx \in \varprog}
  \{ \depconslx \mid  \alphalone \comp \alphacardv(\obsexpcl{x}{T}) \} \\
  & =
  \deplatbigjoin_{\Tr \in \BTr} \;
    \alphalonecc \left(  \bigcup_{\mlsl \in \mlsLat,\varx \in \varprog} \{
    \cardconslxn \mid n = \alphacardv(\obsexpcl{x}{T}) \} \right) \\ 
    & \qquad \quad \justif{with
    $\alphalonecc(\cardconsset) \triangleq \{ \depconslx \mid \cardconslxn \in
      \cardconsset \text{ and }  \alphalone(n)  \} $} \\
    & =  \deplatbigjoin_{\Tr \in \BTr} \;
    \alphalonecc \comp \cardtrop(\Tr) \\
    & =   \justif{$\alphalonecc$ preserves unions} \\
    & \quad %
    \alphalonecc \left(  \cardlatbigjoin_{\Tr \in \BTr} \cardtrop(\Tr)  \right)
    \\  
    & = \alphalonecc \comp \alphacardtr(\BTr)
\end{align*}

Also,
\begin{align*}
  \gammadeptr(\depconsset) & =
  \{ \Tr \mid \deptrop(\Tr) \deplatleq \depconsset \} \\
  & =  \{ \Tr \mid
  \alphalonecc \comp \cardtrop (\Tr) \deplatleq \depconsset \} \\
  & =  \{ \Tr \mid
  \alphalonecc \comp \cardtrop (\Tr) \supseteq \depconsset \} \\
  & =  \{ \Tr \mid \forall \depconslx \in \depconsset, \depconslx \in
  \alphalonecc \comp \cardtrop (\Tr)  \} \\
  & =  \bigg\{ \Tr \mid \forall \depconslx \in \depconsset,  \\
  & \qquad \qquad  \depconslx \in
  \alphalonecc\left(\cup_{\mlsl \in \mlsLat,\varx \in \varprog}
  \{ \cardconslx{\abscardval} \mid \abscardval =
  \alphacardv(\obsexpcl{x}{T}) \; \} \right) \bigg\} \\
  & = \{ \Tr \mid \forall \depconslx \in \depconsset, \alphalone(
  \alphacardv(\obsexpcl{x}{T})) \} \\
  & =  \{ \Tr \mid \forall \depconslx \in \depconsset,
  \alphacardv(\obsexpcl{x}{T}) \leq 1 \} \\
  & = \gammacardtr \comp \gammalonecc(\depconsset) \\
  & \qquad \quad \justif{$ \gammalonecc(\depconsset)
    \triangleq 
    \bigcup_{\mlsl \in \mlsLat, \varx \in \varprog}
    \{ \cardconslxn \mid n = \gammalone( \depconslx \in \depconsset)\}
    $}   
\end{align*}

Therefore, we have $\alphadeptr = \alphalonecc \comp \alphacardtr$, and
$\gammadeptr = \gammacardtr \comp \gammalonecc$, with:
\[
(\psetpsetTraces,\subseteq) \galois{\alphacardtr}{\gammacardtr}
(\cardLat,\cardlatleq)
 \galois{\alphalonecc}{\gammalonecc}
(\depLat,\deplatleq)
\]

\hfill $\qed$

\lemabsvarietydepsound*

\textit{Proof.}

1 --  Derivation of Agreements $\absobsexpdepl{e}{}$ up to $\mlsl$ as an
abstraction of cardinalities up to security level $\mlsl$.

\begin{align*}
& \alphaagreev \comp \obsexphcl{n}{} \comp \gammadeptr (\depconsset) \\
  & \quad =
  \alphalone \comp \alphacardv \comp \obsexphcl{n}{}
  \comp \gammadeptr \comp
  \gammalonecc (\depconsset) \\
  & \quad \impliedby
  \alphalone \comp \absobsexpl{e}{} \comp
   \gammalonecc (\depconsset) 
\end{align*}

Henceforth, we will derive  $\absobsexpdepl{e}{}$  as an abstraction of
cardinalities $\absobsexpl{e}{}$.
This derivation goes by structural induction on expressions.

1.1 -- Case  : $n$

Let $\mlsl \in \mlsLat$, $\depconsset \in \depLat$.
\begin{align*}
  \alphalone \comp \absobsexpl{n}{} \comp
  \gammalonecc (\depconsset)
  & =  \alphalone(1) \\
  & = \true \qquad \justif{$ \triangleq \absobsexpdepl{n}{}$}
\end{align*}

1.2 -- Case : $id$
Let $\mlsl \in \mlsLat$, $\depconsset \in \depLat$.
\begin{align*}
  \alphalone \comp \absobsexpl{id}{} \comp
  \gammalonecc (\depconsset)
  & = \alphalone(\abscardval) \text{ where }
\cardcons{l}{id}{n} \in \gammalonecc (\depconsset) \\
  & =( \depcons{l}{id} \in \depconsset) \qquad \justif{$ \triangleq \absobsexpdepl{id}{}$}
\end{align*}

1.3 -- Case : $e_1 \oplus e_2$
Let $\mlsl \in \mlsLat$, $\depconsset \in \depLat$.
\begin{align*}
  &  \alphalone \comp \absobsexpl{e_1 \oplus e_2}{} \comp
  \gammalonecc (\depconsset) \\
  & \quad =  \alphalone\left(
(\absobsexpl{e_1}{}\comp\gammalonecc (\depconsset)) \times (\absobsexpl{e_2}{}
\comp \gammalonecc (\depconsset)) \right) \\
& \quad \impliedby \alphalone \comp \absobsexpl{e_1}{}\comp\gammalonecc
(\depconsset)
\wedge \alphalone \comp \absobsexpl{e_2}{}\comp\gammalonecc(\depconsset)\\
&  \quad = \absobsexpdepl{e_1}{\depconsset} \wedge \absobsexpdepl{e_2}{\depconsset}
\qquad \justif{$ \triangleq   \absobsexpdepl{e_1 \oplus e_2}{\depconsset}$}
\end{align*}

1.4 -- Case : $e_1 \bcmp e_2$

This case is similar to case 1.3.

1.5 -- Case: conclusion

We conclude by structural induction on expressions. \hfill $\qed$

\theoabsdepsem*

\textit{Proof.}

Recall that we have $\alphadeptr = \alphalonecc \comp \alphacardtr$, and
$\gammadeptr = \gammacardtr \comp \gammalonecc$, with:
\[
\psetpsetTraces \galois{\alphacardtr}{\gammacardtr}
\cardLat \galois{\alphalonecc}{\gammalonecc}
\depLat
\]

Since,

\begin{align*}
&  \alphadeptr \comp \seminshc{c}{} \comp \gammadeptr(\depconsset) \\
& \quad =   \alphalonecc \comp \alphacardtr \comp  \seminshc{c}{} \comp
  \gammacardtr \comp \gammalonecc(\depconsset) \\
  & \quad \deplatleq \alphalonecc \comp \seminshcabs{c}{} \comp
  \gammalonecc(\depconsset) 
\end{align*}

We will continue the derivation of dependences abstract semantics
$\seminshcdepabs{c}{}$ as an abstraction of $\seminshcabs{c}{}$.

We make explicit 2 derivations, for assignments and conditionals. The other
cases are similar to the derivation of the cardinalities abstract semantics.

1 -- Case : $id := e$

\begin{align*}
  &  \alphalonecc \comp \seminshcabs{id := e}{} \comp
  \gammalonecc(\depconsset) \\
  & \quad =  \alphalonecc\left(
  \{ \cardconslxn \in \gammalonecc(\depconsset) \mid \varx \neq id
  \} \cup
  \{ \cardcons{\mlsl}{id}{n} \mid n \triangleq \absobsexpl{e}{} \comp \gammalonecc(\depconsset), 
  \mlsl \in \mlsLat \}
  \right)
   \\
   & \quad = \{ \depconslx \in \depconsset \mid \varx \neq id \} \cup
   \{ \depcons{\mlsl}{id} \mid \absobsexpdepl{e}{} \}
\end{align*}

2 -- Case $\ifcom~ b~ \thencom~ c_1~ \elsecom~ c_2$:

\begin{align*}
  & \alphalonecc \comp
  \seminshcabs{\ifcom~ b~ \thencom~ c_1~ \elsecom~ c_2}{} \comp \gammalonecc(\depconsset) \\
  & \quad = \begin{tabular}[t]{l}
$\; \; \letin{\cardconsset_1 = \seminshcabs{c_1}{} \comp \gammalonecc(\depconsset)}$ \\
$\; \; \letin{\cardconsset_2 = \seminshcabs{c_2}{} \comp \gammalonecc(\depconsset) }$ \\
$\; \; \letin{W = \Mod(\tinyifcomoz)}$ \\
$\; \;  \alphalonecc\left( \bigcup\limits_{\mlsl \in \mlsLat}
\begin{cases}
\projtomlsl{\cardconsset_1} 
\cardlatjoin   
\projtomlsl{\cardconsset_2} 
& \text{if } \absobsexpl{b}{} \comp  \gammalonecc(\depconsset) = 1 \\
\projtomlsl{\cardconsset_1} & \\
\quad \cardlatadd{W,\projtomlsl{\cardconsset}}  \\
\qquad \projtomlsl{\cardconsset_2} & \text{otherwise}
\end{cases}\right)$
  \end{tabular} \\
  & \quad \impliedby
  \begin{tabular}[t]{l}
$ \; \; \letin{\depconsset_1 = \seminshcdepabs{c_1}{ \depconsset}}$ \\
$\; \; \letin{\depconsset_2 = \seminshcdepabs{c_2}{\depconsset} }$ \\
$\; \; \letin{W = \Mod(\tinyifcomoz)}$ \\
$\; \;  \bigcup\limits_{\mlsl \in \mlsLat}
\begin{cases}
\projtomlsl{\depconsset_1} 
\deplatjoin   
\projtomlsl{\depconsset_2} 
& \text{if } \absobsexpdepl{b}{\depconsset} \\
\projtomlsl{\depconsset} \setminus \{ \depcons{\mlsl}{x} \mid x \in W \} 
& \text{otherwise}
\end{cases}$
  \end{tabular} 
\end{align*}

We conclude by structural induction on commands \hfill $\qed$.

\subsection{Precision proof}
\label{app:precision_proof}
\begin{restatable}[]{mylem}{lemvarietycmon}
  \label{lem:lemvarietycmon}
  For all $\mlsl,\mlsl' \in \mlsLat$, for all $\Tr \in \psetTraces$:
  
  \[ \mlsl \mlslatleq \mlsl' \implies
   \obsexpc{e}{\mlsl'}{\Tr}  \subseteq
  \obsexpc{e}{\mlsl}{\Tr} 
 \]
\end{restatable}

\textit{Proof.}

Assume $\mlsl \mlslatleq \mlsl' $.
Then, for all $\altTr \subseteq \Tr$,
\[ \iagree{\altTr}{\mlsl'} \implies
\iagree{\altTr}{\mlsl} \]
Thus,
\[
\{ \altTr \mid \altTr \subseteq \Tr  \text{ and } \iagree{\altTr}{\mlsl'} \}
\subseteq \{ \altTr \mid \altTr \subseteq \Tr  \text{ and } \iagree{\altTr}{\mlsl} \}
\]

Therefore, it holds that:

\[ \obsexpc{e}{\mlsl'}{\Tr} \subseteq \obsexpc{e}{\mlsl}{\Tr} \hfill \qed  \]

\begin{restatable}[]{mycor}{corvarietyhcmon}
    \label{lem:corvarietyhcmon}
  For all $\mlsl,\mlsl' \in \mlsLat$, for all $\BTr \in \psetpsetTraces$:
  \[ \mlsl \mlslatleq \mlsl' \implies
   \obsexphc{e}{\mlsl'}{\BTr}  \subseteq
  \obsexphc{e}{\mlsl}{\BTr} 
 \]
\end{restatable}

\textit{Proof.} This is a direct result from \Cref{lem:lemvarietycmon} and
definition of $\obsexphc{e}{\mlsl}{}$.

\hfill \qed

\begin{restatable}[]{mycor}{coralphadepmon}
  \label{cor:alphadepmon}
  For all $\mlsl,\mlsl' \in \mlsLat$, for all $id$, for all $\BTr \in \psetpsetTraces$:
  \[
  \mlsl \mlslatleq \mlsl' \implies \left(
  \depcons{\mlsl}{id} \in \alphadeptr(\BTr) \implies  \depcons{\mlsl'}{id} \in
  \alphadeptr(\BTr)
  \right)
 \]
\end{restatable}

\textit{Proof.}

Let us assume  $ \mlsl \mlslatleq \mlsl' $.
By \Cref{lem:corvarietyhcmon}, we have
$\obsexphc{id}{\mlsl'}{\BTr}  \subseteq  \obsexphc{id}{\mlsl}{\BTr} $.

  And by monotonicity of $\alphaagreev$, we have
  $ \alphaagreev(\obsexphc{id}{\mlsl'}{\BTr})  \abstruthleq
  \alphaagreev(\obsexphc{id}{\mlsl}{\BTr})$. 

  Thus, if $ \depcons{\mlsl}{id} \in \alphadeptr(\BTr) $, then
  $\alphaagreev(\obsexphc{id}{\mlsl}{\BTr}) = \true$  and also
 
  $\alphaagreev(\obsexphc{id}{\mlsl'}{\BTr}) = \true$, thus
  $\depcons{\mlsl'}{\depconsset} \in   \alphadeptr(\BTr) $.

  \hfill \qed

  \begin{restatable}[]{mycor}{corgammadepmon}
    \label{cor:gammadepmon}
  For all $\mlsl,\mlsl' \in \mlsLat$, for all $id$, for all $\depconsset \in \depLat$:
  \[
  \mlsl \mlslatleq \mlsl' \implies
  \gammadeptr( \depconsset \cup \{ \depcons{\mlsl}{id} \} ) =
  \gammadeptr( \depconsset \cup \{  \depcons{\mlsl}{id} ,  \depcons{\mlsl'}{id}\})
 \]
  \end{restatable}

  \textit{Proof.}
  
  1 --
  Note that $ \depconsset \cup \{  \depcons{\mlsl}{id} ,  \depcons{\mlsl'}{id}\}
  \supseteq
  \depconsset \cup \{ \depcons{\mlsl}{id} \} $, thus
  \[
   \depconsset \cup \{
  \depcons{\mlsl}{id} ,  \depcons{\mlsl'}{id}\}
  \deplatleq 
  \depconsset \cup \{ \depcons{\mlsl}{id} \}
  \]
  Therefore, by monotony of $\gammadeptr$:
  \[
    \gammadeptr( \depconsset \cup \{  \depcons{\mlsl}{id} ,  \depcons{\mlsl'}{id}\} ) 
    \subseteq
    \gammadeptr(\depconsset \cup \{ \depcons{\mlsl}{id} \})
  \]

  2 -- Also, let   
  $\Tr \in  \gammadeptr( \depconsset \cup \{ \depcons{\mlsl}{id} \} )$.
  
  We have
  $ \deptrop(\Tr) \deplatleq \depconsset \cup \{ \depcons{\mlsl}{id} \}$ by
  definition of $\gammadeptr$.

  Thus, $\depcons{\mlsl}{id} \in \alphadeptr({\Tr})$ and also
  $\depcons{\mlsl'}{id} \in \alphadeptr({\Tr})$ by \Cref{cor:alphadepmon}.
  This also means that
  $\deptrop(\Tr) \deplatleq \depconsset \cup \{
  \depcons{\mlsl}{id}, \depcons{\mlsl'}{id} \} $.
  Finally,
  $\Tr \in \gammadeptr( \depconsset \cup \{ \depcons{\mlsl}{id},
  \depcons{\mlsl'}{id} \} )$ and

  \[   
    \gammadeptr( \depconsset \cup \{  \depcons{\mlsl}{id} ,  \depcons{\mlsl'}{id}\} ) 
    \supseteq
    \gammadeptr(\depconsset \cup \{ \depcons{\mlsl}{id} \})
  \]

  This concludes our proof by Case 1 and 2.

  \hfill \qed

Henceforth, we will assume that all $\depconsset \in \depLat$ are well formed,
meaning that $\forall \mlsl,\mlsl' \in \depLat$,
$\depcons{\mlsl}{id} \in \depconsset \implies  \depcons{\mlsl'}{id} \in
\depconsset$.

We conjecture that  this can be proven for the dependence analysis we have derived: given
a well formed initial set of dependence constraints, the analysis always yields a
well formed set of dependence constraints.
For simplicity, we will use \Cref{cor:gammadepmon} to argue that  we can still
augment any set of dependence constraints to ensure it is well formed by adding
the appropriate atomic constraints.
An alternative approach would reduce the set of dependence constraints, and
change slightly the abstract semantics in order to leverage
\Cref{cor:gammadepmon} and guarantee the same precision, but we refrain from
doing so for simplicity.

We consider the constructive version of Hunt and Sands' flow sensitive type system, proposed in
\cite{HS11sd}. 

\begin{restatable}[]{mylem}{lemexpprecision}
  \label{lem:expprecision}
  For all $e$, $\depconsset \in \depLat$, $\Delta \in \varprog \to \mlsLat$,
  $\mlsl \in \mlsLat$ such that $\Delta \vdash e : \mlsl$, it holds that:
  \[ \alphatohs(\depconsset) \dotmlslatleq \Delta 
  \implies \absobsexpdepl{e}{\depconsset} = \true  \]
\end{restatable}

\textit{Proof.}

The proof proceeds by structural induction on expressions.

1 -- Case $n$:

By definition of $\absobsexpdepl{n}$, we have:
\[ \forall \depconsset, \forall \mlsl \in \mlsLat,
 \absobsexpdepl{n}{\depconsset} = \true \]

2 -- Case $id$:

By definition of the type system, we have $\Delta(id) = \mlsl$. Thus:

\begin{align*}
  \alphatohs( \depconsset ) \dotmlslatleq \Delta &
  \implies \mlslatmeet \{ \mlsl' \mid
  \depcons{\mlsl'}{id} \in \depconsset \} \mlslatleq \mlsl \\
  & \implies \justif{since $\depconsset$ is assumed well-formed} \\
  & \qquad \depcons{\mlsl}{id} \in \depconsset \\
  & \implies \absobsexpdepl{id}{\depconsset} = \true
\end{align*}

3 -- Case $e_1 \oplus e_2$:

By definition of the type system, there is $\mlsl_1,\mlsl_2$ such that 
$\Delta \vdash e_1 : \mlsl_1$ and  $\Delta \vdash e_2 : \mlsl_2$, with 
$\mlsl_1 \mlslatjoin \mlsl_2 = \mlsl$.

Thus, by induction on $e_1$ and $e_2$, and assuming $\alphatohs(\depconsset)
\dotmlslatleq \Delta$,  we have:
\[
\absobsexpdep{e_1}{\mlsl_1}{\depconsset} = \true {} \wedge {}
\absobsexpdep{e_2}{\mlsl_2}{\depconsset} = \true \]

Thus, since $\depconsset$ is well formed and $\mlsl_1 \mlslatleq \mlsl$ and 
$\mlsl_2 \mlslatleq \mlsl$, it holds that:
\[
\absobsexpdep{e_1}{\mlsl}{\depconsset} = \true \; \wedge \;
\absobsexpdep{e_2}{\mlsl}{\depconsset} = \true \]

Therefore, also:
\[ \absobsexpdep{e_1 \oplus e_2}{\mlsl}{\depconsset} = \true \]

4 -- Case $e_1 \bcmp e_2$:

This case is similar to Case 3.

5 -- We conclude by structural induction and Cases 1 to 4.

\hfill \qed

Let us denote by $\mlsbot \in \mlsLat$ the bottom element of the lattice
$\mlsLat$. 

\theohsprecision*

\textit{Proof}.

The proof goes by structural induction on commands.
The conditional case explicitly assumes that the modified variables analysis is
precise enough, to enable the simulation of the program counter. 
This can be achieved by collecting variable names in a while language.

1 -- Case $\skipcom$ : this case stems from the premice.

2 -- Case $id := e$ :

Assume $\alphatohs(\depconsset_0) \dotmlslatleq \Delta_0$.

2.1 -- Case : $\varx \neq id$

 Then, for all $\varx \neq id$, $\Delta(\varx) = \Delta_0(\varx)$.
 
Also, 
$\alphatohs(\depconsset)(x) =  
\alphatohs(\depconsset_0)(x) \mlslatleq \Delta_0(\varx)
= \Delta(\varx)$.

2.2 -- Case : $\varx = id$

Otherwise, $\Delta = \Delta_0[id \mapsto \mlsl]$, where 
$\Delta_0 \vdash e : \mlsl$.

By \Cref{lem:expprecision}, since 
$\alphatohs(\depconsset_0) \dotmlslatleq \Delta_0$, we have
$\absobsexpdepl{e}{\depconsset_0} = \true$.

Thus, $\depcons{\mlsl}{id} \in \depconsset$ and :
\[ 
\alphatohs(\depconsset)(id)
\mlslatleq 
\Delta(id)
 \]

2.3 -- Finally, by Cases 2.2 and 2.3, we have :
\[ 
\alphatohs(\depconsset)
\dotmlslatleq \Delta
\]

3 -- Case $c_1;c_2$ :

This case proceeds by induction on both $c_1$ and $c_2$ by remarking that the
type system types both command $c_1$ and $c_2$ in $\mlsbot$.

\clearpage
4 -- Case $\ifcom~(b)~\thencom~c_1~\elsecom~c_2$:

Assume $\alphatohs(\depconsset_0) \dotmlslatleq \Delta_0$.
Let $\mlsl_b, \Delta_1,\Delta_2$ such that:
$\mlsl_b \vdash \Delta_0 \{ c_1 \} \Delta_1$ and 
$\mlsl_b \vdash \Delta_0 \{ c_2 \} \Delta_2$, with
$\Delta = \Delta_1 \dotmlslatjoin \Delta_2$.

Also, let $\Delta_1'$ and $\Delta_2'$ such that:
$\mlsbot \vdash \Delta_0 \{ c_1 \} \Delta_1'$ and 
$\mlsbot \vdash \Delta_0 \{ c_2 \} \Delta_2'$, with:

$\Delta_1 = \Delta_1'[id \mapsto \Delta_1'(id) \mlslatjoin \mlsl_b, \forall id \in
\Mod(c_1)]$, 
$\Delta_2 = \Delta_2'[id \mapsto \Delta_2'(id) \mlslatjoin \mlsl_b, 
\forall id \in \Mod(c_2)]$.

Intuitively, the program counter $pc$ can be simulated by a  modified variables
analysis that is precise enough. For a while language, this can be achieved
simply by collecting variable names.  

Let $\depconsset_1,\depconsset_2 \in \depLat$ such that:
$\seminshcdepabs{c_1}{\depconsset_0} = \depconsset_1$ and 
$\seminshcdepabs{c_2}{\depconsset_0} = \depconsset_2$.

Then, assuming $W = Mod(\ifcom~(b)~\thencom~c_1~\elsecom~c_2)$, we have:

$\depconsset = \bigcup\limits_{\mlsl \in \mlsLat}
\begin{cases}
\projtomlsl{\depconsset_1}
\deplatjoin   
\projtomlsl{\depconsset_2}
& \text{if } \absobsexpdepl{b}{\depconsset}  \\
\{ \depcons{\mlsl}{x} \in \projtomlsl{\depconsset} \mid \\
\quad \qquad x \notin W \} 
& \text{otherwise}
\end{cases}$

4.1 -- By induction on $c_1$, 
we have $\alphatohs(\depconsset_1) \dotmlslatleq \Delta_1'$.

4.2 -- By induction on $c_2$, 
we have $\alphatohs(\depconsset_2) \dotmlslatleq \Delta_2'$.

4.3 -- Assume $\varx \not \in W$, and prove $\alphatohs(\depconsset)(\varx)
\mlslatleq \Delta(\varx)$.

Since $\varx \not \in W$, we have  $\Delta(\varx) = \Delta_0(\varx)$.

Therefore,  
$\alphatohs(\depconsset_0) \dotmlslatleq \Delta_0 $ implies 
$\depcons{\Delta(\varx)}{\varx} \in \depconsset_0$. 

Thus, since $\varx \not \in W$, we have 
 $\depcons{\Delta(\varx)}{\varx} \in \depconsset_1$, and 
 $\depcons{\Delta(\varx)}{\varx} \in \depconsset_2$
(atomic constraints related to variables not explicitly written in $c_1$ are not
discarded from $\depconsset_0$, and likewise for those that are not explicitly
written in $c_2$). 
Thus, $\depcons{\Delta(\varx)}{\varx} \in \depconsset$, meaning that 
$\alphatohs(\depconsset)(\varx) \mlslatleq \Delta(\varx) $.

4.4 -- Assume $\varx \in W$ amd prove
 $\alphatohs(\depconsset)(\varx) \mlslatleq \Delta(\varx)$:

 We have $\mlsl_b \mlslatleq \Delta(x)$ since $x$ is explicitly written in one of the
 branches at least.

 Also, by 4.1 and 4.2, we have 
 $\alphatohs(\depconsset_1)(\varx) \mlslatleq \Delta_1'(x)$, and
 $\alphatohs(\depconsset_2)(\varx) \mlslatleq \Delta_2'(x)$.
Meaning that 
\[ \alphatohs(\depconsset_1 \deplatjoin \depconsset_2)(\varx) 
= \alphatohs(\depconsset_1)(\varx) \mlslatjoin
\alphatohs(\depconsset_2)(\varx)
\mlslatleq \Delta_1'(x) \mlslatjoin \Delta_2'(x) \mlslatleq \Delta(x) \]
Notice that $\Delta_1'$ and $\Delta_2'$ are well formed.  
Thus, exists $\mlsl_\varx $ such that $\mlsl_b \mlslatleq \mlsl_\varx$, such
that $\depcons{\mlsl_\varx}{\varx} \in \depconsset_1 \deplatjoin \depconsset_2$,
and $\mlsl_\varx \mlslatleq \Delta(x)$.

 And since $\forall \mlsl \in \mlsLat$, such that $\mlsl_b \mlslatleq \mlsl$, we
 also have $\absobsexpdep{b}{\mlsl}{\depconsset_0} = tt$ by using \Cref{lem:expprecision}
and $\depconsset_0$ is well formed.

Thus, $\forall \mlsl \in \mlsLat$, such that $\mlsl_b \mlslatleq \mlsl$,
$\projtomlsl{\depconsset} = 
\projtomlsl{\depconsset_1} \deplatjoin \projtomlsl{\depconsset_2} =
\projtomlsl{\depconsset_1 \deplatjoin \depconsset_2} $, i.e.\ 
$\depcons{\mlsl_\varx}{\varx} \in \depconsset$

Thus, 
$\alphatohs (\depconsset)(\varx) \mlslatleq \Delta(\varx)$.

5 -- Case : $\whilecom~(b)~\thencom~ c$

Assume $\alphatohs(\depconsset_0) \dotmlslatleq \Delta_0$.

The output type environment $\Delta$ is defined by:

\[ \Delta =
 \operatorname{lfp} \lambda \Delta_v.  \text{let } \Delta' 
 \text{ s.t. }
 \mlsbot \mlslatjoin \Delta_v(a) \vdash \Delta_v \{ c \} \Delta' 
 \text{ in } \Delta' \dotmlslatjoin \Delta_0
\]

Or written differently, $\Delta$ is given by:
\[ \Delta =
 \operatorname{lfp} \lambda \Delta_v. 
 \text{let } \Delta' \text{ s.t. }
 \mlsbot \vdash \Delta_v \{\ifcom~(b)~\thencom~c~\elsecom~\skipcom\} \Delta'
 \text{ in } \Delta'
 \dotmlslatjoin \Delta_0
\]

Let $(\Delta_n)$ be the sequence defined as 
\[
\Delta_{n+1} =  \text{let } \Delta' \text{ s.t. } 
 \mlsbot \vdash \Delta_n \{\ifcom~(b)~\thencom~c~\elsecom~\skipcom\} \Delta'
 \text{ in } \Delta'
 \dotmlslatjoin \Delta_0
\]

Also, let $(\depconsset_n)$ be the sequence defined as 
$\depconsset_{n+1} = \depconsset_0 \; \deplatjoin \;
\seminshcdepabs{\ifcom~(b)~\thencom~c~\elsecom~\skipcom}{\depconsset_n}$

Then, we prove by induction on $n$ that 
$\alphatohs(\depconsset_n)
 \mlslatleq
 \Delta_n
$.

5.1 -- Case n = 0.

This case holds by assumption  $\alphatohs(\depconsset_0) \dotmlslatleq \Delta_0$.

5.2 -- Case: Assume 
$\alphatohs(\depconsset_n) \mlslatleq \Delta_n$, and prove 
$\alphatohs(\depconsset_{n+1}) \mlslatleq \Delta_{n+1}$.

Let $\Delta'$  such that 
 $\mlsbot \vdash \Delta_n \{\ifcom~(b)~\thencom~c~\elsecom~\skipcom\} \Delta'$

By assumption, we have $\alphatohs(\depconsset_n) \mlslatleq \Delta_n$.
Thus, by using the same proof in Case 4, we have
\[ 
\alphatohs(\seminshcdepabs{\ifcom~(b)~\thencom~c~\elsecom~\skipcom}{\depconsset_n})
\dotmlslatleq
\Delta'
\]

Therefore, 

\[ 
\left(
\alphatohs(\seminshcdepabs{\ifcom~(b)~\thencom~c~\elsecom~\skipcom}{\depconsset_n}) 
\dotmlslatjoin
\alphatohs(\depconsset_0)
\right)
\dotmlslatleq
(
\Delta'
\dotmlslatjoin
\Delta_0
)
\]

Therefore, $\alphatohs(\depconsset_{n+1}) \mlslatleq \Delta_{n+1}$. which proves
that both least fixpoints are equal.

6 -- Finally, we conclude by Cases 1--5, and structural induction on commands.

\hfill \qed

\end{appendices}

\else
\fi
\end{document}

